\documentclass[12pt]{article}
\pdfoutput=1
\usepackage{setspace}

\usepackage{graphicx}
\usepackage{epstopdf}
\usepackage[body={17.5cm, 21cm},right=2cm]{geometry}
\usepackage{amssymb}
\usepackage{amsmath}
\usepackage{dcolumn}

\usepackage{graphicx}
\usepackage{epstopdf}
\usepackage{amsmath}
\usepackage{amsfonts}
\usepackage{amssymb}
\usepackage[usenames]{color}
\usepackage{mathrsfs}
\usepackage{array}

\usepackage{multirow}
\usepackage{psfrag}
\pdfoutput=1

\usepackage[colorlinks,bookmarks]{hyperref}
\definecolor{linkblue}{rgb}{0,0,0.8}
\definecolor{linkgreen}{rgb}{0,0.5,0}

\hypersetup{pdfpagemode=UseNone, pdfstartview=FitH, linkcolor=linkblue, %
            citecolor=linkgreen, urlcolor=linkblue}

\newcommand\nn{\nonumber}

\newcommand\eea{\end{eqnarray}}
\newcommand\bea{\begin{eqnarray}}

\def\beq{\begin{equation}}
\def\eeq{\end{equation}}
\def\d{\partial}

\def\d{\partial}
\def\l{\left(}
\def\r{\right)}

\newcommand{\be}{\begin{equation}}
\newcommand{\ee}{\end{equation}}
\newcommand{\ba}{\begin{align}}
\newcommand{\ea}{\end{align}}
\newcommand{\bg}{\begin{gather}}
\newcommand{\eg}{\end{gather}}
\newcommand{\bseq}{\begin{subequations}}
\newcommand{\eseq}{\end{subequations}}

\newcommand{\vk}{\vec{k}}
\newcommand{\vkp}{\vec{q}}
\newcommand{\vkkp}{\vec{k}-\vec{q}}
\newcommand{\vq}{\vec{q}}
\newcommand{\vp}{\vec{p}}

\newcommand{\vx}{\vec{x}}
\newcommand{\knl}{k_{\rm NL}}

\newcommand{\vkdq}{\vk\cdot\vq}

\def\H{{\cal H}}
\newcommand{\hinvMpc}{h\,$Mpc$^{-1}}
\newcommand{\ktr}{k_{\rm tr}}

 \newcommand{\tknl}{{\tilde k}_{\rm NL}}

\newcommand{\invMpc}{\,h\, {\rm Mpc}^{-1}\,}

\def\Omm{\Omega_{\rm m}}
\newcommand{\tvs}{\text{\textvisiblespace}}

\newcommand{\lp}{\left(}
\newcommand{\rp}{\right)}
\newcommand{\lb}{\left[}
\newcommand{\rb}{\right]}

\def\del3csc{\delta^{(3)}_{l,c_{\rm comb}}}
\def\co{c_{s  (1)}^2}
\def\ct{c_{s  (2)}^2}

\def\kmax{0.5 \invMpc}
\newcommand{\kren}{k_\text{ren}}
\newcommand{\gammai}{(\d\tau)_{\rho_l}}
\newcommand{\tildegammai}{\widetilde{(\d\tau)}_{\rho_l}}

\def\fnt{f_{\rm NT}}
\def\sec#1{Sec.~\ref{#1}}

\begin{document}

\vspace{5mm}
\vspace{0.5cm}
\begin{center}

\def\thefootnote{\fnsymbol{footnote}}

{\Large \bf The Effective Field Theory of Large Scale Structures\\[0.4cm]
 at Two Loops
}
\\[0.8cm]

{\large John Joseph M. Carrasco$^{1}$, Simon Foreman$^{1,2}$,\\[0.5cm] 
Daniel Green$^{1,2}$, and Leonardo Senatore$^{1,2,3}$}
\\[0.5cm]

{\normalsize { \sl $^{1}$ Stanford Institute for Theoretical Physics and Department of Physics, \\Stanford University, Stanford, CA 94306}}\\
\vspace{.3cm}

{\normalsize { \sl $^{2}$ Kavli Institute for Particle Astrophysics and Cosmology, \\ Stanford University and SLAC, Menlo Park, CA 94025}}\\
\vspace{.3cm}

{\normalsize { \sl $^{3}$ CERN, Theory Division, 1211 Geneva 23, Switzerland}}\\
\vspace{.3cm}

\end{center}

\vspace{.8cm}

\hrule \vspace{0.3cm}
{\small  \noindent \textbf{Abstract} \\[0.3cm]
\noindent 
Large scale structure surveys promise to be the next leading probe of cosmological information. It is therefore crucial to reliably predict their observables. The Effective Field Theory of Large Scale Structures~(EFTofLSS) provides a manifestly convergent perturbation theory for the weakly non-linear regime of dark matter, where correlation functions are computed in an expansion of the wavenumber $k$ of a mode over the wavenumber associated with the non-linear scale $\knl$. 
Since most of the information is contained at high wavenumbers, it is necessary to compute higher order corrections to correlation functions. After the one-loop correction to the matter power spectrum, we estimate that the next leading one  is the two-loop contribution, which we compute here. At this order in $k/\knl$, there is only one counterterm in the EFTofLSS that must be included, though this term contributes both at tree-level and in several one-loop diagrams. We also discuss correlation functions involving the velocity and momentum fields. We find that the EFTofLSS prediction at two loops matches to percent accuracy the non-linear matter power spectrum  at redshift zero up to $k\sim 0.6\,\invMpc$, requiring just one unknown coefficient that needs to be fit to observations. Given that Standard Perturbation Theory stops converging at redshift zero at $k\sim 0.1\,\invMpc$, our results demonstrate the possibility of accessing a factor of order 200 more  dark matter quasi-linear modes than naively expected. If the remaining observational challenges to accessing these modes can be addressed with similar success, our results show that there is tremendous potential  for large scale structure surveys to explore the primordial universe.

\noindent 

}
 \vspace{0.3cm}
\hrule
\def\thefootnote{\arabic{footnote}}
\setcounter{footnote}{0}

\vspace{.8cm}

\newpage
\tableofcontents
\newpage
\section{Introduction}

Large scale structures have the potential to become the leading cosmological probe for the early universe.  Our knowledge of the initial conditions scales as the cube of the maximum wavenumber that we can theoretically predict.  This is potentially a tremendous amount of data, and it is therefore essential to understand up to which wavenumber we can reliably predict.

On short scales, the so-called ultraviolet (UV), the density perturbations have become non-linear and have undergone gravitational collapse. A description of this regime likely necessitates the use of $N$-body numerical simulations, with analytical techniques providing guidance.  The situation is very different on large scales, the so-called infrared (IR), where the evolution is quasi-linear. In this regime, non-linearities are weak  and an analytical treatment must be possible. The advantages of an analytical treatment are multifold, from a better understanding of the physics, to a simpler control of the theoretical uncertainties, to potentially a gain in computational efficiency. 

The recently formulated Effective Field Theory of Large Scale Structures (EFTofLSS)~\cite{Baumann:2010tm,Carrasco:2012cv} is the theory that allows us to consistently make predictions for correlation functions at a certain wavenumber $k$ in an expansion in $\delta(k)=\delta\rho(k)/\rho$ or equivalently in powers of $k/\knl$, with $\knl$ defined as the wavenumber corresponding to the non-linear scale, which is indeed defined as the scale at which perturbation theory breaks down. Since we are interested in computing correlation functions for $\delta(k)\ll 1$ or equivalently $k\ll \knl$, this perturbation theory is manifestly convergent. The effective theory differs from the so-called standard perturbation theory (SPT), and relatives (see for example~\cite{Bernardeau:2001qr,Crocce:2005xy} or~\cite{Carlson:2009it}), by additional terms in the equations of motion for the overdensities that parameterize relevant short-distance physics. Approximately, these can be thought of as corrections to the fluid equations of motion in the form of speed of sound, viscosity, and stochastic pressure.  

In the EFT, higher-order corrections to correlation functions of observable quantities involve convolution integrals. These are named loop integrals due to their diagrammatic depiction as graphs with closed loops and whose vertices represent interactions. These integrals are performed formally with a cutoff $\Lambda$ on the maximum wavenumber in the convolution, and the additional parameters introduced in the EFT have the role of canceling the cutoff dependence for physical observables as well as introducing a finite $\Lambda$-independent contribution. This finite contribution encapsulates relevant effects of the underlying microphysical UV theory. After this step, the theory is said to be renormalized, and it is only at this point that the expansion in $k/\knl$ is manifest~\cite{Carrasco:2012cv}. All symmetries in the theory, including those potentially broken by a finite $\Lambda$, are restored in the $\Lambda\to\infty$ limit.  This limit is taken for all calculations performed in this paper.

The EFTofLSS was developed and applied to simulation data in~\cite{Carrasco:2012cv}, showing explicitly how the additional terms  of the EFT remove the cutoff dependence of the loops and how the theory is renormalized. The EFT parameters arise as the decoupling of the microscopic UV physics, which is integrated out. In practice, this formal integration is carried out either by measuring the relevant parameters from observed data, or by extracting them from $N$-body simulations~\footnote{The process of integrating out the UV physics cannot  be done with known analytic techniques because the UV physics is strongly non-linear. In the particle physics context, this is similar to what happens for the QCD Chiral Lagrangian, and it is to be contrasted with what happens for the standard Grand Unified Theories.}. Both approaches were developed in~\cite{Carrasco:2012cv}, where it was shown that the one-loop prediction of the EFTofLSS is in agreement with the dark matter power spectrum to within a percent up to the relatively high scale of $k\sim 0.24 \invMpc$~\footnote{We will see that in this paper the one-loop calculation will agree with the non-linear data up to $k\sim 0.35 \invMpc$. This difference is due only to a difference in the numerical non-linear data that are used in~\cite{Carrasco:2012cv} and here. We take this as a further confirmation of the fact that it is very good to have as much analytical control as possible, which is the point of the EFTofLSS.} { at $z=0$}.   

The solid theoretical setting, the remarkable agreement with observations, the improvement with respect to other currently available techniques (see for example~\cite{Carlson:2009it}), and the self-consistency of the different methods of extracting the parameters of the EFT, give very strong evidence that the EFT is the correct language to make theoretical predictions for Large Scale Structures~(LSS). This becomes even more evident when one tries to make predictions for other toy, `scaling', universes where the initial power spectrum follows a simple power law. In this setup, all techniques other than the EFTofLSS are unable to make predictions, while the EFT can~\cite{senatoretalk,Pajer:2013jj}. By using the renormalization techniques developed in~\cite{Carrasco:2012cv}, Ref.~\cite{Pajer:2013jj} has explicitly verified that indeed the EFTofLSS is able to make predictions for the scaling universes.

From the perspective of the EFTofLSS, the previous approaches like SPT and relatives (see for example~\cite{Bernardeau:2001qr,Crocce:2005xy} or~\cite{Carlson:2009it}) are missing important terms in the equations of motion.  Once these terms are included, perturbation theory may converge for much larger $k$ than previously believed possible.  Given the huge  importance of increasing the window of modes over which we can reliably predict observables, and given the encouraging results of~\cite{Carrasco:2012cv}, we find ourselves well motivated to perform calculations beyond one loop. 

The order in $k/\knl$ at which the various loops and counterterms contribute depends ultimately on the cosmological parameters and the linear power spectrum $P_{11}$ of the universe. By approximating it as a piecewise power law, we are able to reliably estimate the order of the contribution for each $k$-mode. For modes in the range of interest, $k\sim 0.1-0.6\,\invMpc$, the leading contribution after the one-loop term and the counterterm introduced in~\cite{Carrasco:2012cv} is the two-loop term, with no additional counterterms.  In other words, up to two loops in the EFT we find that only one parameter needs be measured: the one associated with the one-loop counterterm already discussed and measured in~\cite{Carrasco:2012cv}.

We compute the two-loop power contribution using the recently developed IR-safe integrand of~\cite{Carrasco:2013sva}, which allows us to compute the two-loop integrals with the necessary precision given finite computational resources. As stressed in~\cite{Carrasco:2013sva}, this is particularly important because the leading (in $k/\knl$) contribution at two loops is degenerate with the one-loop EFT parameter, and is therefore un-calculable, or, equivalently, uninteresting. One therefore needs to extract the subleading contribution, which requires higher numerical precision.

Even though no additional counterterms are required at two-loop order, there are important subtleties. First, the one-loop counterterm contributes to some relevant new tree level and one-loop diagrams that must to be included to perform the two-loop calculation in  a consistent way.

Second, there is a subtlety that concerns the two-loop diagrams. In the current universe, $k^3P_{11}$ becomes quite flat above $k\simeq 3\,\invMpc$, making loop integrals convergent in the UV, and effectively acting as an artificial cutoff. We stress that this is an artificial cutoff because wavenumbers of order $3\,\invMpc$ are believed to be well inside the non-linear regime, where perturbation theory, and therefore loop corrections, are not physically meaningful. The fact that loop integrals are convergent or divergent in the UV should not make any qualitative difference in the non-linear regime, the only difference being in the actual numerical value of some coupling constant.  This is what in quantum field theory is referred to as decoupling. However, this is a property of all physics, and not just of quantum field theory~\footnote{A demystifying example from day-to-day life can help. It is familiar and quite intuitive that pressure waves in various wooden tables will propagate qualitatively in the same way as each other, even as we note the different types of wood the various tables might be made of. The molecular structure of the wood only affects a few constants in the wave equations of motion, such as the velocity of propagation of the waves, the damping rate, etc., but not the qualitative wave behavior. This is what goes under the name of decoupling.}. The presence of this artificial cutoff at $\sim 3\,\invMpc$ introduces some spurious cutoff dependence in our two-loop result that we must remove with a two-loop contribution to the EFT parameter relevant to one loop.  Conveniently, this contribution can be determined simply by requiring that the two-loop contribution be minimal in the $k$-range already well predicted by one loop.  The fact that this requirement can be made makes it very clear that this contribution to the coefficient of the counterterm is a spurious artifact of the integration,  and does not encapsulate any previously undetermined microphysics. Indeed, this phenomenon is again familiar from the renormalization of coupling constants in quantum field theory: when performing a two-loop calculation, the one-loop counterterm changes accordingly so that the renormalized coupling constant does not change.

Third, there is an important conceptual point that we introduce in this paper. In deriving any EFT, it is important to have a hierarchy of scales between the UV physics one is integrating out and the long wavelength physics one is interested in describing. In the limit in which this hierarchy of scales is very large, the EFT is local both in space and time. In the case in which the UV theory is described by non-relativistic collisionless dark matter particles, as in our universe, the situation is a bit peculiar. Since dark matter particles have travelled non-relativistically for a finite amount of time, of order Hubble, the resulting travelled distance is of the order of the non-linear scale. This creates an hierarchy of spatial scales: UV physics is relevant for scales of the order of and shorter than the non-linear scale, while the EFT is valid at spatial distances much longer than this. The EFT is therefore {\it local in space}. On the other hand, the typical time scale of the non-linear modes is of order Hubble, which is the same time scale that controls the long wavelength modes. This means that, upon integrating out the UV modes, the resulting EFT is {\it non-local in time}: the counterterms, instead of being time-dependent numbers, are time-dependent kernels to be multiplied by sources and integrated over time. This leaves us with quite an unusual EFT, local in space, but non-local in time. This is not such a dramatic fact as it appears at first. First, our knowledge of the UV theory allows us to infer that all counterterms are proportional to roughly second derivatives of the gravitational Newtonian potential, a fact that limits the number of possible counterterms. Second, in an iterative solution, the fact that the linear solutions are $k$-independent allows us to parameterize our ignorance of the time-kernels with just a few numbers, as we will describe in detail. The EFT becomes an effective fluid only in the limit in which it is local in time, something that is not parametrically true. As we will find out though, it turns out that the local-in-time approximation for the kernels seems to be a very good numerical approximation. 
 
The bottom line result of our paper is very simple to state: the  EFTofLSS prediction at two loops matches to percent accuracy the non-linear power spectrum up to $k\sim 0.6\,\invMpc$, with just one parameter dependent upon the microphysics. Given that Standard Perturbation Theory stops converging at $k\sim 0.1\,\invMpc$, our results show that we can access a factor of order 220 more  dark matter quasi-linear modes than naively expected.  This represents a fantastic opportunity and a challenge for theorists and observers in the LSS community, as we explain in the concluding discussion. 

The bulk of the paper deals with explaining how this agreement between the EFTofLSS and non-linear data is achieved.  We will review in Sec.~\ref{sec:review} the applicability of perturbation theory to understanding the clustering of matter on large scales.  In Sec.~\ref{sec:scaling} we analyze the linear power spectrum as a piecewise scaling universe to estimate the  relevance of loop and counterterm corrections.  In \sec{sec:twoloop} we carry out the two-loop calculation, present a refinement of the physical meaning of the EFT parameters established in~\cite{Baumann:2010tm,Carrasco:2012cv} given the theory's expected non-locality in time, and compare the results of these calculations with simulated non-linear power spectra.    In \sec{sec:momentum} we present one-loop results for different observables: the momentum power spectrum and the matter-momentum cross spectrum, both of which are related to the matter power spectrum. We also discuss the velocity power spectrum. We close in \sec{sec:discussion} with a discussion of the results and the challenge they pose to the LSS community.  Namely, can the LSS community overcome the technical and conceptual barriers to realizing the promise of such a potential reach into the~UV?

\section{Review of Perturbation Theory in LSS}
\label{sec:review}

 In this section, we briefly review the perturbative approach to the clustering of matter on large scales.  We start from the assumption that dark matter is described by collisionless  dark matter particles described by a Boltzmann equation. By taking moments of the smoothed equation and expectation values over the short modes, one obtains equations of motion for the long-wavelength density and velocity fields, $\delta_l$ and $v_l$ respectively, sourced by an effective stress-energy tensor $\tau^{ij}$, which is just a function of the long-wavelength fields and that encodes the small-scale physics that has been ``integrated out" by the smoothing and averaging procedure. This gives a set of effective equations of motion for the long-wavelength field that goes under the name of EFTofLSS \cite{Baumann:2010tm,Carrasco:2012cv}. The effective stress-energy plays several important roles, and so it is vital that it be retained in the equations of motion. Standard perturbation theory (SPT, \cite{Bernardeau:2001qr})  is recovered in the limit in which this effective stress tensor is neglected. This is however inconsistent as a matter of principle, as can be shown for particular initial conditions for the universe where the initial power spectrum is a simple power law~\cite{senatoretalk,Pajer:2013jj}. It is a quantitative question how much they are important for the specific initial conditions that describe our universe. We will argue that they are quantitatively important in the current universe.

These equations are given by \cite{Carrasco:2012cv} as
\bea
\label{eq:all_equations1}
&&\nabla^2\phi_l=\frac{3}{2}{H}_0^2 \, \Omm \frac{a_0^3}{a}\delta_l \ ,\\ \nonumber
&&\dot\delta_l=-\frac{1}{a\rho_b}\d_i\pi^i_l\ ,\\ \nonumber
&&\dot{\pi}_l^i+4H \pi_l^i+\frac{1}{a} \d_j\left(\frac{\pi_l^j\pi_l^i}{\rho_b(1+\delta_l)}\right)+\frac{1}{a}\rho_b(1+\delta_l)\d^i\phi_l=
-\frac{1}{a} \partial_j \tau^{ij} \ ,
\eea
where $\phi_l$ is the gravitational potential sourced by the smoothed density, $\dot{\tvs}=d\tvs/dt$, $H=\dot{a}/a$, $\Omm$ is the present-day matter fraction, and $a_0$ is the present-day scale factor. $\delta_l$ and $\pi_l$ represent the matter overdensity and momentum fields. 

These equations need to be solved with some initial conditions. Since in the universe non-linearities grow with time, the initial conditions can be taken as linear:
\bea\label{eq:initial}
&&\langle\delta_{l}(t_{\rm in},\vec k)\delta_{l}(t_{\rm in},\vec k)\rangle=(2\pi)^3\delta_D(\vec k+\vec k') P_{11}(t_{\rm in},k)\ , \\ \nonumber
&&\pi_{l, \rm S}(t_{\rm in},k)=-a(t_{\rm in})\, \rho_b(t_{\rm in})\, \dot\delta_l(t_{\rm in},k)\ ,\qquad \pi_{l, \rm V}^i(t_{\rm in},k)=0\ ,
\eea
where all higher correlation functions are taken to vanish. Here we have defined $\pi_{l,\rm S} \equiv \partial_i \pi^i_l$ and $\pi_{l,\rm V}^i=\epsilon^{ijk}\d_j (\pi_l)_k$, and $P_{11}(t_{\rm in},k)$ is the linearly evolved dark matter power spectrum, evaluated at the initial time $(t_{\rm in})$. Truly, at the initial time the higher order correlation functions do not vanish. These are of two kinds: those coming from Inflation, and those coming from the non-linearities of general relativity and of plasma physics. The ones from Inflation are already sufficiently constrained that they are negligible for the power spectrum on which we are focussed in this paper. They can be trivially included in the calculation, but they are not relevant for the scope of this paper, which is to establish the predictivity of the EFTofLSS. The initial higher correlation functions associated with the non-linearities of the equations of motion give a contribution that is suppressed with respect to the ones that we include and that originate from the subsequent evolution by a factor of $D(t_{\rm in})^2/D(t_{\rm final})^2$, where $D$ is the linear growth factor (see~\cite{Fitzpatrick:2009ci} for an early study of how this affects the 3-point function of dark matter). By choosing the initial time early enough, say redshift $z=500$, we can make these sufficiently negligible.

As described in \cite{Carrasco:2012cv}, the overdensity $\delta_l$ and the momentum $\pi_l$ are the natural quantities that appear in the long wavelength theory after we integrate out the short modes and have finite correlation functions. However, one might notice an inconvenient fact in the above equations. By Taylor expanding the term of the form~$\d_i\left( \tfrac{\pi^i_l\pi^j_l}{1+\delta_l}\right)$, we obtain interaction terms with an arbitrarily high number of fluctuations. As it will become obvious later when we explain perturbation theory within the EFTofLSS, this implies that at each higher order, new vertices need to be included. 

This nuisance can be largely avoided by performing the following change of variables in the above equations. We can define a new variable called velocity $\vec v_l$ as
\be\label{eq:velocity_def}
v_l^i(\vec x,t)=\frac{\pi^i_l(\vec x,t)}{\rho_l(\vec x,t)}\ .
\ee
In terms of this new variable, the above equations read:
\bea
\label{eq:all_equations}
&&\nabla^2\phi_l=\frac{3}{2}{H}_0^2 \, \Omm \frac{a_0^3}{a}\delta_l \ ,\\ \nonumber
&&\dot\delta_l=-\frac{1}{a}\d_i\left([1+\delta_l] v_l^i\right)\ ,\\ \nonumber
&&\dot{v}_l^i+H v_l^i+\frac{1}{a} v_l^j\d_jv_l^i+\frac{1}{a}\d^i\phi_l=
-\frac{1}{a}\cdot\gammai{}^{i}
\eea
with the initial conditions properly adjusted from (\ref{eq:initial}), and where we have defined the operator
 \be\label{eq:stressdef2}
\gammai{}^{i}\equiv\frac{1}{\rho_l}\cdot \partial_j \tau^{ij}\ ,
 \ee
 which is obtained by dividing at the same spatial location the short distance stress tensor with the long-wavelength density. $\gammai$ can be thought of a sort of divergence of short-distance stress tensor per unit density. When it is clear from the context, for brevity's sake, we will sometimes refer to this as the short distance stress tensor, though strictly speaking $\tau_{ij}$ and $\gammai{}^{i}$ are different objects.

Ignoring the term proportional to $\gammai$, now the equations involve at most terms cubic in the fluctuations. As we explain later, this implies that the only new vertices that must be included when performing higher order calculations in the EFTofLSS are generated by the operator  $\gammai$, which is a relevant simplification for computational porpuses.

Notice however that this simplification does not come for free. The velocity field $\vec v_l$ as defined in (\ref{eq:velocity_def}) is not a physical (or renormalized) field. In the language familiar from quantum field theory, we can say that $\vec v_l$ is the {\it bare} velocity, to be distinguished from the {\it renormalized} velocity that we introduce next. This definition should be understood perturbatively by Taylor expanding in $\delta_l$ the factor $1/\rho_l$. In doing so, we realize that the definition of $\vec v_l$ involves products of the long wavelength fields $\delta_l$ and $\vec \pi_l$ at the same location. This makes $\vec v_l$ a composite, or contact, operator. This is a definition that is sensitive to the short distance physics~\footnote{For example, we could have defined a different $\tilde v_l^i$ by translating $\rho(\vec x,t)$  by a tiny amount:
\be\label{eq:velocity_def2}
v_l^i(\vec x,t)=\frac{\pi^i_l(\vec x,t)}{\rho_l(\vec x+\delta \vec x,t)}\ .
\ee
If $\delta \vec x$ is much shorter than the distances we care about, $\tilde v^i_l$ should be as good a definition as $v^i_l$.
}, and therefore it requires counterterms to make the UV contribution irrelevant to the correlations involving $\vec v_l$. We can therefore define a renormalized velocity field  $v^i_{l,R}$ schematically as 
\be\label{eq:velocity_def_ren}
v_{l,R}^i(\vec x,t)=v_l^i(\vec x,t)-\int \frac{d a'}{a' \H'} \,K_{\rm v}(a,a')\, \d^i\delta(a', \vec x_{\rm fl})+\ldots\ ,
\ee
where $K_{\rm v}$ is a free function dubbed counterterm, and where $\dots$ represents higher derivative terms (the notation will be precisely explained in Sec.~\ref{subsec:ct}). In the absence of counterterms, correlation functions of $\vec v_l$ receive uncontrolled and possible even infinitely large contributions from short distance physics. $K_{\rm v}$ and the possible higher derivative additional counterterms have the role of  removing this UV dependence in $\vec v_l$, so that correlation functions of $\vec v_{l,R}$ are finite and receive contributions only from modes up to the order of the external wavenumber~\footnote{One potentially confusing point of using the bare velocity instead of the renormalized velocity is that while non-linear corrections for $\vec v_{l,R}$ are small, this does not need to be the case for $\vec v_l$. For example,  in a universe where the power spectrum is of the form $P_{11}\sim k^n/\knl^{n+3}$, one obtains schematically:
\be
\langle v_l(t_{\rm in},\vec k)\,v_l(t_{\rm in},\vec k)\rangle_{\text{non-linear}}\sim(2\pi)^3\delta_D(\vec k+\vec k') \frac{k^2}{\knl^3}\left(\frac{\Lambda}{\knl}\right)^{2n+1} \frac{D(t_{\rm in})^4}{D(t_{\rm final})^4}\ .
\ee
where $\Lambda$ is the cutoff of the convolution integrals. The reason why this expression depends on $\Lambda$ is the fact that $\vec v_l$ is the bare, and not the renormalized, velocity field. If the cutoff were to be very large, then the non-linear correction could be very large and one could question why, as we do next, we can solve these equations and compute correlation functions of $\delta_l$, in an iterative expansion in small $\delta_l$ and $\vec v_l$. The reason why this is allowed can be expressed in a couple of different ways. On one hand, we could work directly with $\vec v_{l,R}$, for which, by definition, all correlation functions are small and the applicability of the Taylor expansion is manifest. Since working with $\vec v_{l,R}$ amounts to working by Taylor expanding also in $\vec v_{l}$ at the same order, the two procedures agree for correlation functions of $\delta$, which is the target of this paper. This procedure is shown in detail in Appendix~\ref{app:velocity}, after we have setup the relevant notation. Another way to justify the Taylor expansion in $\vec v_l$ is to notice that the non-linear corrections are small for small $\Lambda$. Since the final result for $\delta$ is $\Lambda$-independent, this means that the result can be extended to arbitrary large cutoff without needing any further correction.}.

From now on we drop the subscript $_l$ from the fluctuations, as all our expressions will just apply to the long-wavelength fields. It is convenient to work with the divergence and curl of the velocity, defined as $\theta\equiv\d_i v^i$ and $\omega_i\equiv \epsilon^{ijk}\d_j v_k$, and to decompose $v$ as
\beq
v^i = \tfrac{\d^i}{\d^2}\theta - \epsilon^{ijk}\tfrac{\d_j}{\d^2} \omega_k \ .
\eeq
The equations~(\ref{eq:all_equations}) can then be written as
\bea
\label{eq:new_eom}
\nn
&& a\H \delta'+\theta=-\delta \,\theta +\left[ - \tfrac{\d^i}{\d^2}\theta + 
\epsilon^{ijk}\tfrac{\d_j}{\d^2}\omega_k\right] \d_i\delta \ ,\\
\nn
&& a\H \theta'+\H \theta+\frac{3}{2} \H_0^2 \, \Omm  \frac{a_0^3}{a} \delta 
= - \left[ \d_i\tfrac{\d^j}{\d^2}\theta - \epsilon^{jkm}\d_i\tfrac{\d_k}{\d^2}\omega_m  \right]
\left[ \d_j\tfrac{\d^i}{\d^2}\theta - \epsilon^{ikm} \d_j \tfrac{\d_k}{\d^2}\omega_m \right] \\
\nn
&&\qquad\qquad\qquad\qquad\qquad\qquad
+ \left[ - \tfrac{\d^i}{\d^2}\theta  + 
\epsilon^{ijk} \tfrac{\d_j}{\d^2}\omega_k \right] \d_i\theta
- \d_i\,\gammai{}^i \ , \\
&& a\H \omega_i' + \H\omega_i =
\epsilon_{ijk} \d^j \lp \epsilon^{kmn} v_m \omega_n \rp
- \epsilon_{ijk}\d^j \gammai{}^k\ . 
\eea
The prime represents a derivative with respect to the scale factor~$a$:~${\tvs}'=d\tvs/d a$. The linear solution for $\omega_i$ decays with time like $a^{-1}$, so the initial vorticity will steadily decrease with time at this order, allowing $\omega_i$ to be be dominated by whatever is sourced by the non-linear evolution.  Note that in the absence of the stress tensor of the EFTofLSS, the vorticity is not generated at non-linear level. The leading source of the vorticity must come therefore from $\gammai$~\cite{Baumann:2010tm,Carrasco:2012cv}. Later in Sec.~\ref{sec:momentum} we will show that such generation of vorticity is negligible at the order at which we are working. This allows us to set $\omega^i=0$ for the rest of our calculations.

The equations in~(\ref{eq:new_eom}) then simplify to the following form in Fourier space:
\bea
\label{eq:master}
&&a\H \delta(a,\vec k)'+\theta(a,\vec k)= - \! \int \frac{d^3q}{(2\pi)^3}\alpha(\vkp,\vk-\vkp)\;\delta(a,\vk-\vkp)\theta(a,\vkp)\ , \\
\nonumber
&& a\H \theta(a,\vec k)'+\H \theta(a,\vec k)+\frac{3}{2} \H_0^2 \, \Omm  \frac{a_0^3}{a} \delta(a,\vec k)
= - \! \int \frac{d^3q}{(2\pi)^3}\beta(\vkp,\vkkp)\;\theta(\vk-\vkp)\theta(\vkp)
-i\, k_i\;\gammai{}^i(\vec k)\ ,
\eea
where
\be
\alpha(\vk,\vkp)=\frac{\left(\vk+\vkp\right)\cdot\vk}{k^2}\ ,\qquad\beta(\vk,\vkp)=\frac{\left(\vk+\vkp\right)^2 \vec k\cdot\vkp}{2 q^2 k^2}\ .
\ee

These equations are then solved perturbatively around the linear solutions for $\delta(a,\vk)$ and $\theta(a,\vk)$, with the $n$th-order solutions $\delta^{(n)}$ and $\theta^{(n)}$ expressible in terms of integrals over the linear solutions. Typically, the goal is to obtain expressions for different correlation functions of $\delta$ and $\theta$, and these are straightforwardly obtained from the above solutions. Beyond the linear level, these will involve integrations over momenta, which are analogous to loop corrections in quantum field theory and are used to organize calculations of correlation functions.

For example, in the absence of $\tau^{ij}$ (i.e.~in SPT), the matter power spectrum $P(k)$, defined by
\beq
\langle \delta(a,\vk) \delta(a,\vec{k}') \rangle = (2\pi)^3 \delta_{\rm D}(\vk+\vec{k}') P(k)
\eeq
is written as
\beq\label{equ:sptP}
P(k) = \underbrace{P_{11}(k)}_\text{linear}
+ \underbrace{P_{13}(k) + P_{22}(k)}_\text{one-loop corrections}
+ \underbrace{P_{51}(k) + P_{42}(k) + P_{33}^{\rm (I)}(k) + P_{33}^{\rm (II)}(k)}_\text{two-loop corrections}
+ \cdots\ ,
\eeq
where the time-dependence has been dropped for simplicity.
The linear solution $P_{11}$ is obtained separately by evolving the primordial spectrum of perturbations through cosmological time, as done by Boltzmann codes such as CAMB~\cite{Lewis:1999bs}. The one- and two-loop SPT formulas are summarized in Ref.~\cite{Carrasco:2013sva}, and also here in Appendix \ref{app:spt}. The EFTofLSS yields additional terms, which we will describe below. 

Finally, as proven in~\cite{Jain:1995kx,Scoccimarro:1995if}, in equal-time matter correlation functions, all IR divergences must cancel when we sum together all contributing diagrams, provided that the linear matter power spectrum~$P_{11}$ grows in the IR more slowly than $k^{n}$ with $n>-3$  (see \cite{Peloso:2013zw,Kehagias:2013yd} for a recent discussion).  To make these cancelations explicit, we will re-organize the terms in (\ref{equ:sptP}), as described in~\cite{Carrasco:2013sva}.

\section{Dimensional Analysis for the Real Universe}\label{sec:scaling}

We here describe in detail, following~\cite{Carrasco:2012cv,Pajer:2013jj,Carrasco:2013sva}, how we expect perturbation theory in the EFTofLSS to be organized when applied to the real universe.  

When we compute loop corrections to the power spectrum for a given external wavenumber $k$, we integrate over internal momenta, $q_i$.  If one computes the diagrams with an ultraviolet (UV) cutoff, $\Lambda$, and infrared (IR) cutoff, $k_{\rm min}$, each one is guaranteed to be finite.  However, numerically one will find that each term in the loop expansion is roughly of the same order and therefore it does not appear to be an expansion in a small parameter.  

At first sight, this lack of convergence would seem to undermine the very notion of a perturbation theory. One reason that higher loop corrections are not small is that they receive contributions from momenta much larger than the external momentum, $k\ll q_i$.  Nevertheless, even if the loop is divergent as $\Lambda \to \infty$, it does not mean that perturbation theory is breaking down. We can simply apply the procedure that is familiar in quantum field theory: we regularize the loops, for example by cutting off the momentum integrals, and add suitable counterterms to remove the cutoff dependence. The counterterms can be very large, even infinite. What counts for perturbation theory to be well defined is that the {\it sum of loop and counterterms} gives a small overall correction.

In the case of the EFTofLSS, the first counterterms we must add are associated with the Laplacian of the gravitational potential $\Phi.$ In the case the response of the short modes to the long mode was local in time, this would reduce to usual terms of speed of sound and viscosity present in imperfect fluids. 
One chooses these coefficients to cancel the contribution from loop integrals at large momenta.  After performing this procedure, renomalization, we are left only with the contribution from loop momenta of the same order as the external momenta. By dimensional analysis, this scales as some power of $k/\knl$ that grows with the number of loops.  Renormalization therefore ensures that as we increase the loop order, the $\Lambda$-independent piece that remains is suppressed by powers of $k / \knl$ relative to the previous loops.

In addition to cancelling divergences, these counterterms may also give a finite contribution at a given order in $k / \knl$.  These terms arise from physics on scales of order $\knl$ and smaller that we have ``integrated out."  Specifically, these terms give corrections that account for the fact that dark matter is not a pressureless fluid on small scales.  As a result, these coefficients are sensitive to the details of the dark matter on short distances and must be measured from simulations~\footnote{A peculiarity related to systems that virialize at short distances is that modes much shorter than the virialization scale do not contribute to renormalize the finite counterterms~\cite{Baumann:2010tm}. The usual statement of decoupling of short modes in EFT is that short distance physics contributes only to renormalizing some higher derivative terms. For virialized systems, we have a non-renormalization theorem, which is  a stronger statement: modes shorter than the virialization scale do not only decouple, but they do not even renormalize the higher derivative terms.}.  The number of such terms we must include will depend on the order in $k / \knl$ of the computation.

A second reason  for large loop integrals could be contributions from modes much longer that the external momentum: $q_i\ll k$. 
Some quantities, called IR-safe, remain finite loop order by loop order as $k_{\rm min}\to0$ when the matter power spectrum does not grow in the IR faster than $k^{-3}$.  For IR-safe quantities there is no issue  in principle~\footnote{In practice, it can be important to make manifest the IR-safety at the integrand-level~\cite{Carrasco:2013sva,Blas:2013bpa}}. 
One such example is the equal-time matter power spectrum.   

Other quantities may not strictly be IR-safe, and associated large IR contributions represent real physical effects that must be considered.   In some cases, it may be sufficient to resum  the   contributions of IR modes  from all loop orders. Such dynamics should be well described by the perfect fluid  equations, and several techniques have been developed to do this (see for example~\cite{Crocce:2005xy,Bernardeau:2011vy,Anselmi:2012cn,Blas:2013bpa}~\footnote{These resummations should not be confused with the breaking of the fluid equations in the UV and with the necessity of introducing counterterms, as in the EFTofLSS. The resummations apply to IR modes which are well described by the standard perturbative treatment.})

\subsection{The Real Universe as a Scaling Universe}\label{sec:fit}

Performing perturbative calculations for the real universe is complicated by the fact that the linear power spectrum, $P_{11}(k)$, has non-trivial $k$ dependence that does not permit an analytic treatment.  Fortunately, for the equal-time matter power spectrum at given $k$, we know that the contribution from IR modes $q_i \ll k$  must cancel between diagrams.  In addition, the contributions from UV modes, $q_i \gg k$, are degenerate with counterterms and can be ``integrated out".  As a result, to understand the structure of the perturbative expansion, we need only understand the behavior of $P_{11}(k)$ in the vicinity of the $k$ of interest.

It is easiest to illustrate the power of the EFTofLSS in the context of a scaling universe \cite{Pajer:2013jj,Carrasco:2013sva} where 
\beq
P_{11}(k)= (2\pi)^3 \frac{1}{\knl^3}\left(\frac{ k}{\knl} \right)^n  \ . 
\eeq
The advantage of such scaling universes is that their behavior at a given loop order can be determined by dimensional analysis and symmetries.  As we will show in the next subsection, for a given loop order $L$, the result scales as $(k / \knl)^{(3+ n)L} P_{11}$.  Therefore, given $\knl$ and $n$, each order in perturbation theory can be estimated reliably.

\begin{figure}[t]
\begin{center}
\includegraphics[width=.495 \textwidth]{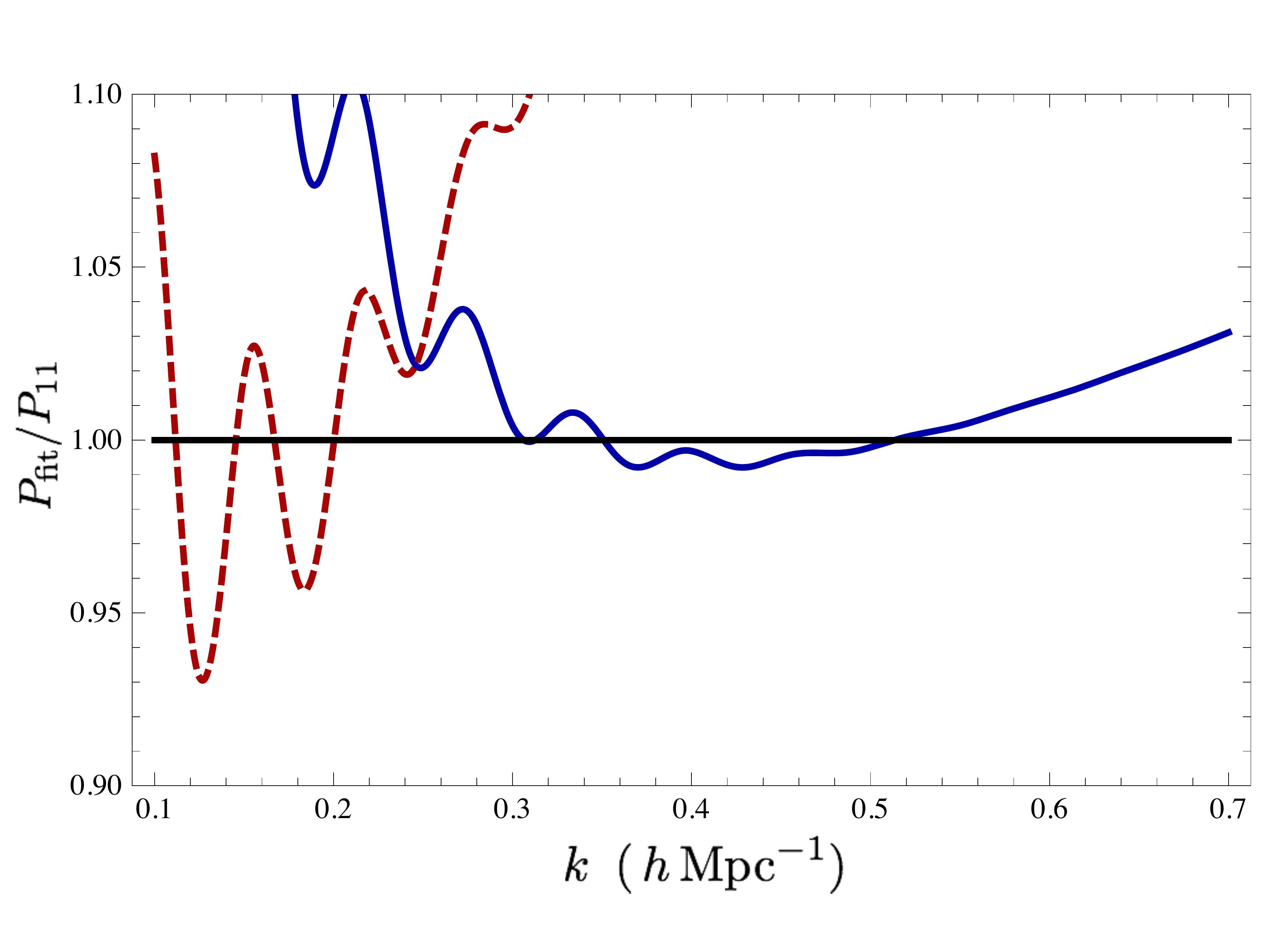}
\includegraphics[width=.495 \textwidth]{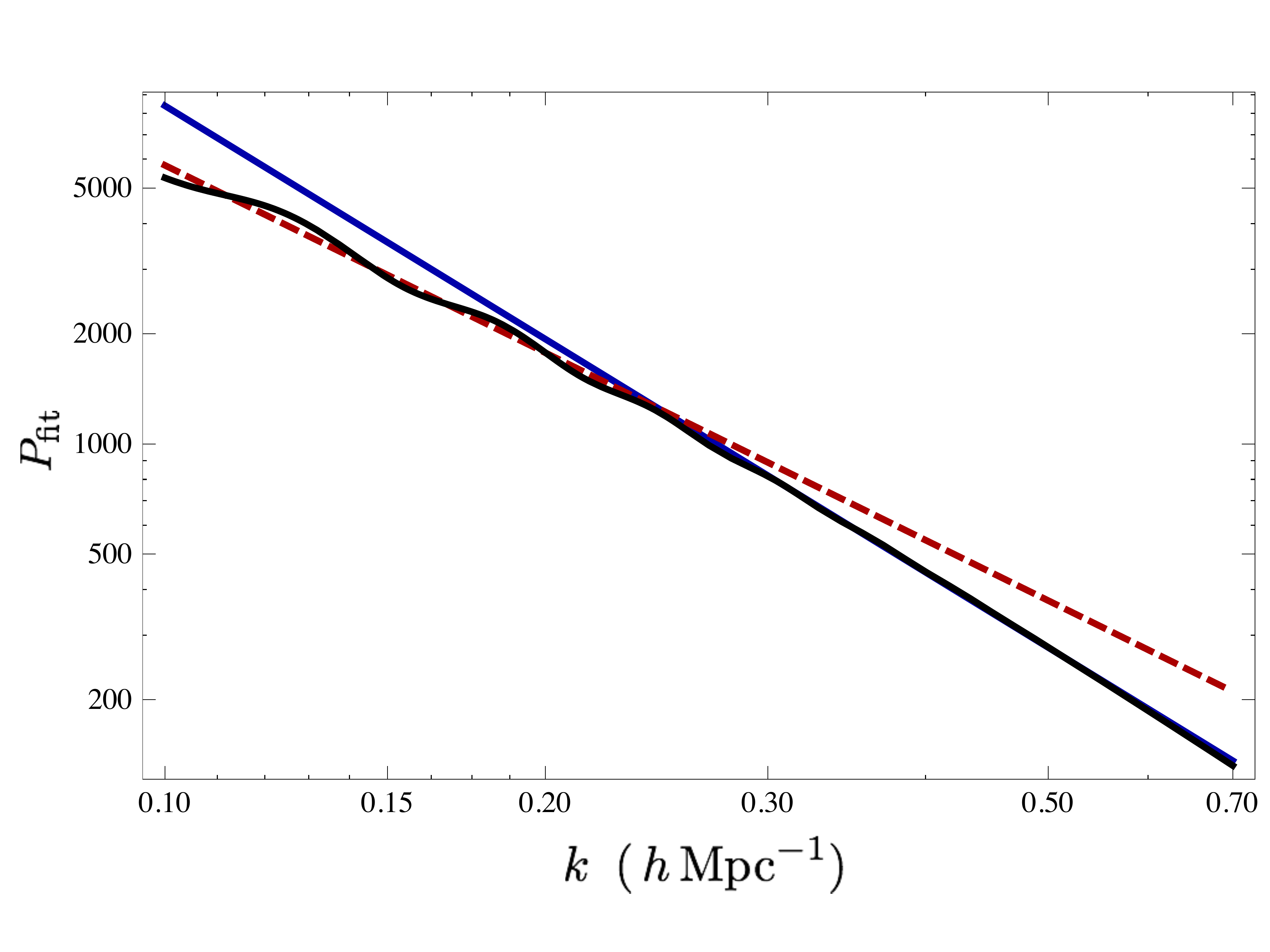}
\caption{ \label{fig:p11fit} \bf{Describing the linear power spectrum.}  \small\it We compare the linear power spectrum $P_{11}$  (black) and two fit scaling descriptions, $P_{\rm fit}\sim\left(\frac{k}{\knl} \right)^n$, with $n\approx-1.70$ using $\knl\approx1.80\hinvMpc$ (dashed, red), and $n\approx-2.12$ using $\knl\approx4.64 \hinvMpc$ (solid, blue). The left plot is shown as a ratio to $P_{11}$ while the right shows the absolute values on a log scale.  We see that a piecewise scaling description using both scales could be a useful approximation for power counting estimates. }
\end{center}
\end{figure}

In order to gain intuition for the calculation in the real universe, it is useful to approximate the linear power spectrum by a sequence of power laws. This approximation will miss important oscillationary features that are present in the matter power spectrum, the  Baryon Acoustic Oscillation, but it should be a good approximation to estimate the size of the various contributions. As our primary interest is pushing $k$ towards the non-linear scale { at $z\sim 0$}, we are interested in the behavior in the range $k \sim 0.1 - 1 \, h \, {\rm Mpc}^{-1}$.  By fitting power laws to the linear power spectrum of the real universe, we find that a useful approximation is
\bea\label{eq:fit}
P_{11}(k)=(2\pi)^3 \left\{ \begin{array}{ll} 
\frac{1}{\knl^3}\left(\frac{k}{\knl}\right)^{-2.1} & \text{ for} \ {k>\ktr} \ , \\
 \frac{1}{\tknl^3}\left(\frac{k}{\tknl}\right)^{-1.7}   & \text{for} \ {k<\ktr} \ ,
\end{array} \right.
\eea
where $\tknl = (\knl^{0.9} \ktr^{0.4})^{1/1.3}$ and $\ktr$ is the transition scale between the two different power-law behaviors.  The results of the fit are shown in Fig.~\ref{fig:p11fit} with the fit parameters
\beq\label{eq:knlest}
\knl = 4.6 \invMpc \qquad \ktr = 0.25 \invMpc \qquad \tknl = 1.8 \invMpc \ .
\eeq
The appearance~\footnote{Our $\knl$ values are large due to the normalization of the power spectrum by $(2 \pi)^3$, rather than the more conventional $2\pi^2$.  This normalization is motivated by the loop counting in the next subsection and by explicit calculations, like those in Appendix~\ref{app:loops}.} of such high scales suggests that the perturbative expansion may be reliable well beyond $k \sim 0.1 \, h \, {\rm Mpc}^{-1}$.

The rest of this section will be devoted to understanding the size of corrections in the real universe given the parameters from the above fit.  For the purpose of understanding the scaling of perturbations (Sec.~\ref{sec:loopscaling}), we will consider the scaling $n=-3/2$ to gain intuition.  However, when it comes to making estimates for the real universe (Sec.~\ref{sec:estimates}), we will focus on $n=-2.1$ for two reasons~\footnote{For some purposes, we could use $n=-2$ to estimate the $k / \knl$ scaling.  However, the best fit value of $\knl$ depends sensitively on $n$.  For example, the fit holding $n=-2$ fixed gives $\knl \sim 3.5 \invMpc$.  The difference between $n=-2.1$ and $n=-2$ is magnified when we raise $\knl$ to large powers.}.  First, from the fit in Fig.~\ref{fig:p11fit}, we see that $n=-2.1$ provides an extremely good fit to $P_{11}$ from $k=0.3 \invMpc$ to $k=0.7 \invMpc$.  Second, we will find that higher order corrections are expected to become important around $k = 0.5 \invMpc$, which is well inside the $n=-2.1$ scaling regime.

\subsection{EFTofLSS in the Scaling Universe}\label{sec:loopscaling}

Let us consider the contributions to the power spectrum from loop momenta satisfying $q_i \gg k$. Depending on the slope of the power spectrum $n$ around a certain wave number, any given loop integral will be either UV convergent or UV divergent. If we take a diagram where $L$ represents the number of loops, $I$ the number of internal lines, and $V$ the number of vertices touching loops~\footnote{ Note that this description does not directly apply to the diagrams in Fig.~\ref{fig:reducible}, which are drawn in a slightly different way. In the above Eq.~(\ref{eq:naivediv}), each vertex is attached to one side of a correlation function. This leads to the various SPT vertices having different kernels and different number of legs. In the way we plot in~Fig.~\ref{fig:reducible}, each SPT vertex is only a cubic vertex, and each line is either a correlation function or a Green's function. The two descriptions are simply related.}, then the naive degree of divergence $D$ is given~\cite{Goroff:1986ep} by
\be\label{eq:naivediv}
D= 3\,L+n I-V \ .
\ee
This formula is easy to understand. The factor of $3L$ counts the contribution from the phase space integrals. The factor of $n I$ accounts for the contribution of the factors of linear power spectra. Finally, the factor of $V$ comes from a decoupling property of the vertices that goes to zero at least as one inverse power of an internal momenta at high internal momenta. Depending on possible additional cancellations, the actual degree of divergence of a diagram can be actually less than $D$. Only for $n\leq-3$ are loops guaranteed to never diverge in the UV~\cite{Goroff:1986ep}. This is so because there is clearly one internal line for each loop, and so $D\leq-V<0$~\footnote{ In the current universe, at high enough wave numbers, $k \gtrsim 3 \,  h \, {\rm Mpc}^{-1}$, the power spectrum approaches $n=-3$. However, this happens at such high wave numbers that the description of dark matter particles as a generalized fluid no longer applies, not to mention the validity of perturbation theory. This is why, even though loops in SPT for the true-universe $P_{11}$ are formally convergent, SPT  does not converge to the true-universe answer. Instead, SPT converges to the resummation of the series of the perfect fluid diagrams. This is not the non-linear solution for the perfect fluid equations, as non-perturbative effects are being neglected; it is furthermore not the solution for the true universe, as dark matter particles do not constitute a perfect fluid.  }. To describe the true universe, it is essential to include counterterms in the fluid equations, and renormalize the theory. This is what the EFTofLSS is all about.

Before proceeding to the general case, let us study the special case of $n=-3/2$.  In this case, Eq.~(\ref{eq:naivediv}) tells us that the two-loop diagram $P_{51}$ is UV divergent, while the other two-loop diagrams are UV convergent. If we regularize it with a cutoff $\Lambda$, dimensional analysis allows us to conclude that, after integration, $P^{\text{I}}_{\text{2-loop}}$ will take the form
\bea\label{eq:p51_div3} \nonumber
&& P^{\text{I}}_{\text{2-loop}}=(2\pi) \left[c^\Lambda_{0} \left(\frac{\Lambda}{\knl}\right)^{2}\left(\frac{k}{\knl}\right)^{1}P_{11}+c^\Lambda_{1} \left(\frac{\Lambda}{\knl}\right)^{1}\left(\frac{k}{\knl}\right)^{2}P_{11}\right. \\ \nonumber
&&\quad\qquad\qquad\qquad +c^{\Lambda}_2 \log \!\left(\frac{k}{\Lambda}\right) \left(\frac{k}{\knl}\right)^{3}P_{11}+c^{{\text{finite}}}_1  \left(\frac{k}{\knl}\right)^{3}P_{11}\\  
&& \left.\qquad\qquad\qquad \quad+ c^{1/\Lambda}_{1} \left(\frac{k}{\Lambda}\right)^{1}\left(\frac{k}{\knl}\right)^{3}P_{11}+{\text{subleading finite terms in }} \frac{k}{\Lambda}\right] \ . 
\eea
where all the coefficients $c^{\dots}{}_{\dots}$ are expected to be numbers of order one.  In the above formula, we have used the fact that the diagrams $P_{42}^{\text{I}}$ and $P_{33}^{\text{I}}$ are not divergent for $n=-3/2$, so the divergent terms come only from $P_{51}^{\text{I}}$ and are therefore proportional to $P_{11}$. The finite parts from  $P_{42}^{\text{I}}$ and $P_{33}^{\text{I}}$ are not proportional to $P_{11}$, but have the same scaling in terms of $k$ and $\knl$ as the term proportional to $c^{{\text{finite}}}_1$, so we have expressed it in that way for simplicity. Here the superscript I  of $P^{\text{I}}_{\ldots}$ (to be distinguished from the ``(I)" and ``(II)" superscripts of $P_{33}$, which denote two separate diagrams) refers to ``irreducible" diagrams, which means that they do not reduce to combinations of lower order diagrams. Roughly speaking, this means that all but one of the loop integrations are nested and are not independent.  Only one azimuthal angular integral is independent, explaining the overall factor of $(2\pi)$ in (\ref{eq:p51_div3}).  We will consider the ``reducible" diagrams\footnote{Here irreducible is not the same as one-particle irreducible (1PI).  Specifically, all non-1PI diagrams are reducible, by our definition, but some 1PI diagrams are also reducible.}  in the next subsection.

Each of the $\Lambda$ dependent terms above needs to have a counterterm that cancels the $\Lambda$ dependence. For example, the second term proportional to $c^\Lambda_{1}$ can be cancelled by the counterterm 
\be\label{eq:two-loop-counter}
P_{\text{2-loop counter}}= (2\pi) c_{\text{counter}}^\Lambda \left(\frac{\Lambda}{\knl}\right)\left(\frac{k}{\knl}\right)^2 P_{11}\ ,
\ee
which comes from the leading order response of the effective stress tensor to a long mode: $\tau_{ij}\supset\delta_{ij} c_s^2 \delta$~\footnote{This counterterm can be written as the usual speed of sound only in the limit in which the response time of the short distance physics to the long wavelength fluctuations is very short. As explained later, this difference is irrelevant if the counterterm is evaluated at tree level.}.
Notice that we multiplied by a factor of $(2 \pi)$ in order to make the scaling match $P^{\text{I}}_{\text{2-loop}}$ with $c_{\text{counter}}^\Lambda = {\cal O}(1)$.  By taking
\be
c_{\text{counter}}^\Lambda =-c^\Lambda_{1} + \delta c_{\text{counter}} \left(\frac{\knl}{\Lambda}\right) \ ,
\ee
with $c^\Lambda_{1} $ and $\delta c_{\text{counter}}$ being $\Lambda$-independent, $c^\Lambda_{1} $ can cancel the UV divergence while $\delta c_{\text{counter}}$ gives a finite contribution.  The first and third terms in (\ref{eq:p51_div3}), proportional to $c^\Lambda_{0}$ and $c^\Lambda_{2}$ respectively, cannot be cancelled by any counterterm available in the EFTofLSS, as such terms would violate the combination of rotation invariance and locality, which requires analyticity in Fourier space. It can be verified numerically that $c^\Lambda_{0} = c^\Lambda_{2} = 0$, as required~\cite{Carrasco:2013sva}. 

By taking $\Lambda$ sufficiently large, we can neglect the terms that depend on $\Lambda$ in a vanishing way as $\Lambda\to \infty$.
Therefore, we have
\be
P^{\text{I}}_{\text{2-loop}}+P_{\text{2-loop counter}}=(2\pi) \delta c_{\text{counter}} \left(\frac{k}{\knl}\right)^2 P_{11}+(2\pi) c^{{\text{finite}}}_1  \left(\frac{k}{\knl}\right)^{3}P_{11}
\ee
We see that even though $P_{51}$ is arbitrarily large, the sum of $P_{51}$ and its counterterm is finite. The divergent term in $P_{51}$  is reabsorbed by the counterterm, and so we identify it with the contribution of the counterterm. Notice that this part is {\it degenerate} with the counterterm, and there is no way to physically distinguish these two contributions. However, the part of $P_{\text{2-loop}}^{\rm I}$ proportional to $c^{{\text{finite}}}_1$ is not degenerate with a counterterm, and is therefore important to calculate. Computing this term is indeed the only reason why we need to compute the full loop. Following the standard jargon of EFT in particle physics, we can call it the ``calculable," or ``finite," part of $P_{\text{2-loop}}$:
\beq\label{eq:p51_div3_finite} 
P^{\text{I}}_{\text{2-loop finite}}=(2 \pi) c^{{\text{finite}}}_1  \left(\frac{k}{\knl}\right)^{3}P_{11}
\eeq
What we have just done is what in the context of particle physics is usually called regularization (for us putting a cutoff to the diagrams), and renormalization (for us adding a counterterm and taking $\Lambda\to\infty$). The most important point here is that the finite, or calculable, contribution of the two-loop diagram scales as $(k/\knl)^{3/2}$ and is smaller than the tree and one-loop contributions when $k \ll \knl$.

\vskip 8pt

One can repeat the exact same logic for general $n$ to determine both the divergent contributions and the remaining finite terms.  Specifically, the divergences are given by 

\bea\label{eq:div}\nonumber
&&P^{\text{I}}_{L\text{-loop diverg.}} =  (2\pi)^4 c^{L\text{-loop diverg.}} \frac{1}{\knl^3}\left(\frac{\Lambda}{\knl}\right)^{(3+n)L-2} \left(\frac{k}{\knl}\right)^2 \left(\frac{k}{\knl}\right)^n+ \text{subleading divergences}\\ \nonumber
&&\qquad\qquad\quad+(2\pi)^4 c^{L\text{-loop diverg.,stoch.}} \frac{1}{\knl^3}\left(\frac{\Lambda}{\knl}\right)^{(3+n)(L+1)-7} \left(\frac{k}{\knl}\right)^4
+ \text{subleading divergences}\ . \\
\eea
The terms in the first line are associated with the UV-divergences of diagrams similar to the one we have discussed so far, such as $P_{13}$ or $P_{51}$, and they are reabsorbed by counterterms originating from the response of the the UV stress tensor to the long wavelength fluctuations $\tau\sim c_s^2\delta+\ldots$.
The terms in the second line come from higher-loop contributions analogous to $P_{22}$. For high enough $n$ or $L$, they also diverge, and the counterterms absorbing their divergences are the stochastic terms of the stress tensor: $\langle\tau_{ij}(\vec k)\tau_{lm}(\vec k')\rangle\sim \delta_D(\vec k+\vec k') (c_s^2)^2\rho_b^2\left[1+\text{terms higher order in $k/\knl$}\right]$, which enters the equations of motion multiplied by four powers of $k/\knl$. The fact that this counterterm starts contributing at order $(k/\knl)^4$ is due to the fact that the correlation of the stress tensor of the UV modes is expected to be Poisson-like distributed~\cite{Carrasco:2012cv}. For canceling higher order divergences, these kinds of counterterms might mix, i.e.\ one could consider the response of the stochastic term to a long wavelength mode.  
The finite contribution at a given loop order is instead given by 
\beq\label{eq:finite}
P^{\text{I}}_{L\text{-loop finite}}= (2\pi)^4 c^{L\text{-loop finite}}\frac{1}{\knl^3}\left(\frac{k}{\knl}\right)^{(3+n)L} \left(\frac{k}{\knl}\right)^n+\text{subleading in } \frac{k}{\knl} \ .
\eeq

All the terms that depend on $\Lambda$, including $P_{L\text{-loop diverg.}}$, can be removed by adding counterterms in the equations of motion of $\delta$.   As in the above example, the leading divergence can always be removed by including the appropriate counterterm (e.g.~$\tau^{ij}\supset\delta^{ij} c_s^2 \delta$).
 In this sense, the only meaningful term is $P_{L\text{-loop finite}}$.

A new counterterm is required for any coefficient that depends on $\Lambda$.  By reversing this logic, we can then determine the possible $\Lambda$-dependence of a given order in SPT by comparing to the available counterterms.  The scaling of the counterterms is determined both by the number of loops at which they are evaluated, $L$, and the derivatives of the counterterm, $2 M$, where $M$ is a positive integer.  Therefore, the general scaling of the finite contribution of a given counterterm is given by
\be\label{eq:finitecounter}
P^{\text{I  }}_{ L, M} = (2\pi)^4 c^{\rm finite \, counterterm}_{L,M} \frac{1}{\knl^3}\left(\frac{k}{\knl}\right)^{(3+n)L}\left(\frac{k}{\knl} \right)^{2 M} \left(\frac{k}{\knl}\right)^n\ .
\ee
It should be emphasized that, because $M \geq 1$, the expansions in (\ref{eq:finite}) and (\ref{eq:finitecounter}) are not the same (for general $n$).  This is the origin of the predictive power of the EFTofLSS: terms (\ref{eq:finite}) that do not arise in (\ref{eq:finitecounter}) cannot be altered by short distance physics and are uniquely predicted by the EFT.  On the other hand, the coefficients appearing in (\ref{eq:finitecounter}) encapsulate relevant microphysics from the underlying UV theory and must always be measured from non-linear data, as their values are not predicted by the EFT.  

\subsection{Reducible Diagrams}

\begin{figure}[t]
\begin{center}
\includegraphics[width=0.7\textwidth]{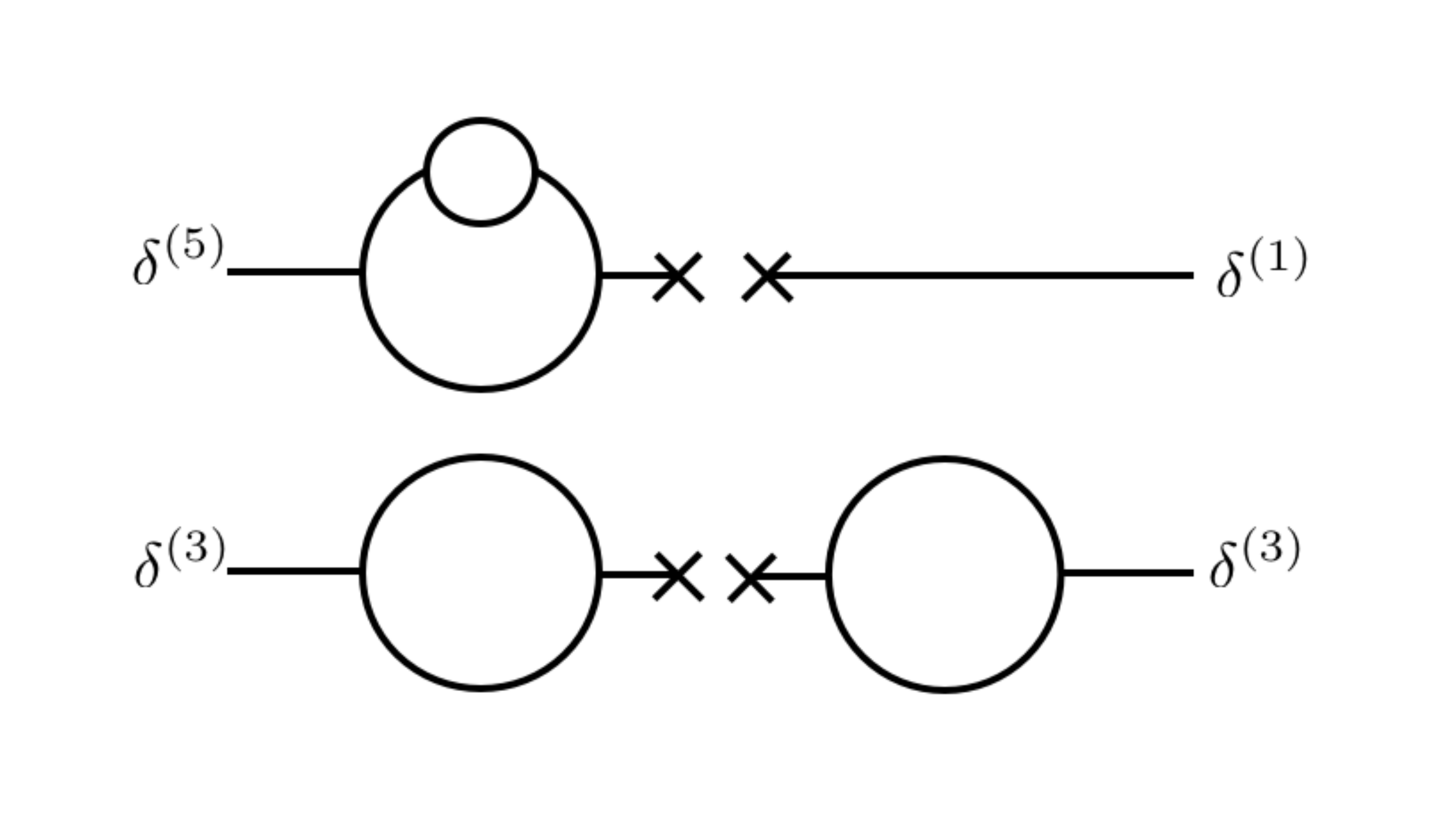}
\caption{\label{fig:reducible} \small\it {\bf Examples of reducible two-loop diagrams.}  The upper diagram is included in $P_{51}$ and is 1PI.  The lower diagram is included in $P_{33}$ and is not 1PI.  
}
\end{center}
\end{figure}

In the previous section, we focused on the irreducible contributions at a given loop order.  These are associated with diagrams that cannot be reduced to combinations of lower loop diagrams. In our way of organizing perturbative calculations, we do not compute just irreducible diagrams, but rather all diagrams at a given loop order.  At two loops there are important reducible diagrams.  Because the loop integrals of reducible diagrams can be independent, these contributions scale as
\beq
P_{\text{2-loop finite}}^{\text{R}} \sim (2\pi)^2 \left[ \left(\frac{k}{\knl}\right)^{(3+n)2} P_{11}(k) + \ldots \right] \ .
\eeq
These terms are enhanced relative to the diagrams by a factor of $(2\pi)$, due the the additional trivial integration.  Some examples of such diagrams are shown in Fig.~\ref{fig:reducible}.

Because of the extra factors of $(2 \pi)$, one might worry that as we compute higher loop diagrams such contributions overwhelm the irreducible contributions.  If true, it would suggest that perturbation theory breaks down at $k \sim \knl / (2\pi)^{1/(3+n)} \ll \knl$.  The diagrams with the most independent angular integrals are simply products of one-loop integrals, as shown in Fig.~\ref{fig:reducible}.  However, every loop level $L$ is combinatorially suppressed relative to linear due to time dependence, and these diagrams are further combinatorially suppressed relative to other contributions at that loop level due to the relatively small number of contractions that can conspire to generate these isolated loop-diagrams. It is these combinatorial factors which ensure that such contributions are negligible at high $L$ protecting our perturbation theory to $k \sim \knl$.

Another feature of having independent loop integrals is that a given diagram may be {\it more} divergent that the naive estimate made in (\ref{eq:naivediv}).  The reason is that there can be sub-diagrams that are divergent on their own, where (\ref{eq:naivediv}) assumed all the loop momenta diverge simultaneously.  These divergences are associated with the divergences of the lower loop diagrams.  As a result, the $\Lambda$ dependence must be canceled by the same counterterm that was introduced at that loop order.  No new counterterm is required.

By applying the logic in reverse, we see that when a counterterm is introduced at one loop it will also contribute at two loops like a reducible diagram.  We must therefore be careful to keep the one-loop counterterms explicit in higher loop calculations if they are relevant.  The enhancement by factors of $2\pi$ holds both for the reducible diagrams and the associated lower loop counterterms.  This feature will be crucial for understanding the real universe at two loops. 

\vskip 8pt
For the real universe, there are two lessons from the reducible diagrams. First, at low loop order, the reducible diagrams may be larger than the irreducible ones.  Second, counterterms introduced at lower orders in perturbation theory might have to be  included, with the same coefficient, at higher orders to account for the contributions from reducible diagrams.

\subsection{Estimating Corrections in the Real Universe}\label{sec:estimates}

In the previous subsections, we showed how to estimate the size of different contributions to the matter power spectrum in a scaling universe.  Since we will be calculating up to two loops, we need only consider the terms that are larger than the largest three-loop terms we are intentionally neglecting.  To estimate the size of various terms, we will use the approximate scaling universe behavior from Sec.~\ref{sec:fit}:
\beq
P_{11}(k) \sim \frac{(2\pi)^3}{\knl^{0.9} k^{2.1}} \ , \qquad \knl \sim 4.6 
{ \invMpc} \ .
\eeq
The first thing we would like to estimate is where the one and two-loop contributions become important.  At one loop this is straightforward because there is no subtlety regarding reducible diagrams and we find
\beq
\label{eq:est1Crossing}
\frac{P_{\text{1-loop finite}}}{P_{11}} \sim \left( \frac{k}{0.6 \invMpc} \right)^{0.9} \ .
\eeq
The above expression holds for $k> \ktr \sim 0.25 \invMpc$ and shows that at $k = \ktr$, we have $\frac{P_{\text{1-loop}}}{P_{11}} \sim 0.5$.  At two loops, the leading contribution is from reducible diagrams which scale  roughly as 
\beq
P_{\text{2-loop}}^{\text{R}} \sim \alpha(2 \pi)^2 \left(\frac{k}{\knl}\right)^{1.8} P_{11}\, ,
\eeq where 
\beq
\alpha \equiv \frac{P^{\text{finite}}_{13} }{ P^{\text{finite}}_{\text{1-loop}}}  \,.
\eeq
The $\alpha$-dependence is estimated using the fact that the reducible diagrams go as as products of $P_{13}$ integrals, yet we find it convenient to compare with $P_{\text{1-loop}}= (P_{13} + P_{22})$ scaling. If $P^{\text{finite}}_{13} \sim P^{\text{finite}}_{22}$, it would be reasonable to take $\alpha \sim 1/2$.  

Because $P_{11} \sim P_{\text{1-loop}}$ for $k>\ktr$ we should consider the correction to the total one-loop power spectrum
\beq
{ \frac{P_{\text{2-loop finite}}^{\text{R}}}{P_{11} + P_{\text{1-loop finite}}}|_{k =\ktr} \sim 0.14 \,  \alpha\ . }
\eeq
Therefore, for $k \simeq  \ktr$, we expect that two-loop diagrams are roughly a 10 percent correction to the power spectrum and are a larger correction for $k>\ktr$.

There are a number of counterterms that could also be relevant at the scales of interest. Since we will not be including any three-loop diagrams, we  should only include any counterterms that contribute more than $P_{\text{3-loop}}$ in the range of interest.  First we must determine the largest $k$ at which we can ignore the contributions from
\beq
P_{\text{3-loop finite}}^{\text{I}} \sim (2 \pi) \left( \frac{k}{\knl} \right)^{2.7} P_{11}(k) \quad \text{ and} \quad P_{\text{3-loop finite}}^{\text{R}} \sim 2 \frac{\tfrac{1}{2} P_{13} P_{\text{2-loop}}^{\text{I}} }{P_{11}}\sim \alpha (2\pi)P_{\text{3-loop finite}}^{\text{I}} \ .
\eeq
For $\alpha > (2\pi)^{-1}$, $P_{\text{3-loop}}^{\text{R}} > P_{\text{3-loop}}^{\text{I}}$.  We will assume this is the case and use $P_{\text{3-loop}}^{\text{R}}$ for our estimates.  By definition, $P^{\text{finite}}_{\text{3-loop}} < P^{\text{finite}}_{\text{2-loop}}$ for $k< \knl$.  However, this is not a sufficient condition to ensure that $P_{\text{3-loop}}$ is a negligible correction.  Instead, we should compare to $P_{\text{non-linear}} \sim P_{11} +P_{\text{1-loop finite}}+ 2(2\pi) \co (k/\knl)^2P_{11}+P^R_{\text{2-loop finite}}$  to find
\bea
\frac{P_{\text{3-loop finite}}^{\text{R}}}{P_{\text{non-linear}} }|_{k =0.4 \invMpc} \sim \alpha \, 0.02 \ ,
 \qquad \frac{P_{\text{3-loop finite}}^{\text{R}}}{P_{\text{non-linear}} }|_{k =0.5 \invMpc} \sim \alpha \, 0.03  \ .
\eea
In order to obtain these estimates, we have used the value of $\co\sim5.5$ that we obtain after fitting the non-linear power spectrum at one loop (see next section). This is consistent, as we are estimating higher order contributions and therefore we should use all what we know from lower orders. We conclude from these estimates that we should include any terms which are $>$ 3 percent corrections for $k \lesssim 0.4 - 0.5 \invMpc$.

Additional counterterms that we could potentially include have the following scaling:
\bea
P_{\text{counter}} &\sim& (2\pi) \co \frac{k^2}{\knl^2} (2 P_{11}(k)+2 P^{\text{finite}}_{\text{1-loop}}(k))+ ( 2 \pi \co )^2 \frac{k^4}{\knl^4} P_{11}(k) + 2 (2\pi) \ct \frac{k^2}{\knl^2} P_{11}(k)+ \nonumber \\
&& (2\pi) \co \frac{k^2}{\knl^2} P^{\text{finite}}_{\text{2-loop}}(k) + 2 \kappa \frac{k^2}{\knl^2} P^{\text{finite}}_{\text{1-loop}}(k) + 2 (2\pi) \lambda \frac{k^4}{\knl^4} P_{11}(k) + \ldots \ .
\eea
Here $k^2 P^{\text{finite}}_{\text{1-loop}}$ and $k^2 P^{\text{finite}}_{\text{2-loop}}$ schematically represent the scaling of the counterterms evaluated at higher loops, but the detailed form will be different.  We will perform the detailed calculations in the next section.  We have also introduced $\kappa \sim \lambda \sim {\cal O}(1)$ which are new counterterms that would need to be fit to the non-linear data. The factors of $2 \pi$ associated with $ \ct $, $\kappa$ and $\lambda$ are determined by the irreducible diagram to which they are associated.  Since both are new counterterms at two loops, they must have the same scaling as $P^{\text{I}}_{\text{2-loop}}$, rather than $P^{\text{R}}_{\text{2-loop}}$.  

Comparing these terms to $P_{\text{3-loop}}^{\text{R}}$, using $\co =5.5$ found in the next section, we have
\bea
\alpha \frac{4 \pi \co  \frac{k^2}{\knl^2}  P_{\text{1-loop finite}}(k)}{P_{\text{3-loop finite}}^{\text{R}}} 
	&\sim& \left( \frac{k}{3 \times 10^{-5} \invMpc}  \right)^{0.2}    
	\sim 6.8 \, \left( \frac{k}{0.5 \invMpc}  \right)^{0.2}  \\
\alpha \frac{(2 \pi \co)^2  \frac{k^4}{\knl^4}  P_{11}(k)}{P_{\text{3-loop finite}}^{\text{R}}} 
	&\sim&    \left( \frac{k}{0.35 \invMpc} \right)^{1.3} 
	\sim 1.5 \left( \frac{k}{0.5 \invMpc}  \right)^{1.3} \\
\alpha \frac{4 \pi \co  \frac{k^2}{\knl^2}  P_{\text{2-loop finite}}(k)}{P_{\text{3-loop finite}}^{\text{R}}} 
	&\sim&    \left( \frac{k}{0.53 \invMpc} \right)^{1.1} 
	\sim 0.92 \left( \frac{k}{0.5 \invMpc}  \right)^{1.1}\\
\alpha \frac{ 2 \kappa \frac{k^2}{\knl^2}  P_{\text{1-loop finite}}}{P_{\text{3-loop finite}}^{\text{R}}} 
	&\sim&    \kappa \left( \frac{k}{1600 \invMpc}  \right)^{0.2} 
	\sim 0.20 \, \kappa \, \left( \frac{k}{0.5 \invMpc}  \right)^{0.2}  \\
\alpha \frac{ 4 \pi \lambda \frac{k^4}{\knl^4}  P_{11}(k)}{P_{\text{3-loop finite}}^{\text{R}}} 
	&\sim&   \lambda \, \left( \frac{k}{11 \invMpc}  \right)^{1.3} 
	\sim 0.018 \, \lambda \, \left( \frac{k}{0.5 \invMpc}  \right)^{1.3}  \ .
\eea
We see that we should include the first two terms on the list, in addition to $ 2(2\pi) (\co+\ct) \frac{k^2}{\knl^2} P_{11}$.  Comparing these to $P_{\text{2-loop}}^{\text{R}}$ we have
\bea
{ \alpha  \frac{4 \pi \co  \frac{k^2}{\knl^2}  P_{\text{1-loop finite}}(k)}{P_{\text{2-loop finite}}^{\text{R}}}   }
	&\sim& \left( \frac{k}{0.53 \invMpc}  \right)^{1.1}   
	 \sim 0.92 \, \left( \frac{k}{0.5 \invMpc}  \right)^{1.1}  \\
 { \alpha \frac{(2 \pi \co)^2  \frac{k^4}{\knl^4}  P_{11}(k)}{P_{\text{2-loop finite}}^{\text{R}}}  }
 	&\sim&    \left( \frac{k}{1 \invMpc} \right)^{2.2} 
	\sim 0.21 \left( \frac{k}{0.5 \invMpc}  \right)^{2.2} \ .
\eea
Including only these counterterms, we should expect agreement to with a few percent up to $k \sim 0.5 \invMpc$ (depending somewhat on how small we can take $\alpha$).  

\subsection{Summary of Estimates for Scaling Universes}

In this section, we have modelled the linear power spectrum of the real universe as a piecewise-scaling power spectrum.  The goal was to use our knowledge of the scaling universe to separate the physical contributions at each loop order from those degenerate with corrections to the generalised fluid equations of motion.  This procedure of renormalization is crucial for achieving a convergent perturbative expansion and for understanding the predictive power of the EFTofLSS.

In the process, we obtained estimates for the range of $k$ where two-loop effects become important and where three-loop corrections become necessary.  These estimates suggest that the two-loop EFTofLSS is potentially reliable up to $k \sim 0.5 \, h \ {\rm Mpc}^{-1}$.  In order to achieve such accuracy, we will need to adjust the counterterm that scales as $(k/\knl)^2 P_{11}(k)$ and introduce the previously determined counterterms that scale as $(k/\knl)^2 P_{\text{1-loop}} (k)$ and $(k/\knl)^4 P_{11}$.

\section{The Matter Power Spectrum at Two Loops}\label{sec:twoloop}

Now that we have understood the scaling behavior of the SPT diagrams and the behavior of the counterterms up to three loops, we are prepared to calculate the two-loop contribution to the matter power spectrum at $z=0$ as calculated in the EFTofLSS. The basic strategy we will follow is to determine the parameters of the EFT by fitting to a non-linear power spectrum obtained from $N$-body simulations.

Based on the piecewise scaling model for $P_{11}(k)$, we established several results that will be crucial to the analysis in this section:
\begin{itemize}
\item The tree-level counterterm, proportional to $\co$, that is determined by the one-loop power spectrum, must be included in one-loop diagrams.  These terms are comparable to the reducible two-loop diagrams.
\item We are required to adjust the coefficient of the tree level counterterm from $\co$ to $\co+\ct$, in order to cancel the leading divergence at two loops. 
\item In principle, we can determine $\ct$ without fitting to the non-linear data, since the finite part of the counterterm was determined by the one-loop measurement.   

\item The three-loop contribution should be negligible up to $k\sim \kmax$.
\item No additional counterterms, beyond the three mentioned above, should contribute up to $k\sim \kmax $. All the terms we include are all computable in terms of the single coefficient~$\co$.
\end{itemize}
Based on these observations, we should find agreement between the non-linear data and the two-loop EFT up to $k \sim 0.5 \invMpc$  by considering only the parameter $c_s^2$ (that receives a one-loop and two-loop contribution).

It is worth emphasizing that because $\ct$ is introduced to remove the UV-dependence from the two-loop integrals, we can actually determine $\ct$ without fitting to non-linear data.  It can be determined by simply subtracting the excess in $P_{\text{2-loop}}$ at scales where the finite terms are expected to be negligible.  Therefore, although we have two parameters that are not included in SPT, $\co$ and $\ct$, only $\co$ is determined by fitting the one-loop power spectrum to data and $\ct$ is determined from perturbation theory~\footnote{We could determine $\co$ explicitly by imposing, similarly to what is usually done in particle physics, that the one-loop prediction for the power spectrum equals the non-linear spectrum at some renormalization scale~$k_{\rm ren}$. Then, we could compute $\ct$ completely from perturbation theory, by imposing that the one- and two-loop predictions are equal at~$k_{\rm ren}$. This yields the following expression for $\ct$:

\be
\ct(\kren) = \frac{  P_\text{2-loop}(k_{\rm ren}) + (2\pi) \co(k_{\rm ren}) P_{\text{1-loop}}^{(c_{\rm s})}(k_{\rm ren}) }
{2(2\pi)(k_{\rm ren}^2/\knl^2)P_{11}(k_{\rm ren})} 
+{ \pi [\co(k_{\rm ren})]^2 \frac{k_{\rm ren}^2}{\knl^2} }\ .
\ee
In practice, when applying this procedure, one has to average over $k_{\rm ren}$ in order to remove some noise that we believe is due to our available non-linear data. When doing this, we obtain the same results as with the method described in the text.}. Thus, there is no need to fit $P_{\text{2-loop}}$ to the non-linear data to make predictions up to $k \sim \kmax$~\footnote{A similar analysis can be extended to higher redshifts. We leave this to future work.}.

\subsection{Deriving the counterterms}
\label{subsec:ct}

The counterterms we require for our one- and two-loop calculations of the power spectrum arise from the effective stress-energy, $\tau^{ij}$, or equivalently from the divergence of the stress tensor per unit mass $\gammai$, which we introduced in Sec.~\ref{sec:review}. By the equivalence principle, the short modes we have smoothed over can only be influenced by the long gravity modes via tidal effects, so we should write $\gammai$ as an expansion in powers and derivatives of $\partial_i \partial_j \phi$, where $\phi$ is the gravitational potential sourced by~$\delta$. Furthermore, the short modes can depend in a similar way on the gradient of the long velocity field $\d_iv^j$ (or equivalently one can use the long momentum field $\pi^j$), and on the overdensity $\delta$, though the dependence on $\delta$ can be removed exactly by using the Poisson equation to express $\delta$ as a local function of $\phi$~\cite{Baumann:2010tm,Carrasco:2012cv}. The usual subtleties that are discussed in Sec.~\ref{sec:review} about  the fact that the velocity field is a contact operator apply here unchanged. However, we should also be careful at what coordinate values we evaluate $\phi$ and $v^i$ when they appear in $\gammai$, for the following important reason. 

The EFTofLSS is local in space, but is non-local in time.  Specifically, we are integrating out modes with $k \gtrsim \knl$, but all these modes have slow time-dependence, on the order of a Hubble time.  As a result, we have integrated out physics that has long range correlations in time.  In this sense, the behavior of $\phi$ and $\vec v$ along a fluid element's entire path should influence its current state. This can be encoded by writing each term in $\gammai$ as a convolution with some (unknown) time-dependent kernel:
\beq
\label{eq:setensor}
\gammai{}^i =
\int d\tau' \kappa_1(\tau,\tau') \, \d^i \partial^2 \phi(\tau',  \vx_{\rm fl} )
+ \cdots,
\eeq
where the ellipses denote higher powers and derivatives of $\d_i \d_j \phi$ and of $\d_j v_i$~\footnote{As we will explain later, for the level of accuracy we reach in this paper, we will be interested only in the linear counterterm that is leading-order in derivatives, and we can neglect higher order ones.  Since, as will show in Sec.~\ref{sec:vorticity},  we can neglect vorticity, the velocity can appear in the linear counterterm only in a combination equivalent to $\gammai^i \supset \d^i (\d_j v^j)$. By the continuity equation, $\d_i v^i=-a\,\dot\delta-\d_i(\delta\, v^i)$, and therefore it is redundant.  In fact, at any order we can remove $\d_i v^i$ using the equations of motion.  So, the linear dependence in $\vec v$ is equivalent to a linear dependence in $\phi$ plus quadratic terms. We use this fact to neglect the velocity counterterm for the rest of the paper and just work with a linear counterterm in $\d_i\d^2\phi$.}. The partial derivatives are evaluated with respect to $\vx$ but the gravitational potential $\phi$ is evaluated along the path $\vx_{\rm fl}[\tau,\tau']$ of a fluid element, defined recursively by
\beq
\vx_{\rm fl}[\tau,\tau'] = \vx - \int_{\tau'}^{\tau} d\tau'' \vec{v}(\tau'',\vx_{\rm fl}[\tau,\tau''])\ .
\eeq
The specific form of $\vx_{\rm fl}$ ensures that the equations of motion are diffeomorphism-invariant\footnote{In the Newtonian limit, in which we are working, the diffeomorphisms  relevant for us manifest themselves as generalized galilean transformations, where we shift the $v^i(\vec{x},\tau) \to v^i(\vec{x},\tau) - u^i(\tau)$ and $\phi \to \phi + (\H u^i(\tau) + \partial_\tau u^i(\tau)) \cdot x^i$.  For a recent discussion, see \cite{Kehagias:2013yd}.}  and therefore that IR divergences cancel between diagrams for the equal time matter power spectrum. We use the conformal time $\tau$ because 
the expressions with $\tau$ rather than $t$ have cleaner behavior under diffeomorphism transformations. 

The equations of motion then become
\bea
\label{eq:master2}
\nonumber
&&a\H \delta'+\theta= - \! \int \frac{d^3q}{(2\pi)^3}\alpha(\vkp,\vk-\vkp)\delta(\vk-\vkp)\theta(\vkp)\ , \\
\nonumber
&&  a\H \theta'+\H \theta+\frac{3}{2} \H_0^2 \, \Omm  \frac{a_0^3}{a} \delta
= - \! \int \frac{d^3q}{(2\pi)^3}\beta(\vkp,\vkkp)\theta(\vk-\vkp)\theta(\vkp) \\
&&\qquad\qquad\qquad\qquad\qquad\qquad\quad
+ k^2 \int \frac{da'}{a'\H(a')}   \kappa_1(a,a')\; [ \d^2\phi(\tau',\vec x_{\rm fl})]_{\vec k}  
\eea
 where $\kappa_{1}(a,a')\equiv \kappa_{1}(\tau[a],\tau[a'])$ and where we use the labelled brackets $ [ f(\vec x_{\rm fl}(\vx))]_{\vec{k}}$ to mean that we take the Fourier transform of the given function $f$ and evaluate it at the momentum~$\vec k$: $[f(\vec x_{\rm fl}(\vx))]_{\vec{k}}\equiv\int d^3x\; e^{- i\vec x\cdot \vec k} f(\vec x_{\rm fl}(\vx))$.   
When these are solved for non-SPT terms at tree level, there is no subtlety with the non-locality in time incorporated in the $\kappa$ kernels, because our uncertainty regarding these kernels is the same as our uncertainty in $\co$ in the first place. However, at one loop, we must add up several diagrams with the same combination of kernels but different integrands, and therefore the relative coefficients of these diagrams are sensitive to the precise form of the $\kappa$'s.
For this paper, the term explicitly displayed in Eq.~(\ref{eq:setensor}) is sufficient to supply the necessary counterterms for a two-loop calculation of the matter power spectrum. At this order, we can therefore use the Poisson equation to rewrite $\d^2\phi$ in terms of $\delta$, transforming the second equation of motion into 
\bea
\nn
&& a\H \theta'+\H \theta+\frac{3}{2} \H_0^2 \, \Omm  \frac{a_0^3}{a} \delta
= - \! \int \frac{d^3q}{(2\pi)^3}\beta(\vkp,\vkkp)\theta(\vk-\vkp)\theta(\vkp) \\
&&\qquad\qquad\qquad\qquad\qquad\qquad\quad
+ k^2 \int \frac{da'}{a'\H(a')} K(a,a') [\delta(a',\vec x_{\rm fl}) ]_{\vk} \ ,
\eea
where
\beq
K(a,a') \equiv \frac{3}{2} H_0^2 \, \Omm \frac{a_0^3}{a'}\, \kappa_1(a,a') \ .
\eeq

We will now make some simplifying assumptions. These are not strictly necessary for us to proceed, but they simplify the algebra with little loss of numerical accuracy. First, we will use the approximation that the time evolution in terms shared with SPT is given by $\delta^{(n)}(a',\vk) = [D_1(a')/D_1(a)]^n \, \delta^{(n)}(a,\vk)$.   From here, we can reduce the dependence on $K(a,a')$ to a few functions of time:
\beq\label{eq:cndef}
c_n(a) \equiv \int \frac{da'}{a'\H(a')} K(a,a') \frac{D_1(a')^n}{D_1(a)^n} \ .  
\eeq
We now make the following ansatz for the time-dependence of each perturbative solution:
\bea
\label{eq:deltaexp}
\delta(a,\vk) &=& \sum_{n=1}^\infty \, [D_1(a)]^n \delta^{(n)}(\vk)
+ \sum_{n=1}^\infty \, [D_1(a)]^{n+2} \tilde{\delta}^{(n)}(\vk)\ , \\
\label{eq:thetaexp}
\theta(a,\vk) &=& -\H(a) f \sum_{n=1}^\infty \, [D_1(a)]^n \theta^{(n)}(\vk)
-\H(a) f \sum_{n=1}^\infty \, [D_1(a)]^{n+2} \tilde{\theta}^{(n)}(\vk)\ ,
\eea
where $\delta^{(n)}(\vk)$ and $\theta^{(n)}(\vk)$ are the standard SPT solutions, written in terms of kernels $F_n$ and $G_n$ (see, e.g.,~\cite{Bernardeau:2001qr}), while $\tilde{\delta}^{(n)}(\vk)$ and $\tilde{\theta}^{(n)}(\vk)$ are new solutions involving factors of $c_n(a)$, determined by similar relations to their SPT counterparts. 

We take
\bea
&&\langle\delta^{(1)}(\vec k)\delta^{(1)}(\vec k')\rangle=(2\pi)^3\delta_D(\vec k+\vec k') P_{11}(k) \ , \qquad \theta^{(1)}(a,\vec k)=- a{\cal H}\delta^{(1)}(a,\vec k)\ , \\ \nonumber
\eea
while $\delta^{(n>1)}$ and $\theta^{(n>1)}$ are taken to be proportional to $(\delta^{(1)})^n$~\footnote{This ansatz does not correctly implement the initial conditions in Eq.~(\ref{eq:initial}), by an amount proportional to the initial non-linearities of $\delta$ and $\theta$, which at late times are suppressed with respect to the non-linearities induced by the gravitational dynamics by a factor proportional to $D(t_{\rm in})/D(t_{\rm final})$ or by $D(t_{\rm in})/D(t_{\rm final})\cdot f_{\rm NL}$ for non-Gaussianities from inflation. Since Eq.~(\ref{eq:initial}) is not correct by the same amount, which is negligible, the mistake we make here is irrelevant. In case it was needed to implement correctly the initial conditions, this can be done by either assigning non-trivial correlation functions to $\delta^{(1)}$ or equivalently by assigning terms non-proportional to $(\delta^{(1)})^n$ to $\delta^{(n)}$. See~\cite{Fitzpatrick:2009ci} where this procedure was implemented to implement the effects from the early phase of radiation domination and Hubble re-entry.}.

In order to solve the equations of motion consistently at each order, we assume that 
\beq
\label{eq:cntdep}
c_n(a)  =  \bar c_n ( 9 D_1(a)^{2} \H^2 f^2)
\eeq
where $f \equiv \partial \log D_1/\partial \log a$ \footnote{Note that the assumption~(\ref{eq:cntdep}) is not fully physically justified, and rather is made out of convenience. To avoid making this assumption, one should in principle measure these parameters as a function of time. We defer this to future work.} and $\bar c_n$ are constants with units of $k^{-2}$. This also ensures that the time-dependence of the lowest-order non-SPT diagram has appropriate time-dependence to cancel the $\Lambda$-dependent part of $P_{13}$ at all times. The factor of 9 is inserted for later convenience.

\begin{figure}[t]
\begin{center}
\includegraphics[width=0.7\textwidth]{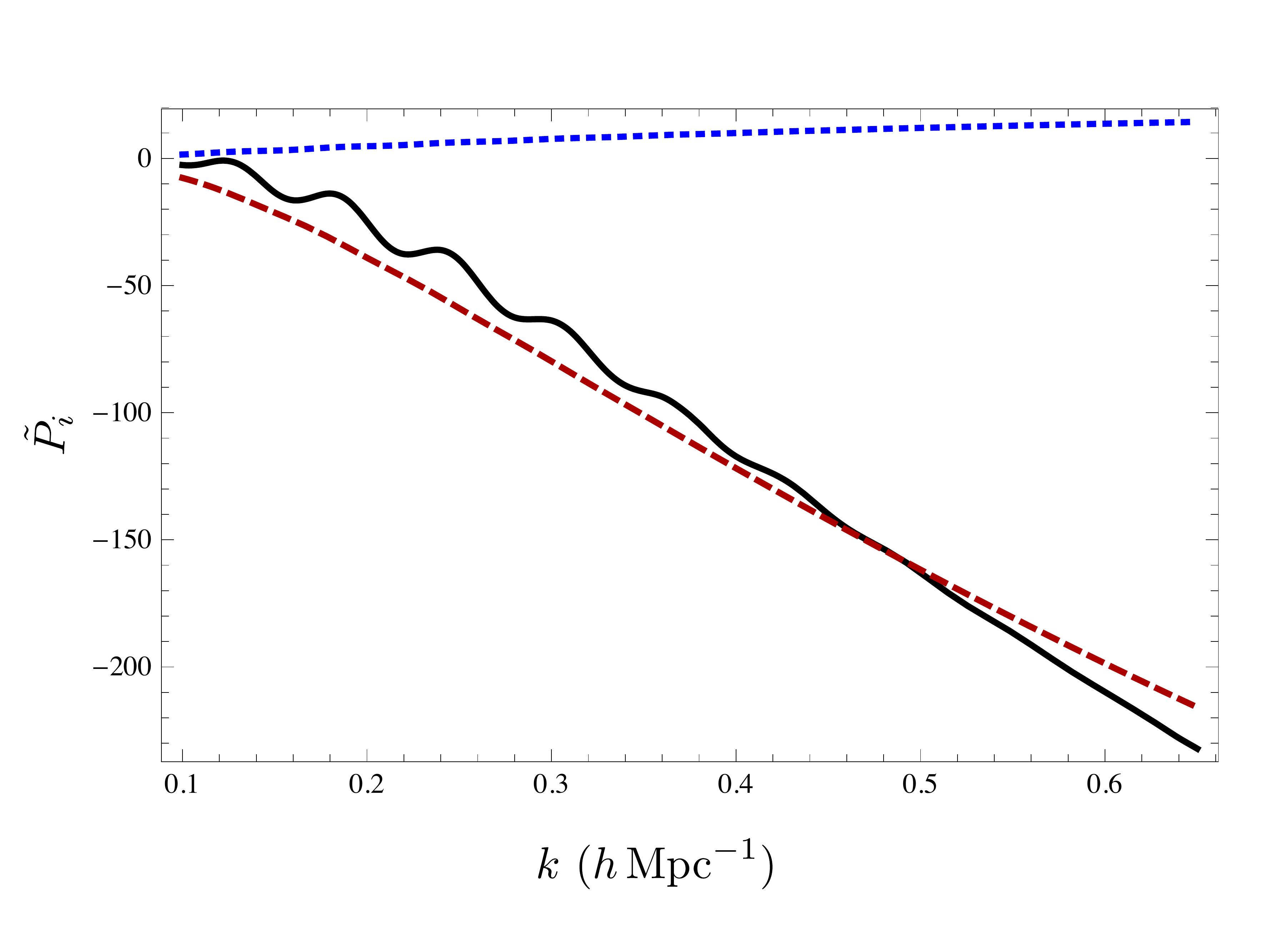}
\caption{\label{fig:funcs} {\bf Components of the one-loop EFT correction.} \small\it  There are three functions $\tilde{P}_{1,2,3}$ (given by Eqs.~(\ref{eq:ptilde1}) to~(\ref{eq:ptilde3}) in App.~\ref{app:solutions}) relevant to the one-loop perturbative correction to the matter power spectrum.   We plot these functions $\tilde{P}_{1,2,3}$ as solid (black), dashed (red) and dotted (blue) lines respectively.  Note that $\tilde{P}_3$ is subdominant for much of the range. 
}
\end{center}
\end{figure}

The full details of these solutions, as well as formulas for the one-loop diagrams needed in a two-loop power spectrum calculation, can be found in Appendix~\ref{app:solutions}. For an arbitrary kernel, $K(a,a')$, each $c_n(a)$ in (\ref{eq:cndef}) is an independent parameter.  Therefore, there can be no cancellations between the diagrams proportional to $\bar c_m$ and $\bar c_n$ when $n \neq m$, and we can split this one-loop contribution into three terms:
\beq
(2 \pi) c_{\rm s}^2 P_{\text{1-loop}}^{(c_{\rm s})}(k) \equiv
\bar c_1 \tilde P_1(k) +\bar c_2 \tilde P_2 (k) +\bar c_3 \tilde P_3(k) \ .
\eeq
A plot of these three functions is shown in Fig.~\ref{fig:funcs}.

Without making any further assumptions or without measuring these parameters as a function of redshift, one cannot determine the relationship between $\bar c_{n >1}$ and $\bar c_1 \propto \co$ that we measure at one loop.  Since we do not have any constraints on $K(a,a')$ we will use a simple ansatz that recovers the assumed time-dependence (\ref{eq:cntdep}) while allowing us to parametrize the severity of non-locality in time with some fixed power $p$:
\beq
\label{eq:Kansatz}
K(a,a') =\frac{(2 \pi )\co}{\knl^2}  \left [9 (p+1) \H(a)^2 f(a)^2 \, a'\H(a')\frac{\partial D_1(a')}{da'} \frac{D_1(a')^{p-1}}{D_1(a)^{p-2}} \right] \ .
\eeq
Using this ansatz, we can evaluate equation (\ref{eq:cndef}) to find 
\be\label{eq:mathcing}
\bar c_n = \frac{p+1}{p+n}\frac{(2\pi) \co}{\knl^2}\ . 
\ee
For general $p$, we therefore have {\beq
(2 \pi) \co P_{\text{1-loop}}^{(c_{\rm s}, p)}(k) \equiv \frac{(2\pi) \co}{\knl^2} [ \tilde P_1(k) + \tfrac{p+1}{p+2}  \tilde P_2 (k) + \tfrac{p+1}{p+3} \tilde P_3(k) ] \ .
\eeq
The local case, $K(a,a') \propto \delta(a-a')$ is captured by the limit $p \to \infty$, where all the coefficients are equal.  Results for different choices of $p$ are shown later on the right-hand side of Fig.~\ref{fig:results}.

\subsection{Results}\label{sec:results}

We will now compare our calculations with the results of simulations. Specifically, we use the Coyote interpolator~\cite{Heitmann:2008eq,Heitmann:2009cu,Lawrence:2009uk,Heitmann:2013bra} to generate a non-linear matter power spectrum with cosmological parameters $h=0.7136$, $\Omega_{\rm m}=0.258$, $\Omega_{\rm b}=0.0441$, $n_{\rm s}=0.963$, and $\sigma_8=0.796$. We use CAMB~\cite{Lewis:1999bs} to generate our linear power spectrum, and the recursive implementation of the kernels given in the Copter library~\cite{Carlson:2009it}, but combined to be IR-safe as described in~\cite{Carrasco:2013sva}, with all numerical integrations performed using Monte Carlo integration routines from the CUBA library~\cite{CUBA}, to compute $P_\text{1-loop}$ and $P_\text{2-loop}$.

First, let us outline the procedure schematically:  in order to determine $\co$, we consider the one-loop EFT prediction,
\beq
\label{eq:peft1loop}
P_{\text{EFT-1-loop}} = P_{11} + P_{\text{1-loop}} -  { 2\, (2\pi)} \co \frac{k^2}{\knl^2} P_{11} \ .
\eeq
We determine $\co$ by fitting $P_{\text{EFT-1-loop}}$ to the non-linear power spectrum at low $k$, where $P_{\text{1-loop}}$ is expected to be reliable.  After determining $\co$, the two-loop power spectrum is given by
\beq\label{equ:peft2loop}
P_{\text{EFT-2-loop}}  = P_{11} + P_{\text{1-loop}} +P_{\text{2-loop}}
 -{ 2\,(2 \pi)}  (\co+\ct) \frac{k^2}{\knl^2} P_{11}
 + (2\pi)  \co P_{\text{1-loop}}^{(c_{\rm s}, p)}
 +  (2\pi)^2 c_{s(1)}^4 \frac{k^4}{\knl^4} P_{11} \ .
\eeq
The purpose of $\ct$ is to cancel the $(\tfrac{k}{\knl})^2 P_{11}$ dependence of $P_{\text{2-loop}}$ that arises from loop momenta with $q \gg k$.  Because this contribution is larger than $P^{\text{finite}}_{\text{2-loop}}$, we can determine it by comparing to $P_{\text{EFT-1-loop}}$ in the region where $P^{\text{finite}}_{\text{2-loop}}$ is negligible.  By doing so, we can determine all the parameters in $P_{\text{EFT-2-loop}}$ without ever fitting the two-loop power spectrum to the non-linear data directly.

However, implementing the above procedure is somewhat challenging.  It is easy to measure $\co$ and $\ct$ when they contribute significantly to the power spectrum, namely above $k \sim 0.1 \invMpc$.  However, in this regime, it is difficult to determine, a priori, at which range of $k$ the contribution from $P^{\text{finite}}_{\text{2-loop}}$ can be ignored (which is required for both measurements to be valid).  If one works at $k \ll 0.1 \invMpc$, one can safely use ${ 2(2\pi)}(\co+\ct) (\tfrac{k}{\knl})^2 P_{11} \gg P^{\text{finite}}_{\text{2-loop}}$.  However, because  ${ 2(2\pi)}(\co+\ct) (\tfrac{k}{\knl})^2 P_{11} \ll P_{11}$, one requires very high precision non-linear data to make the measurement of~$\co$.

In practice, it appears that the real universe is much better behaved than one would have naively expected.  As we discussed in Sec.~\ref{sec:fit}, in the regime $0.1 \invMpc < k < 0.3 \invMpc$, the universe behaves much like a scaling universe with $n = -1.7 \sim -3/2$.  As we show in Appendix~\ref{app:loops}, in the $n=-3/2$ universe $P^{\text{finite}}_{\text{2-loop}}$ is smaller than our loop counting would suggest by a factor of~5.  As a result, we can trust $P_{\text{EFT-1-loop}}$ up to $k\sim \ktr$, which is slightly higher scale than our naive counting would suggest.  Therefore, in the range $0.1 \invMpc < k < 0.25 \invMpc$, we can safely measure $\co$ and $\ct$ by implementing the above procedure.  This is very fortunate because the error on available non-linear data is too large to apply to above procedure for $k \ll 0.1 \invMpc$. 

\vskip 8pt

We determine $\co$ from a least-$\chi^2$ fit of the $P_{\text{EFT-1-loop}}$ to the Coyote power spectrum over the range $k \sim 0.15 - 0.25 \invMpc$ with $\Delta k \sim 0.005 \invMpc$.  From the fit, we find~\footnote{In the convention of~\cite{Carrasco:2012cv}, we have
\be
c_{\rm comb}^2(a_0){}^{\rm there}= \co{}^{\rm here} \times 9 (2\pi)\frac{{\cal{H}}_0^2}{c^2 \knl^2} \frac{D'(a_0)^2 a_0^2}{D(a_0)^2} \ ,
\ee
where $D_1(a){}^{\rm here}=D(a)^{\rm there}/D(a_0){}^{\rm there}$. }
\beq  
\co =( 1.62 \pm {  0.03}) \times \frac{1}{2\pi} \lp \frac{\knl}{\invMpc} \rp^{2} \qquad (\text{1-$\sigma$}) .  
\eeq
The result of the fit is shown in Fig.~\ref{fig:1looperror}.  
 \begin{figure}[t]
\begin{center}
\includegraphics[width=0.7\textwidth]{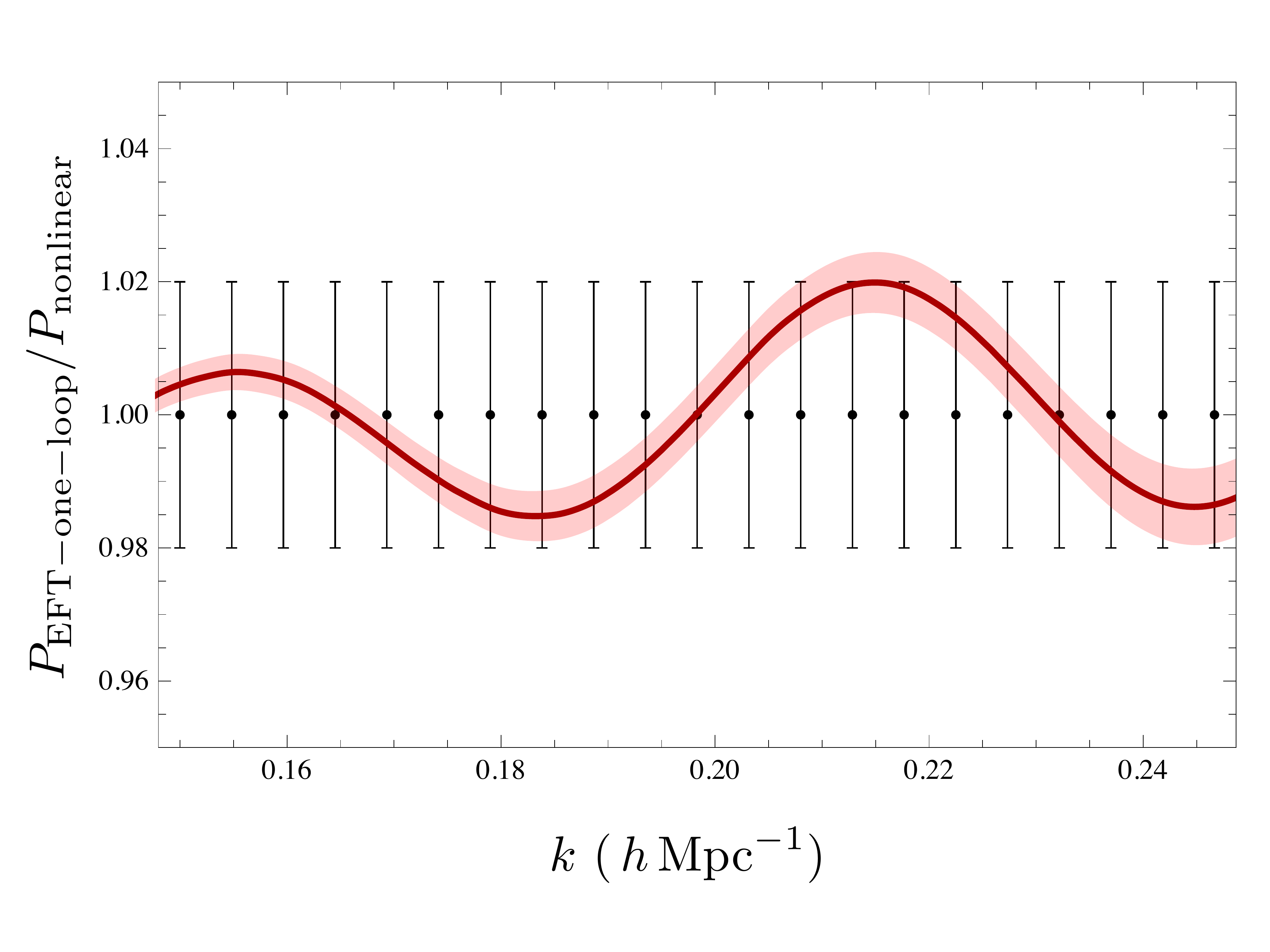}
\caption{\label{fig:1looperror} \small\it 
{\bf Fit range for one-loop EFT.} We plot  $P_{\text{EFT-1-loop}}$ normalized to non-linear data over the range to which the sole EFT parameter is fit.  The red curve is the best fit value of $\co$ and the red band shows the 2-$\sigma$ error on $\co$.  We also show the data points that are fit, along with their 2-$\sigma$ errors (assuming 1 percent error on all points). 
}
\end{center}
\end{figure}

Having measured $\co$, we can now fit $\ct$ to $P_{\text{EFT-1-loop}}$.  In performing this fit, we must make some assumption about $P_{\text{1-loop}}^{(c_{\rm s}, p)}$.  For now, we will take the $p \to \infty$ limit, which corresponds to the assumption that the effective stress tensor $\tau^{ij}$ is completely local in time.  We now fit $P_{\text{EFT-2-loop}}$ to $P_{\text{EFT-1-loop}}$ over the range $k \sim 0.15-0.25 \invMpc$.  Here we are using the expectation that $P^{\text{finite}}_{\text{2-loop}}$ is negligible over this range, such that the dominant source of error is our uncertainty in~$\co$.  Using this fitting procedure, we find that 
\beq
\ct = (-3.316 \pm { 0.002}) \times \frac{1}{2\pi} \lp \frac{\knl}{\invMpc} \rp^{2} \qquad (\text{1-$\sigma$}) .  
\eeq
The error bar for $\ct$ is much smaller than $\co$ because we can determine it without using the non-linear data.  In fact, for a scaling universe we can determine $\ct$ exactly.

It is important to note that $\ct$ is not a physical parameter but exists solely to cancel the {\em unphysical}  UV contributions from $P_{\text{2-loop}}$.  For this reason, the size of the coefficient and its negative sign carry no particular significance on their own.  This is illustrated in Figure~\ref{fig:terms}.  If we were to consider every term that contributes to $P_{\text{EFT-2-loop}}$ separately, many of the terms appear to be of the same order.  However, when grouped according to the terms which are expected to show large cancellations, the size of each group decreases with loop order~\footnote{Notice that in Figure~\ref{fig:terms},  the one-loop contribution becomes larger than tree-level at $k=0.45\invMpc$.  This is a wavenumber quite smaller than $\knl$ from~(\ref{eq:knlest}).  Indeed this is as expected from (\ref{eq:est1Crossing}), which was predicting within~${\cal O}(1)$ the scale at which one-loop contribution starts dominating the tree level term. This fact does not imply that the non-linear scale, or the scale where the EFT fails, is $k=0.45\invMpc$. Higher loops are indeed smaller than the one-loop term at that scale. Roughly speaking, the most reducible contribution of each $L$- loop order scales as $\frac{(2\pi)^L}{L!}\left(\tfrac{k}{\knl}\right)^{0.9 L}$. The factors of $(2\pi)$ make the contribution of the one-loop term anomalously large, while the diagrams from higher loops are hierarchically suppressed either by the factorial or by missing factors of~$(2\pi)$, as can already be seen for the two-loop term on the right in the same Figure~\ref{fig:terms}.}.

 \begin{figure}[t]
\begin{center}
\includegraphics[width=.495 \textwidth]{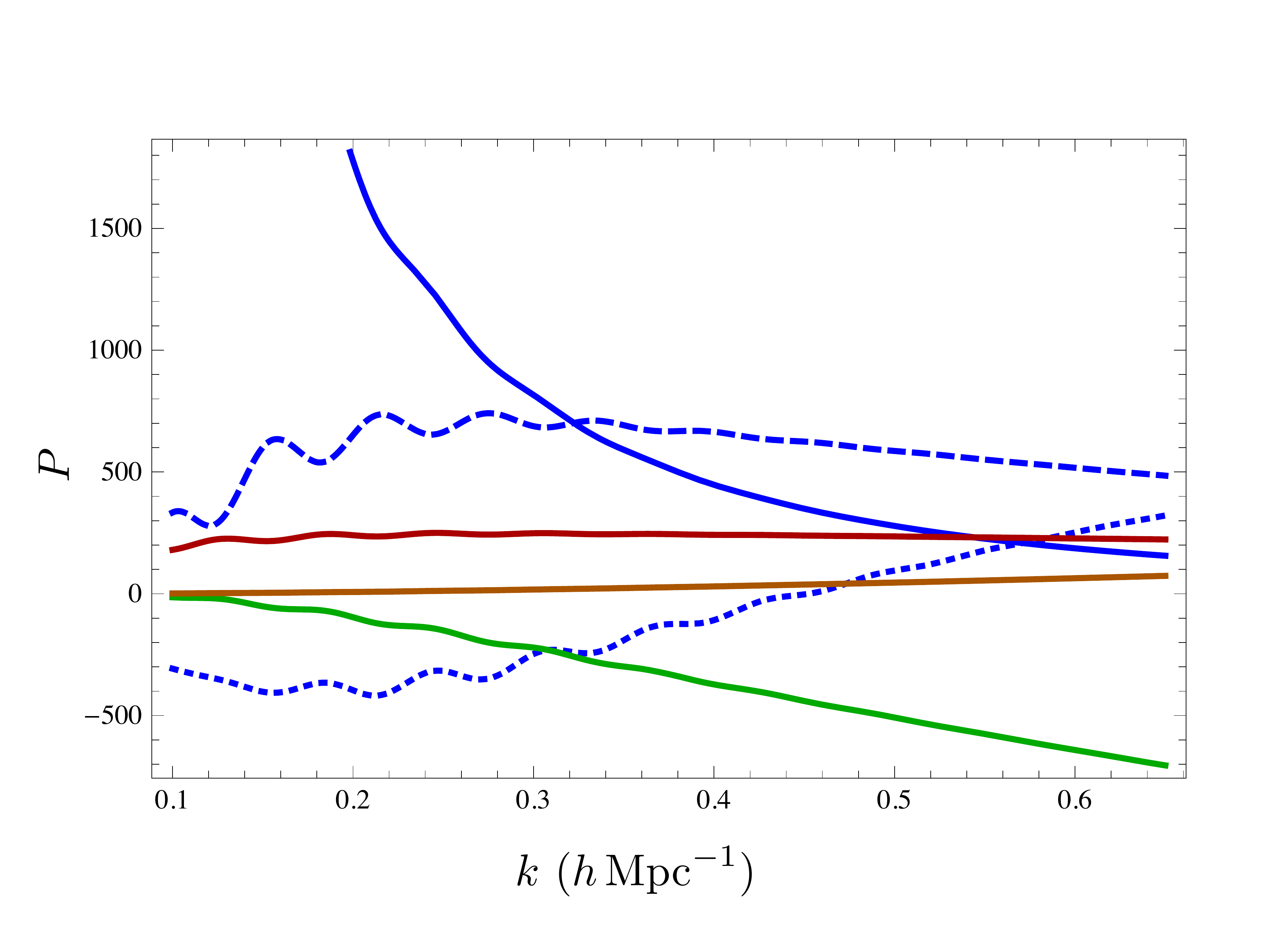}
\includegraphics[width=.495 \textwidth]{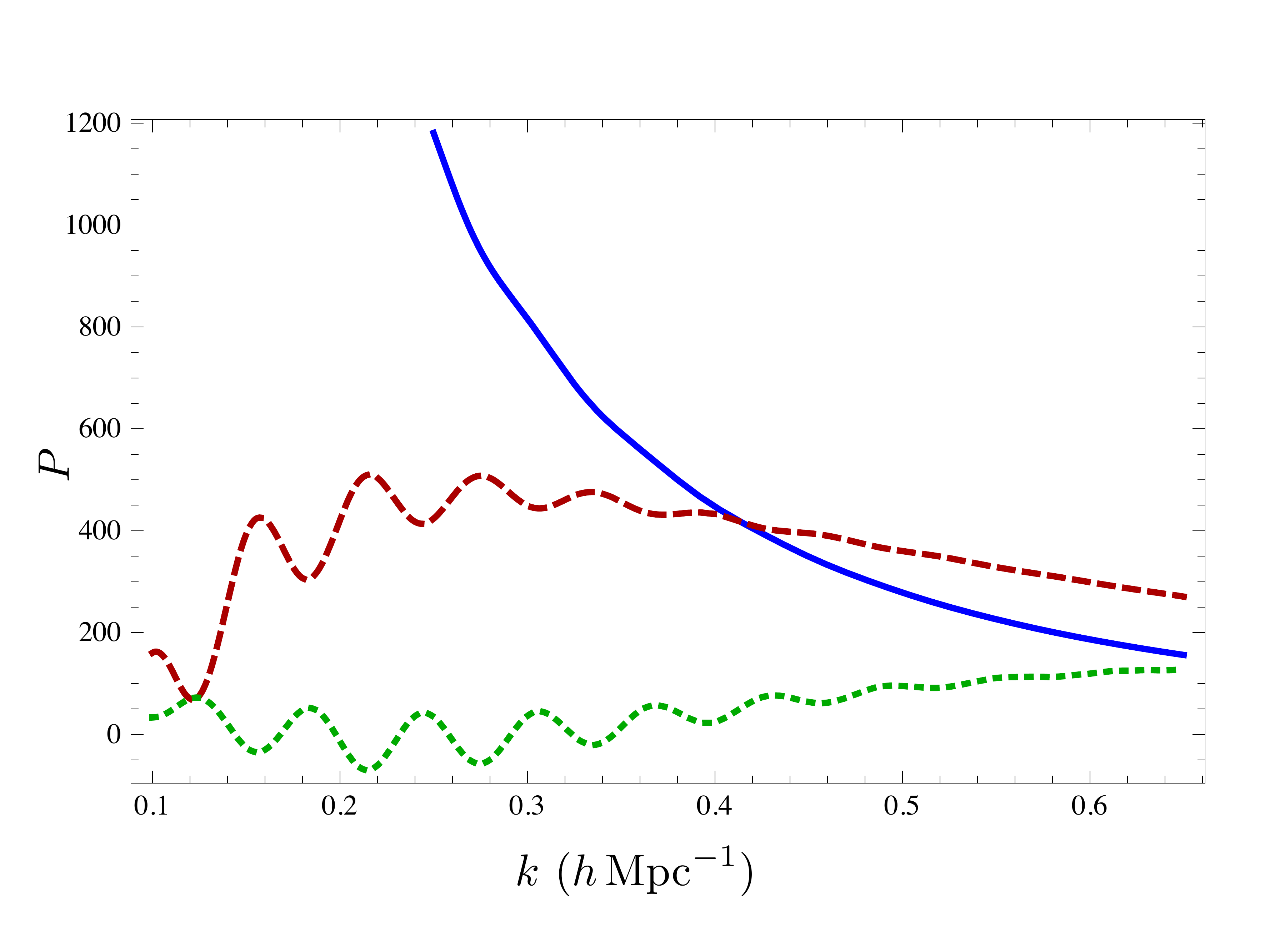}
\caption{ \label{fig:terms} \small\it  {\bf Contributions to the power specturm.} {\bf Left:}  We show every term appearing in Eq.~(\ref{equ:peft2loop}) separately.  We plot $P_{11}$, $P_{\text{1-loop}}$ and $P_{\text{2-loop}}$ in solid, dashed and dotted blue respectively.  The contributions of $-2(2\pi) (\co+\ct) \frac{k^2}{\knl^2} P_{11}$, $ (2\pi) \co P_{\text{1-loop}}^{(c_{\rm s}, p)}$ and $(2\pi)^2c_{s(1)}^4 \frac{k^4}{\knl^4} P_{11}$ have been plotted in red, green and orange respectively. Notice that many of the terms are of the same order of magnitude. {\bf Right:}  We plot the sums of terms appearing in Eq.~(\ref{equ:peft2loop}) that appear together at each loop order: tree-level ($P_{11}$), one-loop ($P_{\text{1-loop}}-{2\,(2 \pi)}  \co \frac{k^2}{\knl^2} P_{11}$) and two-loop (everything else) in solid blue, dashed red and dotted green respectively.  Notice now that each group of terms is smaller than the previous group, as required for a consistent perturbative expansion.
}
\end{center}
\end{figure}

In additional to statistical errors, we also have theoretical uncertainties due to the higher orders terms we are neglecting.  This includes three-loop SPT, $\co$ and $\ct$ insertions in two-loop diagrams, and higher-order counterterms.  These contributions were estimated in Sec.~\ref{sec:estimates}.  The two largest uncertainties are due to
\bea
P_{\text{3-loop}}^{\text{R}} &\sim& \alpha  (2 \pi)^2 \left( \frac{k}{\knl} \right)^{2.7} P_{11}(k) \ ,
\eea
and
\bea 
4 \pi \co  \frac{k^2}{\knl^2}  P^{\rm I}_{\text{2-loop}}(k)&\sim& 3.24 \frac{k^2}{(\invMpc)^2} (2 \pi)  \left( \frac{k}{\knl} \right)^{1.8} P_{11}(k) \ .
\eea
In Fig.~\ref{fig:2looperror}, we show the best fit  $P^{\text{finite}}_{\text{2-loop}}$ normalized to non-linear data.  We have included these two sources of uncertainty as a series of red bands.  The outermost and innermost bands (at $k = 0.6 \invMpc$) are from $P_{\text{3-loop}}^{\text{R}} $ with $\alpha= 1$ and $1/2$ respectively.  The middle band is given by $4 \pi \co  \frac{k^2}{\knl^2}  P_{\text{2-loop}}(k)$.  Our target is 1\% agreement between the power spectrum of the EFTofLSS and the non-linear data, so in the figure, the dotted black line shows the 2-$\sigma$ limit associated with 1\% agreement with the non-linear data, that is estimated to have a 1-$\sigma$ error of 1\%.
 \begin{figure}[th!]
\begin{center}
\includegraphics[width=0.7\textwidth]{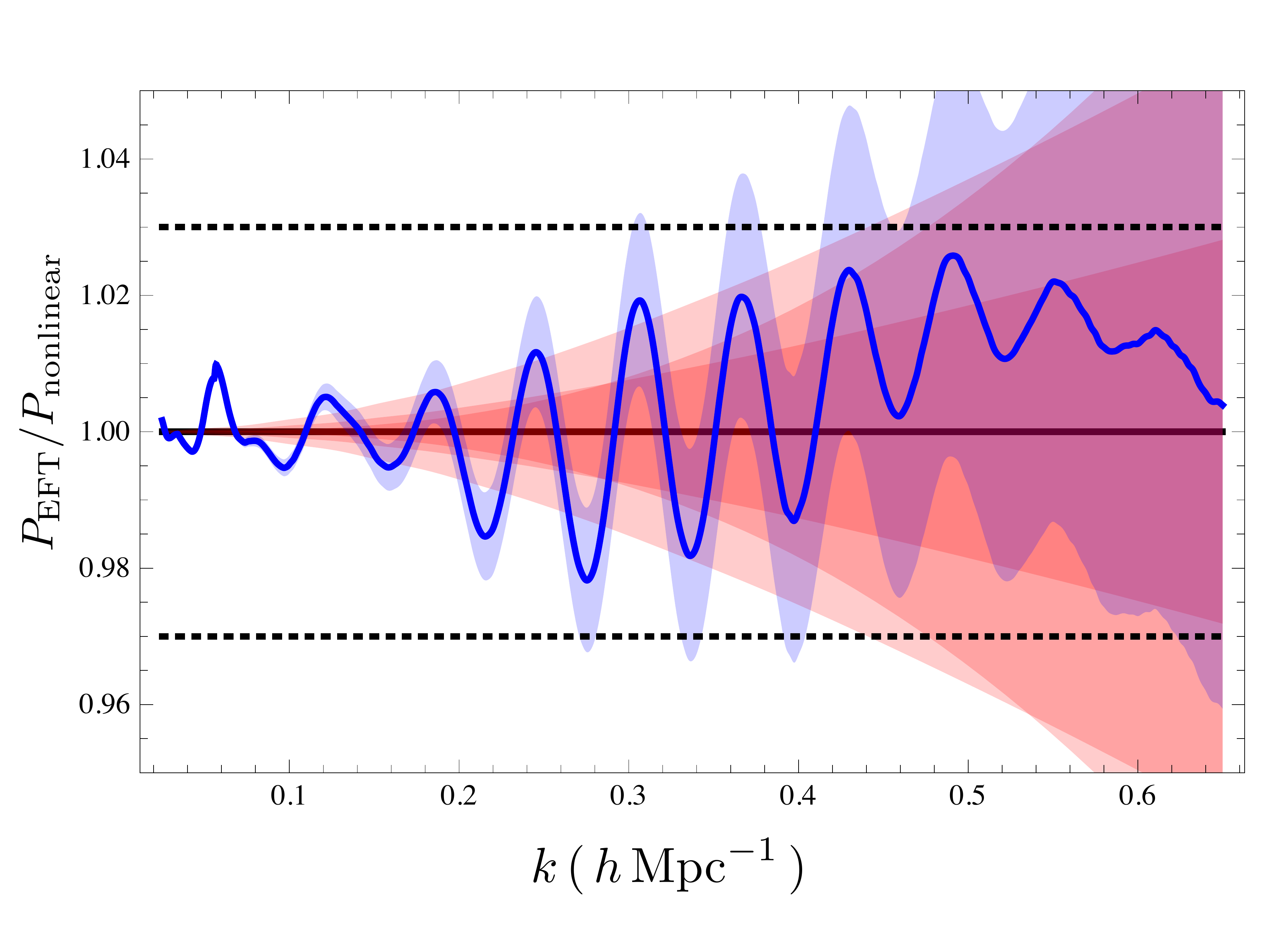}
\caption{\label{fig:2looperror} \small\it 
{\bf Effectiveness of EFT at two loops.}  We plot  $P_{\text{\rm EFT-2-loop}}$ normalized to non-linear data.  The blue curve is the best fit value of $\co$ (with $\ct$ fit to $P_{\text{\rm EFT-1-loop}}$, as described in the main text) and the blue band shows the 2-$\sigma$ error on $\co$ and $\ct$.  The black solid line shows the non-linear data. The red shaded regions are different error estimates described in Sec.~\ref{sec:results}.   The dotted black line is the 2-$\sigma$ limit associated with 1\% agreement with the non-linear data, that we take to have a 1-$\sigma$ error of 1\%. We find that it is possible to obtain 1\% agreement with the non-linear power spectrum after having fit only one new parameter, $\co$, and furthermore that the agreement stretches well past the range, $0.15\invMpc < k < 0.25\invMpc$, where the parameter was fit. The EFTofLSS at two loops is in percent agreement with simulations up to~$k\simeq0.6\invMpc$.
}
\end{center}
\end{figure}

 \begin{figure}[t]
\begin{center}
\includegraphics[width=.495 \textwidth]{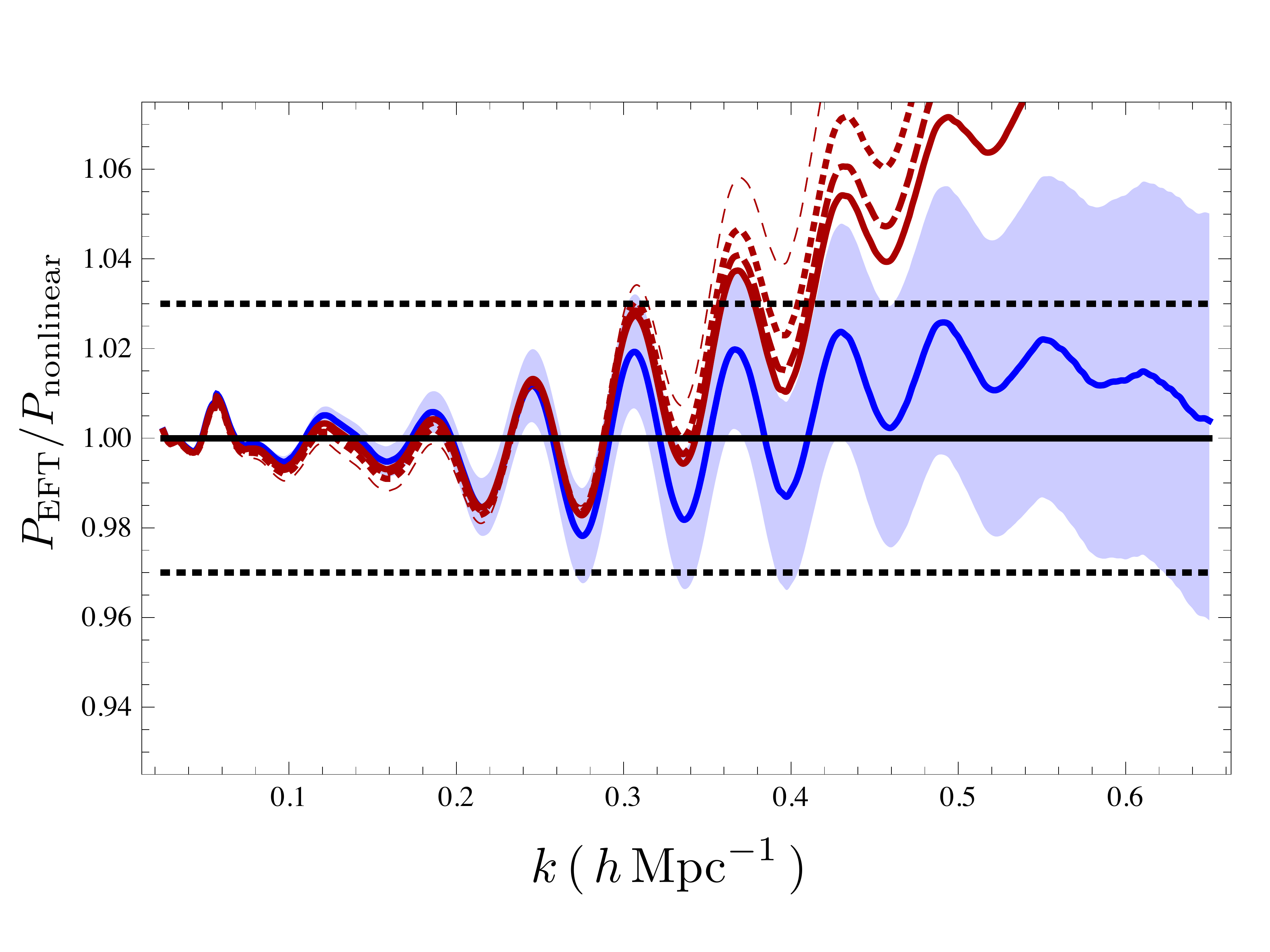}
\includegraphics[width=.495 \textwidth]{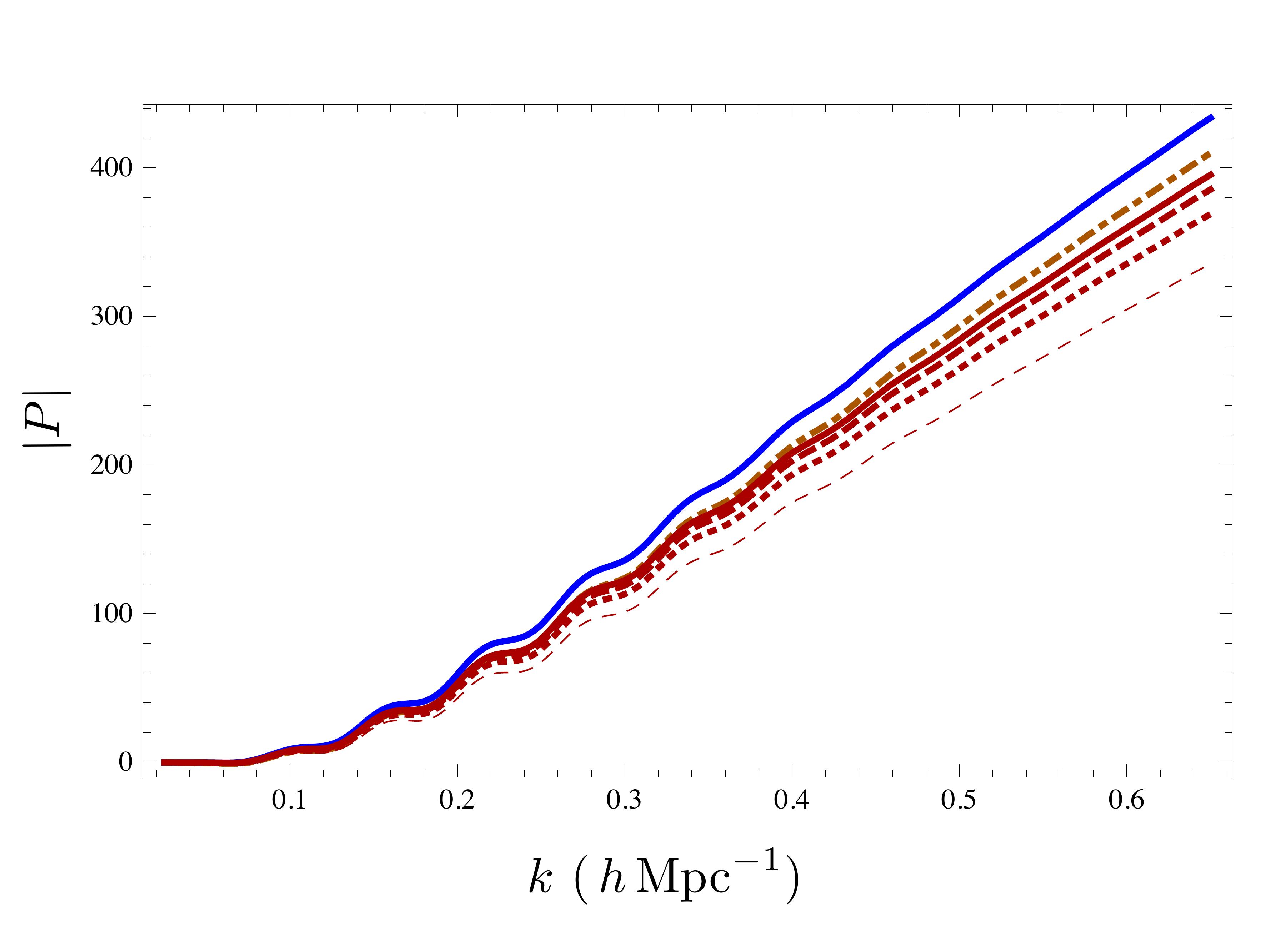} 
\caption{ \label{fig:results} \small\it 
{\bf Effectiveness of a local kernel at two loops.}  We show the effects of varying the non-locality of the kernel by changing the parameter $p$ in $P_{\text{1-loop}}^{(c_{\rm s}, p)}$. On both the left and the right, the blue solid curve corresponds to $p\to\infty$ (i.e.\ the local kernel), while $p=0,1,2,3$ are shown in red with light dashed, dotted,  dashed, and solid curves respectively.   {\bf Left:} In the left figure we plot  $P_{\text{EFT-2-loop}}$ with the various values of $p$, as well as the non-linear data (solid black).  Note the effectiveness of the local kernel for describing the non-linear power spectrum.  The dotted black line is the 2-$\sigma$ limit associated with 1\% agreement with the non-linear data, that we take to have a 1-$\sigma$ error of 1\%, as in Fig.~\ref{fig:2looperror}.  The red curves show the best fit results for the various values of $p$, using the measured value of $c_s$ at one loop as an input and matching $\ct$ to the one-loop data.  The blue shaded region shows that 2-$\sigma$ error on the local model.    {\bf Right:} The right figure shows $| \knl^2 P_{\text{1-loop}}^{(c_{\rm s}=1, p)}|$ for the different values of $p$, along with $2 k^2 P_{\text{1-loop}}$  (the orange dot-dashed curve). For $p \geq 3$, the best fit curves are within the 2-$\sigma$ errors of the local model up to $k \sim 0.5 \invMpc$.
}
\end{center}
\end{figure}

Although we have been careful throughout to estimate where  $P_{\text{EFT-3-loop}}$ becomes important, there is still uncertainty in the precise order one factors.  As a result, which of the three-loop bands (if any) represents the real breakdown of the two-loop prediction cannot be determined at this level.  Alternatively, we could estimate the maximum $k$ up to which we can trust our calculation using the scale at which $P_{\text{EFT-2-loop}}$ deviates from the non-linear data~\footnote{ One possible concern in using this criterion of convergence is that we might be over fitting. But this is clearly not the case for two reasons.  First, we are fitting only one parameter in the range $0.15 \invMpc<k<0.25 \invMpc$, which is quite below the values of $k$ where our calculation begins to fail. Second, we expect the breakdown of the two-loop prediction to be quite close to the value of $k$ where the three-loop term becomes important, which is indeed the case.}.
We see from Fig.~\ref{fig:2looperror} that the prediction of the EFTofLSS at two loops agrees with the non-linear data at redshift $z=0$ at percent level up to  $k\simeq 0.6\invMpc$. This is a very remarkably high wavenumber given the results of former perturbative calculations, as we discuss next.
 
Now, let us remove our assumption of locality in time.  As we discussed in the previous section, the EFTofLSS is not local in time, but our knowledge of its non-locality is limited.  We can check how our results depend on the assumption by repeating the above procedure for different values of $p$.  This is shown in Fig.~\ref{fig:results}.  We see that for $p \geq 3$, the best fit curves are within the 2-$\sigma$ errors of the local model up to $k \sim 0.5 \invMpc$.  In that sense, we did not require any strict use of locality to achieve this level of agreement with the non-linear data~\footnote{Nevertheless, this result seems to indicate that the counterterms of the EFTofLSS are numerically well described by the local-in-time approximation. This indeed can potentially be justified by noticing the following. Modes that are shorter than the virialization scale do not contribute to renormalize the counterterms. However, there is some room in $k$-space between the non-linear scale and the virialization scale. This implies that the time scale for virialized objects is about a factor of order 5 faster than Hubble, which is the time scale of modes in the linear regime. Since most of the phase space is at high wavenumbers, the weight of the modes of high $k$ up to the virialization scale is expected to matter more.  This might suggest that an expansion in local-in-time counterterms by perturbatively including time-derivatives could be feasible to some level of approximation. We leave the exploration of this possibility to future work.}.

Finally, we would like to see how much of an improvement we have made by going to two loops in the EFTofLSS.  Fig.~\ref{fig:SPT} shows the results of the one and two loop EFT compared to various orders in SPT.  The improvement over SPT is dramatic both for one and two loops.  By going from one to two loops, we see the agreement with non-linear data has been pushed from $k \sim 0.3 \invMpc$ out to $k \sim 0.5-0.6 \invMpc$ (and possibly beyond).  Given the cubic scaling of the number of available modes, this corresponds to a factor of 6 improvement from one to two loops with the EFTofLSS, and about a factor of 200 with respect to SPT.

   \begin{figure}[t]
\begin{center}
\includegraphics[width=.65 \textwidth]{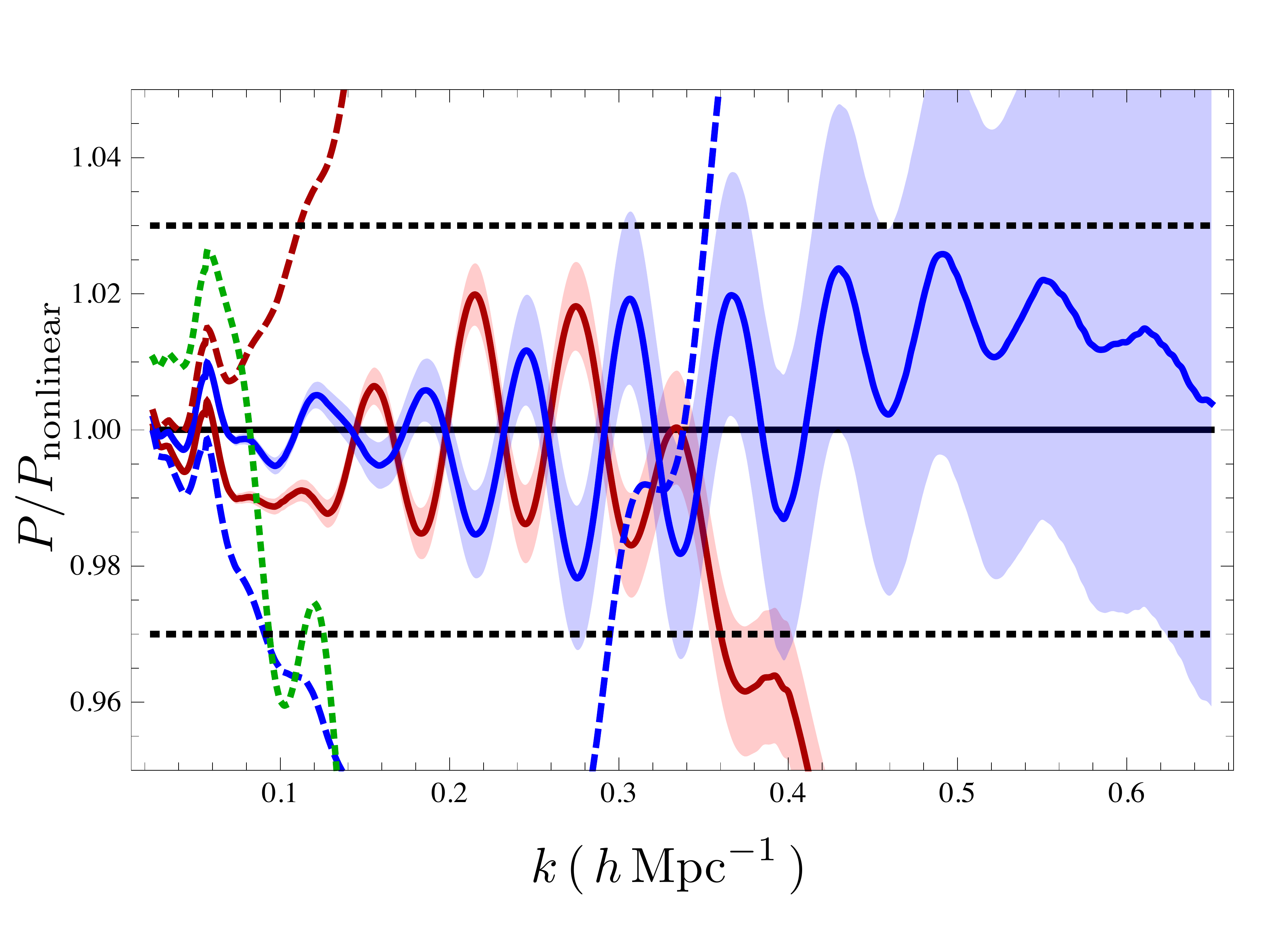}
\caption{ \label{fig:SPT} \small\it   {\bf Relevance of higher order perturbative correction, normalized.}  We plot  one-loop EFT (solid red), two-loop EFT (solid blue) and linear EFT (equivalent to linear SPT, dotted green), normalized to the non-linear data  (solid black).  We also show one-loop (dashed red) and two-loop (dashed blue) SPT, normalized to non-linear data.  The dotted black line is the 2-$\sigma$ limit associated with 1\% agreement with the non-linear data, that we take to have a 1-$\sigma$ error of 1\%, as in Fig.~\ref{fig:2looperror}.  The red and blue bands show the 2-$\sigma$ errors on the one- and two-loop EFT parameters respectively.   See Fig.~\ref{fig:SPTA} for unnormalized plot.  }
\end{center}
\end{figure}

  \begin{figure}[t]
\begin{center}
\includegraphics[width=.7 \textwidth]{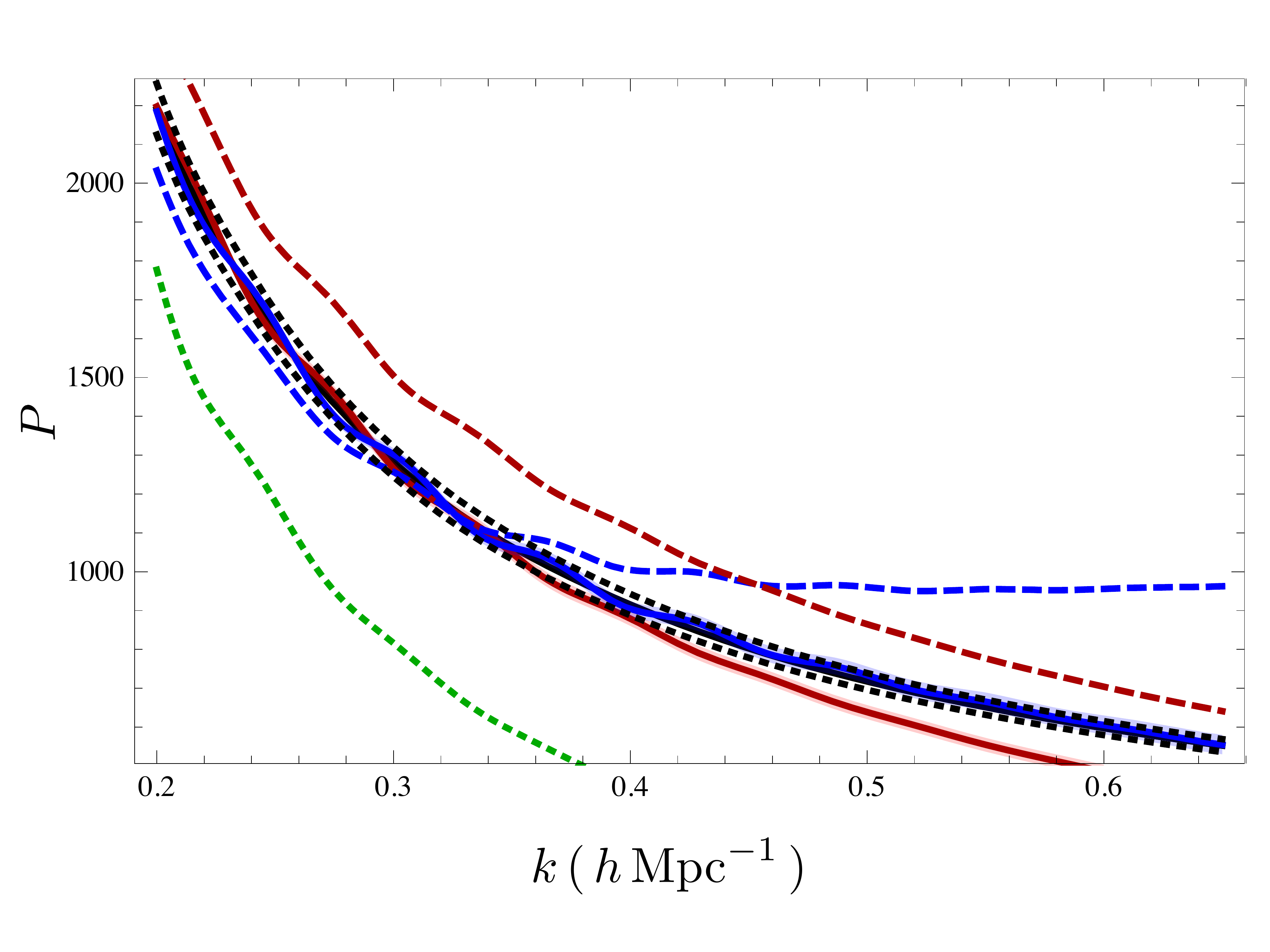}
\caption{ \label{fig:SPTA} \small\it  {\bf Relevance of higher order perturbative corrections, unnormalized.}   We plot two-loop EFT (solid blue),  one-loop EFT (solid red),  linear EFT (equivalent to linear SPT, dotted green), as well as one-loop SPT (dashed red), two-loop SPT (dashed blue),   and non-linear data (solid back).   Note the non-linear data (solid black line) is almost indistinguishable from the two-loop EFT prediction (solid blue line).  We make the agreement clear by including the dotted black lines of the 2-$\sigma$ limit associated with 1\% agreement with the non-linear data, as in Fig.~\ref{fig:2looperror}.  The strong agreement of the two-loop EFT with non-linear data is evident. See Fig.~\ref{fig:SPT} for normalized plot. }   
\end{center}
\end{figure}

Furthermore, it is instructive to see the manner in which the EFT achieves its improvement in going from one loop to two loops. The one-loop data are fit in the range $0.15\invMpc < k < 0.25\invMpc$ to determine $\co$, and they begin to deviate from the non-linear data at about $k\simeq0.35\invMpc$. By adding the two-loop terms, which entails no new fitting parameter in practice, we improve the reach of the fit up to $k\sim 0.5- 0.6\invMpc$. Notice how the two-loop SPT term is large in that range. This means that all the EFT terms that enter at two loops, and that are predicted in terms of the one-loop counterterm, are essential to improve the reach of the fit. We take this as a strong confirmation of the goodness of the EFT description of the dark matter clustering.

\section{Correlation Functions Involving Momentum at One Loop\label{sec:momentum}}
\label{sec:pipi}

As we have seen, the EFTofLSS allows one to compute power spectra and higher-point functions (such as bi- or trispectra) of quantities relating to the dark matter distribution, such as $\delta$ and the momentum (or mass-weighted velocity) $\pi^i \equiv \rho v^i$. These predictions will involve one or more EFT parameters (such as $\co$), and therefore, one could measure these parameters using one observable and then use these measurements to obtain predictions for other observables.

In this section, we put this procedure into practice by measuring $\co$ from a matter power spectrum measured from $N$-body simulations, and using it to predict the power spectrum of the divergence, or scalar, part of} momentum, $\pi_{\rm S} \equiv \partial_i \pi^i$, as well as the cross spectrum between $\delta$ and~$\pi_{\rm S}$. Notice that $\pi^i$ has also a vorticity component $\pi_{\rm V}^i=\epsilon^{ijk}\d_j \pi^k$. This term is vanishingly small at linear level, as vector modes decay in the early universe. Interesting, this term is not sourced by the leading order $\co$-like terms that represent the linear response of the short scale stress tensor from the long modes. It is however sourced by non-linear terms in its equations of motion.
Here we focus on $\pi_S$,  whose predictions we compare to measurements from the same set of simulations. 

Specifically, we use simulations by Okumura {\it et al.}~\cite{Okumura:2011pb}, based on a flat $\Lambda$CDM model with $\Omega_{\rm b}h^2=0.0226$, $\Omega_{\rm m} h^2=0.1367$, $h=0.7$, $n_{\rm s}=0.96$, and $\sigma_8=0.807$.  Strictly speaking, as we will explain later, what enters in the computation of the momentum is not only ${c}_1(a)$ at a given redshift but also its time derivatives at the same redshift.  In principle, by measuring the matter power spectrum at various redshifts, we could reconstruct the time derivative of ${c}_1$.  We believe that this requires precise sampling of the $N$-body simulations as a function of redshift, something that is not available to our collaboration currently. We therefore leave this exploration for future work. What we will do is to compare the prediction of the power momenta as obtained by {\it assuming} that the time dependence of ${c}_1(a)$ is the same as the one given in (\ref{eq:cntdep}), so that the ${c}_1$ counterterm has the same time dependence as  $P_{13}$, with furthermore the approximate treatment of the time dependence that we use in this paper. We will then also show the prediction of the power spectrum by allowing for the time dependence of ${c}_1(a)$ to be different from the one inferred from~(\ref{eq:cntdep}), something that is physically motivated because ${c}_1$ encapsulates the contribution of the short distance physics which cannot be described by a perturbative treatment. We stress that by measuring the matter power spectrum as a function of redshift, the momentum power spectrum could be predicted without any new parameter.

\subsection{On the Velocity Vorticity}
\label{sec:vorticity}

At this point, it is however important to clarify an outstanding issue, which is the size of the vorticity of the velocity $\omega^i=\epsilon^{ijk}\d_j v_k$ (not to be confused with the vorticity of momentum $\pi_{\rm V}$), which has been set to zero from the beginning~\footnote{While this paper was being written, Ref.~\cite{Mercolli:2013bsa} appeared which discusses related ideas.}. As we noticed in  (\ref{eq:new_eom}), the velocity-vorticity will be sourced by the stress tensor of the EFT~\cite{Baumann:2010tm,Carrasco:2012cv}. This puts into evidence another crucial difference with respect to SPT, for which the velocity-vorticity cannot be generated at any order. We repeat here the velocity-vorticity equation for convenience:
\be\label{eq:omega_eq}
a\H \omega_i' + \H\omega_i =
\epsilon_{ijk} \d^j \lp \epsilon^{kmn} v_m \omega_n \rp
- \epsilon_{ijk}\d^j \gammai{}^{k}\ . \ee
There are two possible sources for velocity-vorticity from $\gammai$: one is the response of $\gammai$ from long wavelength fluctuations, the other is the stochasticity of $\gammai$. Let us consider first the contribution from the response. The leading terms considered in (\ref{eq:setensor}) do not contribute to the velocity-vorticity equation because we can replace $\d^2\phi$ with $\delta$ by using Poisson's equation, and then we are left with a symmetric tensor contracted with $\epsilon_{ijk}$. The same accidental cancellation does not affect higher order terms. The leading one is of the form
\beq
\label{eq:setensor-higher}
\gammai{}^{i}\supset \ldots+
\int \frac{da'}{a'\H(a')}  \frac{da''}{a''\H(a'')}   \; \kappa_{1,v}(a,a',a'')\;\frac{\d_j}{\knl^2}\left[\d_l \partial^j \phi(\tau',  \vx_{\rm fl} ) \;\d^l\d^i  \phi(\tau'',  \vx_{\rm fl} )\right]
+ \cdots .
\eeq 
The factors of $\H$ suppressing the additional fluctuating long wavelength fields beyond the linear one that appear in the stress tensor are chosen so that there is no suppression if the added long wavelength fields are evaluated at the non-linear scale.  $\kappa_{1,v}(a,a',a'')$ is a kernel of size of order~one, with support of order one Hubble time.
By solving the equations iteratively, we obtain a contribution to the power spectrum of $\omega^i$ by contracting two of these $\epsilon^{ijk}\d_j(\gammai)_k$ contributions, one from each side. For a scaling universe we obtain
\be\label{eq:omega-higher}
\langle\omega^i\omega^j\rangle_{\gammai\text{ higher order}}\sim \delta_D(\vec k+\vec k')\; \delta^{ij}\; c_v^4  \frac{H^2}{\knl^3}\left(\frac{k}{\knl}\right)^{7+2n}\ ,
\ee
where $c_v^2$ is an order one number defined from $\kappa_{{1,v}}$ analogously to the parameters $c_s^2$ that appeared earlier in the computation of the matter power spectrum. It is nice to notice why this term contributed to generating vorticity: this term is  higher derivative than the leading one, which is analogous to a pressure term. So this higher derivative term is analogous to a viscous term, explaining the subscript $v$ in $\kappa_{1}$ and $c_v^2$. Remembering the fit in~(\ref{eq:fit}), in the current universe this contribution goes as $(k/\knl)^{2.8}$ for $0.2\invMpc\lesssim k\lesssim 0.6\invMpc$, while it becomes steeper at lower $k$'s $(k/\knl)^{3.6}$ for $0.1\invMpc\lesssim k\lesssim 0.2\invMpc$, to slowly asymptote  to $(k/\knl)^{9}$ for $k$'s smaller than the equality scale. Notice that $\langle\omega^i \theta\rangle$ and $\langle\omega^i \delta\rangle$ vanish because of rotation and/or parity invariance.

Another contribution comes from the stochasticity of the stress tensor, as pointed out in~\cite{Carrasco:2012cv}. The two point function of the fluctuations of the stress tensor is expected to be Poisson-like distributed~\cite{Carrasco:2012cv}
\be
\langle\Delta\tau(\vec k)\Delta\tau(\vec k')\rangle\sim\delta_D(\vec k+\vec k')\; \rho_b^2  c_{st}^4\left(\frac{H}{\knl}\right)^4 \frac{1}{\knl^3}\ .
\ee
where $c_{st}^2$ is expected to be an order one number. This implies that the correlation function of stochastic term of $\gammai$ is obtained by adding two additional factors of $k$~\footnote{This is just to comply with the index structure, and because the division by $\rho_l$, being a short distance effect, can only add relative contributions with higher derivative terms.}, to obtain
\be
\langle\Delta\gammai^i(\vec k)\Delta\gammai^j(\vec k')\rangle\sim\;\delta^{ij}\;\delta_D(\vec k+\vec k')\;  c_{st}^4\left(\frac{H}{\knl}\right)^4 \frac{k^2}{\knl^3}\ .
\ee
The induced $\omega$ power spectrum goes as
\be\label{eq:omegastoch}
\langle\omega^i\omega^j\rangle_{\gammai\;\text{ stoch}}\sim\delta_D(\vec k+\vec k') \delta^{ij}\; c_{st}^4  \frac{H^2}{\knl^3}\left(\frac{k}{\knl}\right)^{4}\ .
\ee
Notice that the power of the stochastic contribution is independent of $n$. This contribution is smaller  than one from (\ref{eq:omega-higher}) at large $k$'s: $k\gtrsim 0.05-0.1\invMpc$. But as we move to lower and lower $k$'s, it becomes the leading one.  

So far we have talked about the {\it bare} $\vec \omega$. The truly observable and well defined vorticity is the {\it renormalized} $\omega_{R}^i=\epsilon^{ijk}\d_j v_{R,k}$. If we add suitable local counterterms to the velocity field as in (\ref{eq:velocity_def_ren}), we obtain a relationship at the level of the vorticity of the form
\be
\omega^i_{R}=\omega^i+\epsilon^{ijk}\frac{\d_j}{\H(a)}\int \frac{da'}{a'\H(a')}  \frac{da''}{a''\H(a'')}   \; \bar\kappa_{1,v}(a,a',a'')\;\frac{\d_l}{\knl^2} \partial^2 \phi(\tau',  \vx_{\rm fl} ) \,\d^l\d_k \phi(\tau'',  \vx_{\rm fl} )
+\Delta\omega^i
\ee
where $\Delta\omega^i$ is a stochastic term, and $ \bar\kappa_{1,v}$ is a function of order one with width of order one. The second term leads to a contribution parametrically equal to the one in (\ref{eq:omega-higher}). The stochastic term for $\omega^i$ originates from the stochastic local counterterm for the renormalized velocity $v^i_R$. Symmetries allow us to write the leading one as
\be
\langle\Delta v^i(\vec k)\Delta v^j(\vec k')\rangle\propto (2\pi)^3\delta_D(\vec k+\vec k') \delta^{ij} ,\quad\Rightarrow\quad \langle\Delta \omega^i(\vec k)\Delta \omega^j(\vec k')\rangle\propto (2\pi)^3\delta_D(\vec k+\vec k') (k^i k^j-\delta^{ij} k^2 )\ .
\ee
In terms of slope, this is the leading term. However, it is quite unclear what the scale suppressing this operator is. It is easy to verify that no counterterm of this form is required by perturbation theory (while a term $\propto k^4$ is required), which means that this term has no divergent coefficient, but only a finite one. This furthermore suggests that the scale suppressing this operator might not be $\knl$, but rather some higher wavenumber associated to the resolution of the experiment or of the simulation. This makes it interesting to consider the subleading power, which is generated by (\ref{eq:omega-higher}). In the region $0.1\lesssim k\lesssim 0.6 \invMpc$ this is predicted to be of the form $(k/\knl)^{(7+2n)}\sim(k/\knl)^3$, with $n\simeq -2.1$ from  $0.3 \invMpc\lesssim k\lesssim 0.6 \invMpc$ and $n\simeq -1.7$ for  $0.1 \invMpc\lesssim k\lesssim 0.3 \invMpc$. These  results seem to be in agreement with measurements in simulations~\cite{Pueblas:2008uv}.

By plugging into the equations for $\delta$, we can see that this non-vanishing velocity-vorticity gives a contribution to the matter power spectrum that scales roughly as
\be
\langle\delta\delta\rangle_{\text{from }\omega^i}\sim \delta_D(\vec k+\vec k')\frac{1}{\knl^3}\left(\frac{k}{\knl}\right)^{{\rm Max}[10+3n,5+n]}\ .
\ee
depending if it is the viscous-like or the stochastic term that dominates. For $n\simeq -2$, this is smaller than the two-loop contribution by at least a factor of $(k/\knl)^3$. This justifies the approximation of setting $\omega^i$ to zero. 

A final comment on the power spectrum of the velocity-vorticity when at low-$k$'s is dominated by the stochastic term.  Notice that a Fourier-space correlation function that is an analytic function in $k$-space times the $\delta$-function of momentum conservation, as the one we have for the velocity-vorticity, corresponds to a very local correlation function in real space. In this case it is of the form of $(\d^2_{\vec x})^2\delta(\vec x-\vec x')$. This means that the term is UV dominated, as it is confirmed here by the fact that we obtain this term from a counterterm. While the slope of the power spectrum is predicted in the EFT, the actual coefficient is not predicted at all. Observationally and/or in simulations, these terms might depend on the way objects are measured, and these are, by their own nature, UV sensitive.

We are now ready to look at the momenta power spectra. Notice that $\pi_V$ will be generated at much lower order in $k/\knl$, as it is sourced already by the non-linear terms of SPT.

\subsection{Predictions for Correlation Functions Involving Momentum}
\label{subsec:momprediction}

It is straightforward to find an expression for the divergence of momentum if we recall the second equation of motion in Eq.~(\ref{eq:all_equations}):
\beq
\dot\delta=-\frac{1}{a}\d_i\left([1+\delta] v^i\right)\ .
\eeq
By Fourier-transforming this equation, we find that
\beq
\pi_{\rm S}(a,\vk) = - a \H  \rho_{\rm b} \, \delta'(a,\vk)\ ,
\eeq
where $\delta'\equiv \partial\delta/\partial a$. Therefore,
\begin{align}
\langle \pi_{\rm S}(\vk,a) \pi_{\rm S}(\vk',a) \rangle 
&= \lp a \H  \rho_{\rm b} \rp^2 \langle \delta'(\vk,a) \delta'(\vk',a) \rangle\ , \\
\langle \delta(\vk,a) \pi_{\rm S}(\vk',a) \rangle 
&= - \lp a \H  \rho_{\rm b} \rp \langle \delta(\vk,a) \delta'(\vk',a) \rangle\ ,
\end{align}
and we can use the perturbative solutions for $\delta(\vk,a)$ (see App.~\ref{app:solutions}) to expand the right-hand sides  up to the desired order. At one loop, we get  
\begin{align}
\nn
P_{\pi_{\rm S}\pi_{\rm S}}(a,k) &= \lp a\H \rho_{\rm b} \rp^2 \lp
[D_1'(a)]^2 P_{11}(k)
+ [D_1(a) D_1'(a)]^2 \left\{ 3P_{13}(k) + 4P_{22}(k) \right\} \frac{ }{ } \right. \\
&\qquad\qquad\qquad\qquad\left. 
- 6 [D_1(a) D_1'(a)]^2 \bar{c}_1 k^2 P_{11}(k) \rp \ , \label{eq:ppipi-eds} \\
\nn
P_{\delta\pi_{\rm S}}(a,k) &= - \lp a\H \rho_{\rm b} \rp \lp
D_1(a) D_1'(a) \, P_{11}(k)
+ 2[D_1(a)]^3 D_1'(a) \left\{ P_{13}(k) + P_{22}(k) \right\} \frac{ }{ } \right. \\
&\qquad\qquad\qquad\qquad\left. 
- 4[D_1(a)]^3 D_1'(a) \, \bar{c}_1 k^2 P_{11}(k) \rp \ . \label{eq:pdpi-eds}
\end{align}
Here we assumed that the time dependence of $\bar{c}_1=(2\pi)\co/\knl^2$ is the same as the one from~(\ref{eq:cntdep}). We should emphasize that the power spectra involving momenta depend sensitively on the assumptions made about the time-dependence of the $c_n(a)$ functions defined in Sec.~\ref{subsec:ct}, since we must take time derivatives of $\delta(\vk,a)$. In particular, if the time-dependence of $c_1(a)$ is not given by Eq.~(\ref{eq:cntdep}), but has some different form, which we could write as
\be
c_1(a) = \fnt(a) \times \bar{c}_1 ( 9 D_1(a)^2 \H^2 f^2) \ ,
\label{nontrivialParam}
\eeq
then Eq.~(\ref{eq:ppipi-eds}) would contain an additional term involving $\fnt'(a)$. We will see below that the assumption~(\ref{eq:cntdep}) yields predictions for $P_{\pi_{\rm S}\pi_{\rm S}}$ and $P_{\delta\pi_{\rm S}}$ of sufficient accuracy, given the quality of the simulation data we use. However, since this assumption (along with the time-dependence of the perturbative solutions for $\delta$ and $\theta$, Eqs.~(\ref{eq:deltaexp}) and~(\ref{eq:thetaexp})) is not well-controlled, possible deviations from it should be investigated, using the exact analytical time-dependence described in Ref.~\cite{Carrasco:2012cv}, and measuring the power spectrum of matter from $N$-body simulations at various redshifts. We leave this for future work.

Furthermore, notice that $P_{\pi_{\rm S}\pi_{\rm S}}$, contrary to $P_{\delta\delta}$ and $P_{\delta\pi_{\rm S}}$, is a quantity that depends  on the very long wavelength power spectrum (i.e. not a so-called IR-safe quantity). For this reason, it might well be that results will improve after the IR modes have been properly resummed, along the lines of the techniques derived in~\cite{Crocce:2005xy,Bernardeau:2011vy,Anselmi:2012cn,Blas:2013bpa}. We leave this as well to future work.

\subsection{Results for Correlation Functions Involving Momentum}
\label{subsec:momresults}

\begin{figure}[t]
\begin{center}
\includegraphics[width=0.7\textwidth]{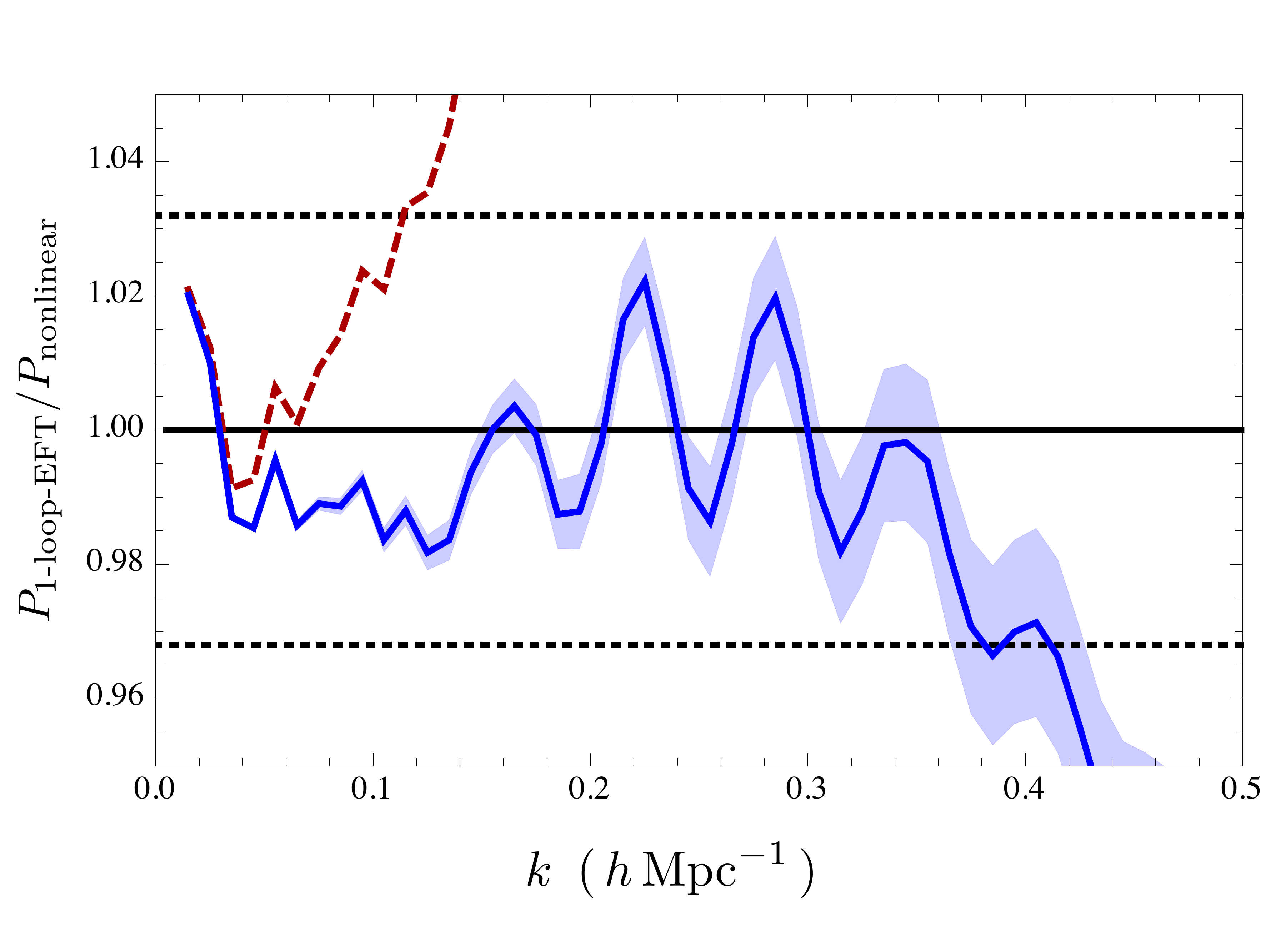}
\caption{\label{fig:vlah_dd_fit_z0} \small\it
{\bf Effectiveness of the one-loop EFT matter power spectrum compared to a simulation with available momentum data.}  Here we plot the ratio of  the one-loop EFT prediction to the non-linear matter power spectrum from the simulations of Okumura {\it et al.}~\cite{Okumura:2011pb}, who are able to provide momentum data as well.  The blue band shows the range of the prediction based on 1-$\sigma$ error on the fitted value of $\co$, while the red dashed line shows the one-loop SPT prediction. The dotted black lines corresponds to 1\% agreement with the data, which itself is estimated to have 1\% statistical and 2\% systematic uncertainties, added in quadrature. The reach in $k$ is comparable as with the Coyote~data, but the size of the oscillations is a bit larger.  See Fig.~\ref{fig:vlah_pp} for using the EFT at one-loop to predict the momentum divergence power-spectrum, and Fig.~\ref{fig:vlah_dp} for the cross-spectrum of $\delta$ and $\pi_{\rm S}$. 
}
\end{center}
\end{figure}

First, we fit the one-loop EFT prediction for the matter power spectrum, Eq.~(\ref{eq:peft1loop}), to the $z=0$ spectrum obtained from Okumura {\em et al.} As in Sec.~\ref{sec:results}, we use a least $\chi^2$ procedure for $0.15 \hinvMpc < k < 0.25 \hinvMpc$ to determine the best-fit value of $\co$, sampling the measured power spectra every $0.01\hinvMpc$. The reported errors on the power spectrum are based on the variance between realizations, but they do not account for systematic errors in the simulations, which are likely to be at the few-percent level. Because of this, we use a 2\% errorbar on each point (assumed to be uncorrelated between different points), unless the reported error is greater than this. Using this procedure, we obtain \footnote{ While the nonlinear spectra obtained from the cosmology of Sec.~\ref{sec:results} and here differ by roughly $\sim$1\% in the range we fit over, the two linear power spectra differ by $\sim$3\% in this range. This suggests that the relative difference of the power spectra is parametrically of the same order as the difference in the cosmological parameters, but with order one numerical coefficients relating the two, maybe due to statistical or systematic mistakes, or to non-linear corrections. This is consistent with the observation that the two values of $c_{s(1)}$  differ by twice as much as the difference in the cosmological parameters: 8\% vs 4\%.  On top of this, this discrepancy in the values of $c_{s(1)}$ is just 1.5$\sigma$, so a shift of just $4\%$ brings the two values within 1$\sigma$ of each other.}
\beq
\co =( 1.76 \pm 0.09) \times \frac{1}{2\pi} \lp \frac{\knl}{\invMpc} \rp^{2} \qquad (\text{1-$\sigma$}) .  
\eeq
This fit is shown in Fig.~\ref{fig:vlah_dd_fit_z0}. Notice that the reach of the fit at high $k$ is comparable to the reach of the one-loop results using the Coyote data (see Fig.~\ref{fig:SPT}), as it should be. However, the size of the oscillations is a bit larger. We take this as an indication that the numerical errors in the simulations we are using here are a bit larger than the purely statistical errors reported by the authors.

\begin{figure}[t]
\begin{center}
\includegraphics[width=0.7\textwidth]{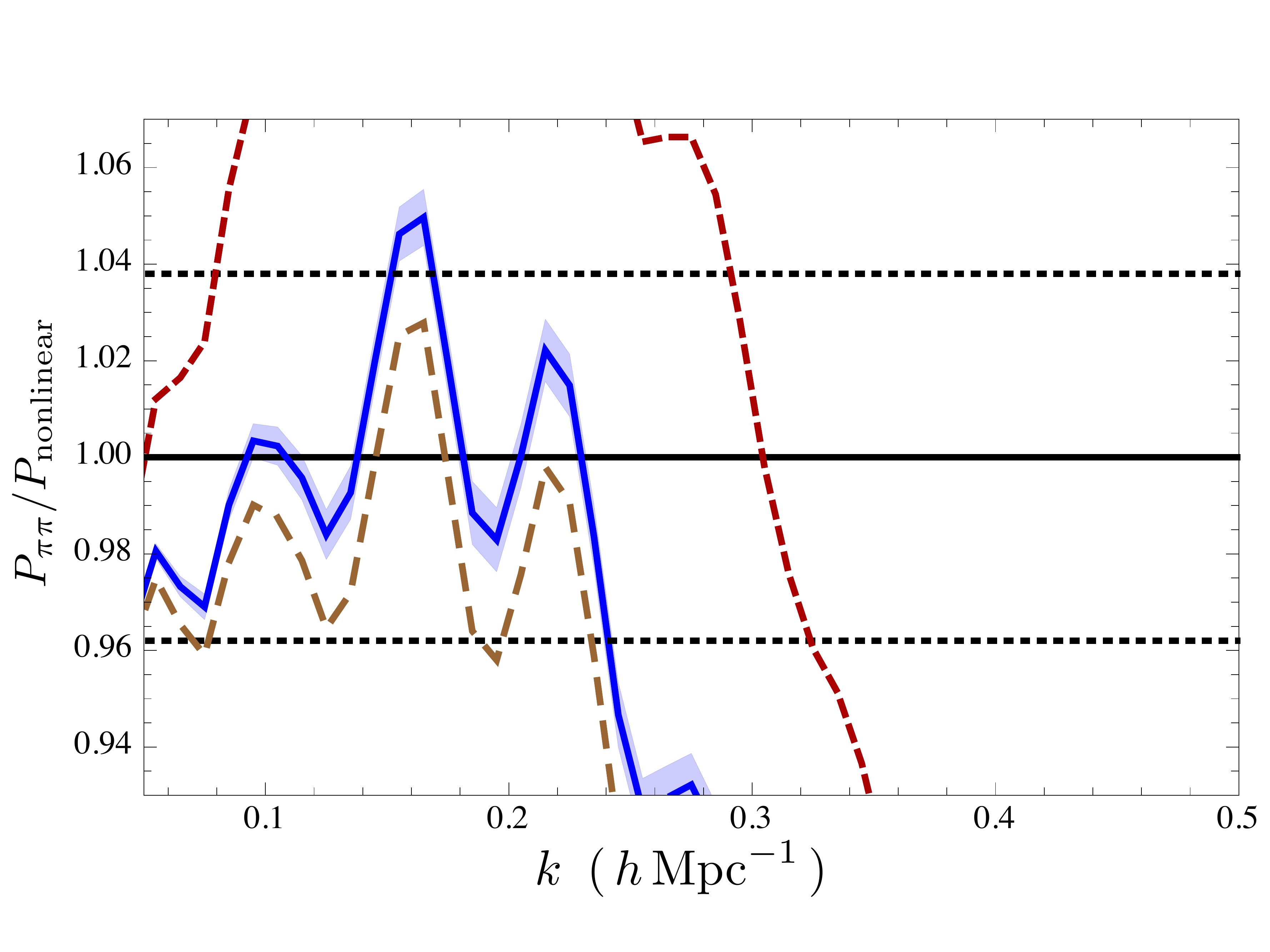}
\caption{\label{fig:vlah_pp} \small\it  
{\bf  Effectiveness of the one-loop EFT prediction of the momentum divergence power spectrum.}   Plot of the one-loop EFT prediction for the momentum divergence power spectrum $P_{\pi \pi}$ normalized to measurements from the simulations of Okumura {\it et al.}~\cite{Okumura:2011pb}. The blue band shows the range of the prediction based on 1-$\sigma$ error on the value of $\co$ that has been fit to the matter power spectrum. The dotted black lines corresponds to 1\% agreement with the data, which itself is estimated to have 2\% statistical and 2\% systematic uncertainty (added in quadrature).
The red dashed line shows the one-loop SPT prediction, while the brown long-dashed line shows the EFT prediction without assuming $\left.\fnt'\right|_{a=1}=0$, and instead fitting for it directly to data for $P_{\delta\pi_{\rm S}}$. It is possible that the (relatively) low reach in $k$ is due to the lack of IR-safety of this quantity.
}
\end{center}
\end{figure}

\begin{figure}[t]
\begin{center}
\includegraphics[width=0.65\textwidth]{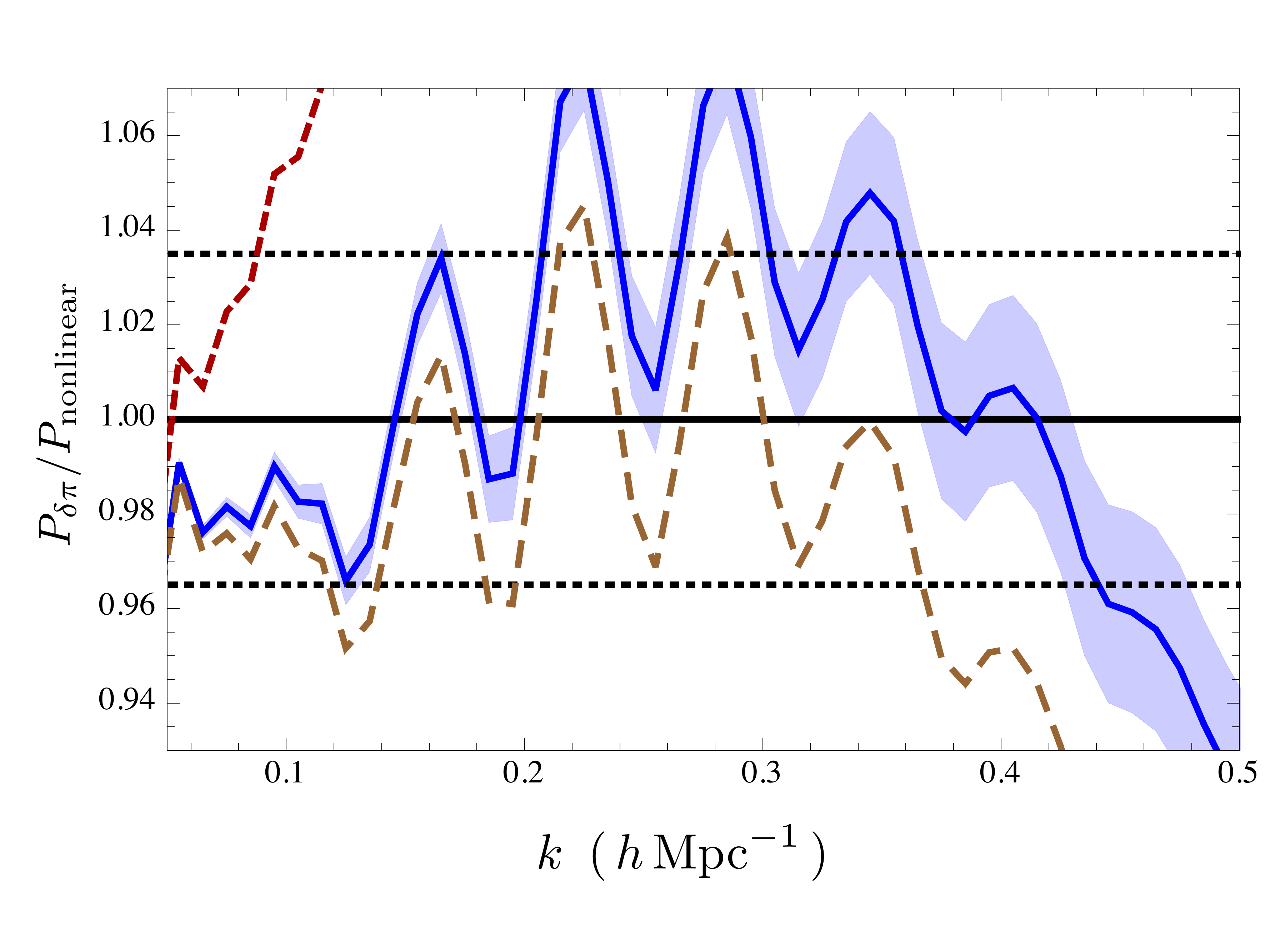}
\caption{\label{fig:vlah_dp} \small\it   
{\bf Effectiveness of the one-loop EFT prediction for the cross spectrum.} Comparison of the one-loop EFT prediction for the cross spectrum of $\delta$ and $\pi_{\rm S}$ to measurements from the simulations of Okumura {\it et al.}~\cite{Okumura:2011pb}. The blue band shows the range of the prediction based on 1-$\sigma$ error on the value of $\co$ that has been fit to the matter power spectrum. The red dashed line shows the one-loop SPT prediction, while the brown long-dashed line shows the one-loop EFT prediction without assuming $\left.\fnt'\right|_{a=1}=0$, and instead fitting for it directly to the $P_{\pi\delta}$ data. The dotted black lines corresponds to 1\% agreement with the data, which itself is estimated to have 1.5\% statistical and 2\% systematic uncertainty (added in quadrature). The reach of the fit in $k$ up to about $0.4-0.5 \invMpc$ is comparable to the reach for  the matter power spectrum at one loop, as it should be. However, the size of the oscillations is worryingly large. It is evident that the fits improve relevantly by allowing $\fnt'(a)\neq 0$. 
}
\end{center}
\end{figure}

We then use this value for $\co$ in the other one-loop predictions, Eqs.~(\ref{eq:ppipi-eds}) and~(\ref{eq:pdpi-eds}), and compare to the measured spectra in Figs.~\ref{fig:vlah_pp} and~\ref{fig:vlah_dp}. For the momentum power spectrum, we obtain 2\% agreement with estimated errors on the data up to $k\sim 0.24 \invMpc$, a considerable improvement over the one-loop SPT prediction (also shown). It might be that the reach is limited by some IR divergences~\footnote{This is also the reason why we do not study explicitly $\pi_{\rm V}$ here: it requires slightly more work to derive the one-loop results, and given that it is not IR safe, the reach might be limited by IR-issues rather than UV-issues, which is our focus here.}, and we therefore expect the situation to improve  for the matter-momentum cross spectrum, which instead is IR-safe. Indeed in this case the reach of the fit in $k$ improves considerably up to $k\sim 0.4-0.5\invMpc$. This is a value comparable to the reach of $P_{\delta\delta}$ at one loop, as it should be. However, there are worrisomely large oscillations at lower $k$'s. This may arise from non-trivial time-dependence of the EFT parameters, as discussed in the previous section. If, instead of taking $\fnt(a)=1$, we allow it to have some nontrivial form, this amounts to making the replacement
\beq
\bar{c}_1 \to \bar{c}_1 \left[ \fnt(a) + \gamma \fnt'(a) \frac{D_1(a)}{D_1'(a)} \right]
\eeq
in Eqs.~(\ref{eq:ppipi-eds}) and~(\ref{eq:pdpi-eds}), where $\gamma=1/3$ in~(\ref{eq:ppipi-eds}) and $\gamma=1/4$ in~(\ref{eq:pdpi-eds}). If we then fit the quantity in square brackets directly to $P_{\delta\pi_{\rm S}}$ at $z=0$, and use the fact that $\fnt(1)=1$ by definition, we find
\beq
\fnt'(1) \approx 0.27\ ,
\eeq
which is a small, but sizeable, correction.
The $P_{\delta\pi_{\rm S}}$ curve using this value (the brown dashed line in Fig.~\ref{fig:vlah_dp}) exhibits much better agreement with non-linear data (out to $k\sim 0.4 \invMpc$). The  $P_{\pi_{\rm S}\pi_{\rm S}}$ curve shows mild improvement. We take this as an indication that it would be worthwhile to explore this time-dependence further in the future. { In particular, the favor of the fit towards $\fnt'>0$ is consistent with the following physical interpretation.  As discussed, the $\bar c_1$ counterterm is supposed to encapsulate the effect at long wavelength of the short distance physics that has become non-linear. From an argument based on virialization, the modes that most contribute to the $\bar c_1$ are expected to be the ones around the non-linear scale. Since the power spectrum evaluated at that scale becomes less and less steep as we move back in the past, from $n\simeq -2$ to $n\simeq -3$, we expect that $\bar c_1$ decreases faster than if the slope would not change. Just to make the reasoning very clear, in the limit that the power spectrum gots to zero above some $k_{\rm cut}$, we would expect $\bar c_1$ to drop rapidly to zero as, as move backwards in the past, the $\knl$ scale becomes shorter than $k_{\rm cut}$.  Since in the case of unchanging slope we would expect a time dependence as in~(\ref{eq:cntdep}), we expect $\fnt'>0$ in the true universe.}

It would be additionally worthwhile to compare with precise $N$-body simulations at various redshifts, in particular as the size of the oscillatory features is larger than when we compare to Coyote data. As with the prediction of the power spectrum at two loops, these results represent a strong confirmation of the internal consistency of the EFTofLSS, as only one (or at most two) unknown parameters are able to predict several different observables to much greater accuracy and reach in wavenumber.




\section{Discussion}
\label{sec:discussion}

The bottom line result of our paper is very simple to state: the  EFTofLSS prediction at two loops matches to percent accuracy the non-linear power spectrum up to $k\sim 0.6\,\invMpc$, with just one parameter dependent upon the microphysics. Given that Standard Perturbation Theory stops converging at $k\sim 0.1\,\invMpc$, our results show that we can access a factor of order 220 more  dark matter quasi-linear modes than naively expected. Similar conclusions seems to be shared by the study of the momentum power spectrum, that we have now computed at one loop. This represents a fantastic opportunity and a challenge for theorists and observers in the LSS community, as we now explain.

It is fair to say that we will probably not understand in an analytic way more modes than the ones that are understandable for the dark matter clustering alone. In truth, halos and baryonic physics might prevent us for exploiting all these modes.  
Therefore, we should consider the cosmological information contained in dark matter clustering as an upper bound to the amount of cosmological information extractable from LSS surveys. However, with this in mind, we emphasize that all forecasts that previously stopped at the maximum wavenumber $k_{\rm max}\sim 0.1\,\invMpc$ (due to a belief that dark matter clustering could not be predicted at higher scales), can be potentially extended, now with the EFT, to $k_{\rm max}\sim 0.6\,\invMpc$.  This is a huge gain in information. To put forth a concrete example, limits on inflationary non-Gaussianities of the equilateral~\cite{Creminelli:2005hu} or orthogonal~\cite{Senatore:2009gt} kind, assuming a cosmic variance limited experiment and assuming the same scaling as in the linear regime, go as $k_{\rm max}^{-3/2}$. Preliminary forecasts for these parameters for the Euclid survey~\cite{Sefusatti:2007ih}, using $k_{\rm max}\simeq 0.1\,\invMpc$ at redshift $z=0$, give constraints of order $\Delta f_{\rm NL}^{\rm equal,ortho}\sim 10$, 
which is about a factor of 7 improvement from the Planck's limits~\cite{Ade:2013ydc}. If we naively rescale as stated the above limits by the $k_{\rm max}$ that we find in this paper for the dark matter clustering, we find
\be\label{eq:guess}
\Delta f_{\rm NL}^{\rm equil.,\,ortho.}\sim 10\quad\rightarrow\quad\Delta f_{\rm NL}^{\rm equil.,\,ortho.}\sim 1/2\ ,
\ee
which is about a factor of 150 improvement with respect to the Planck's limits. 

This is a fantastic improvement. To understand how big this improvement is, we should remember that in going from the WMAP satellite to the Planck satellite, limits on inflationary non-Gaussianities improved by a factor of three or so. In terms of the scale suppressing the operators in the Effective Field Theory of Inflation~\cite{Cheung:2007st}, roughly $\sim(\d\pi)^3/\Lambda^2$, absence of non-Gaussianity in Planck implies that we simply raised the scale $\Lambda$ by a mere factor of about 1.7 with respect to the scale implied by WMAP. Clearly, a further improvement of 12 as implied by our rescaling would be a major improvement. In particular, it would allow us to constrain the operators of the EFT of Inflation to be comparable to the size that can be expected in slow roll inflation; in a sense, allowing us to discover that the inflation that happened in our universe was of the slow roll kind. This would be a major discovery that would be realized in absence of detection. Of course, improving the limits by a factor of 150 means also that there would be a relatively good chance at discovering non-gaussianities, something that would be of tremendous importance. 

In this discussion, we have focussed only on non-Gaussianities, as we consider them to be one of most important probes of the inflationary physics. Of course similar rescalings should apply to the forecasts for neutrino masses, dark energy fluctuations, tilt and running of the spectral index, et cetera\,.

We stress that our rescaling of the limits on non-Gaussianities should be taken just as estimates: we neglect noise, shot noise, corrections in scaling due to mild non-linearities, biases, redshift space distorsions and baryonic physics. Furthermore the results of our paper about the dark matter clustering should be further confirmed using analysis of $N$-body simulations and by computing additional observables, such as the 3-point function, that are less simple to reproduce. But we believe our results do represent a significant challenge for the LSS community: given that, thanks to the EFTofLSS, it seems that we can understand dark matter clustering to much higher wavenumber than previously believed, can we dominate all other sources of error to improve the limits as much as seen in Eq.~(\ref{eq:guess})? If the answer will be `yes', then LSS surveys will be an incredibly powerful new probe of the physics of the early universe.


\section*{Acknowledgments}
We thank  Asimina Arvanitaki, Tom Abel, Tobias Baldauf, Daniel Baumann, Guido D'Amico, Savas Dimopoulos, Lance Dixon, Eiichiro Komatsu, Rafael Porto,  Emiliano Sefusatti, Steve Shenker, Slava Rychkov, Filippo Vernizzi, Zvonimir Vlah,  and especially Francis Bernardeau, Uro\v{s} Seljak and Matias~Zaldarriaga for useful conversations.  We thank Tobias Baldauf, Uro\v{s} Seljak and Zvonimir Vlah for providing us with the detailed data of the Okumura {\it et al} simulations that we use for the momentum calculation. We would like to thank Academic Technology Services at UCLA for computer support, as well as Stuart Marshall and KIPAC Computing for computer support. We also thank the KITP Santa Barbara for hospitality. J.J.M.C.~is supported by the Stanford Institute for Theoretical Physics and the NSF grant no. PHY-0756174 and a 
grant from the  John Templeton Foundation.  S.F. is partially supported by the Natural Sciences and Engineering Research Council of Canada.   D.G.~is supported in part by the Stanford ITP and by the U.S. Department of Energy contract to SLAC no.\ DE-AC02-76SF00515.  L.S. is supported by DOE Early Career Award DE-FG02-12ER41854 and by NSF grant PHY-1068380.  The opinions expressed in this publication are those of the authors and do not necessarily reflect the views of the John Templeton Foundation.

\appendix

\section*{Appendix}

\section{One and Two-Loop Results in the Scaling Universe}\label{app:loops}

In this appendix we collect some results from calculations in the scaling universe.  For the $n=-2$ scaling universe, the one-loop diagrams are given by~\footnote{Our conventions for $\knl$ differ from \cite{Pajer:2013jj}
by $\knl^{\text{ours}} =(4\pi)^{1/(3+n)}\; \knl^{\text{theirs}}$. 
} \cite{Pajer:2013jj}
\bea
P^{\text{finite}}_{13} &=& (2 \pi ) \frac{5 \pi^2}{56} \frac{k}{\knl} P_{11}(k) \simeq 0.88 \times (2\pi) \frac{k}{\knl} P_{11}(k) \\
P^{\text{finite}}_{22} &=&(2\pi) \frac{75 \pi^2}{392} \frac{k}{\knl} P_{11}(k)\simeq 1.88 \times (2\pi) \frac{k}{\knl} P_{11}(k) \ .
\eea 
For comparison, the two-loop result is given by
\begin{align}
\nn
P_\text{2-loop}^{n=-2}(k) &=  \lb (2\pi) 3.5 \, \text{log}(k/\Lambda)
+ (2\pi)^2 \, 3.8 \rb  \frac{k^2}{\knl^2} P_{11}
 + (2\pi)^2 \, 1.5 \frac{k^3}{\Lambda\, \knl^2} P_{11} - (2\pi)^2 \, 27.5 \frac{k_{\rm min} \, k}{\knl^2} P_{11} \\
&= \frac{(2\pi)^5}{ \knl^3}  \left[ \frac{3.5}{2\pi} \log(k /\Lambda) + 3.8 
+ 1.5 \frac{k }{\Lambda} - 27.5 \frac{k_{\rm min}}{k} \right]
\end{align}

For the $n=-3/2$ scaling universe, the results are slightly more surprising.  At one loop, one has
\bea
P^{\text{finite}}_{13} &=& (2 \pi ) \frac{1984 \pi}{6615} \frac{k^{3/2}}{\knl^{3/2}} P_{11}(k) \sim 0.94 \times (2\pi) \frac{k^{3/2}}{\knl^{3/2}} P_{11}(k) \\
P^{\text{finite}}_{22} &\sim& - 0.464 \times (2\pi) \frac{k^{3/2}}{\knl^{3/2}} P_{11}(k) \ .
\eea 
We see that there will be a significant cancellation between $P_{22}$ and $P_{13}$ in computing $P_{\text{1-loop}}$.  These results become more dramatic when we consider two loops,
\begin{align} 
\nn
P_\text{2-loop}(k) &= - (2\pi) \, 2.3 \frac{\Lambda \, k^2}{\knl^3} P_{11}
- (2\pi)^2 \, 0.2 \frac{k^3}{\knl^3} P_{11} 
+ (2\pi)^2 \, 0.6 \frac{k^{7/2}}{\Lambda^{1/2} \, \knl^3} P_{11} \\
&= \frac{(2\pi)^5}{ \knl^3}  \lp \frac{k}{\knl} \rp^{3/2} \left[ - \lp  \frac{2.3}{2\pi} \rp \frac{\Lambda}{k} 
- 0.2 + 0.6 \frac{k^{1/2} }{\Lambda^{1/2}} \right] \ .
\end{align}
The small size of the $\Lambda$-independent term might be surprising, given our loop counting we expected ${\cal O}(2 \pi)$, but the small number can be understood as cancellations due to the relative sizes of $2 P_{22}$ and $P_{13}$, which is the combination that appears in reducible diagrams not yet heavily suppressed at two loops.

\section{SPT Formulas up to Two Loops}\label{app:spt}

The loop corrections to the power spectrum in SPT are conventionally written in terms of separate diagrams, which themselves are integrals of factors of $P_{11}$ times symmetrized kernels $F_n^{\rm (s)}$ and $G_n^{\rm (s)}$ that can be obtained from recurrence relations (see, e.g.,~\cite{Bernardeau:2001qr}). The separate diagrams also have different properties in the UV. However, for numerical evaluation, it is extremely useful that all diagrams at each loop order are combined into a single integrand, in order that IR divergences cancel between the different diagrams {\em before} integration. The resulting integrand is called IR-safe~\cite{Carrasco:2013sva}. For convenience, we include both the separated and IR-safe formulas in this appendix.

\subsection{One Loop}

The one-loop correction can be written as
\be
P_\text{1-loop} = P_{13}+P_{22}
\ee
where
\begin{align}
\nn
P_{13}(k) &= \int \frac{d^3q}{(2\pi)^3} \; 6 P_{11}(k) P_{11}(q) \; F_3^{(\rm s)}(\vk,\vq,-\vq)\ , \\
P_{22}(k) &= \int \frac{d^3q}{(2\pi)^3}\; 2 P_{11}(q) P_{11}(|\vk-\vq|) \left[ F_2^{(\rm s)}(\vq,\vk-\vq) \right]^2 \ ,
\end{align}
with the kernels $F_2^{\rm (s)}(\vq,\vk-\vq)$ and $F_3^{\rm (s)}(\vk,\vq,-\vq)$ given by
\begin{align}
F_2^{\rm (s)}(\vq,\vk-\vq) &=
\frac{k^2 \l 7\vkdq + 3q^2 \r - 10 \l \vkdq \r^2}{14q^2|\vk-\vq|^2}, 
\end{align}
\begin{align}
\nn
F_3^{\rm (s)}(\vk,\vq,-\vq) &=
\frac{1}{|\vk-\vq|^2} \left[ \frac{5k^2}{126} - \frac{11\vkdq}{108}
+ \frac{7( \vkdq )^2}{108k^2} - \frac{k^2 ( \vkdq )^2}{54q^4}
+ \frac{4( \vkdq )^3}{189q^4} \right. \\
\nn
&\qquad \left. - \frac{23k^2 \vkdq}{756q^2}
+\frac{25( \vkdq )^2}{252q^2} - \frac{2 ( \vkdq )^3}{27k^2q^2} \right] \\
\nn
&\quad +\frac{1}{|\vk+\vq|^2} \left[ \frac{5k^2}{26} + \frac{11\vkdq}{108}
- \frac{7(\vkdq)^2}{108k^2} - \frac{4k^2(\vkdq)^2}{27q^4}
- \frac{53(\vkdq)^3}{189q^4} \right. \\
&\qquad \left. + \frac{23k^2 \vkdq}{756q^2} 
- \frac{121(\vkdq)^2}{756q^2} - \frac{5(\vkdq)^3}{27k^2q^2}\right] \ .
\end{align}

The IR-safe formula is~\cite{Carrasco:2013sva}
\bea
\nn
P_{\text{1-loop IR-safe}} 
&=& \int \frac{d^3q}{(2\pi)^3}
\left[ 6 P_{11}(k) P_{11}(q) \, F_3^{(\rm s)}(\vk,\vq,-\vq) \right. \\
\nn
&&\qquad\qquad \left.
+ 2 P_{11}(q) P_{11}(|\vk-\vq|) \left[ F_2^{(\rm s)}(\vq,\vk-\vq) \right]^2 \Theta(|\vk-\vq|-q) \right. \\
&&\qquad\qquad \left.
+ 2 P_{11}(q) P_{11}(|\vk+\vq|) \left[ F_2^{(\rm s)}(-\vq,\vk+\vq) \right]^2 \Theta(|\vk+\vq|-q) \right]\ .
\eea

\subsection{Two Loops}

In separated form, there are four 2-loop diagrams (see for example~\cite{Carlson:2009it}):
\beq
P_{\text{2-loop}} = P_{15}+P_{24}+ P_{33}^{\rm (I)}+ P_{33}^{\rm (II)}
\eeq
where
\begin{align}
\label{eq:2loopints}
\nn
P_{51}(k) &= \int \!\! \frac{d^3 p}{(2\pi)^3} \int \!\! \frac{d^3 q}{(2\pi)^3} \, 30
F_5^{(\rm s)}(\vk,\vkp,-\vkp,\vp,-\vp) P_{11}(k) P_{11}(q) P_{11}(p)\ , \\
\nn
P_{42}(k) &=  \int \!\! \frac{d^3 p}{(2\pi)^3} \int \!\! \frac{d^3 q}{(2\pi)^3} \,
24 \, F_2^{(\rm s)}(\vkp,\vk-\vkp) F_4^{(\rm s)}(-\vkp,\vkp-\vk, \vp,-\vp) 
P_{11}(q) P_{11}(p) P_{11}(|\vk-\vkp|)\ , \\
\nn
P_{33}^{\rm (I)}(k) &= \int \!\! \frac{d^3 p}{(2\pi)^3} \int \!\! \frac{d^3 q}{(2\pi)^3} \,
 9 F_3^{(\rm s)}(\vk,\vq,-\vq) F_3^{(\rm s)}(-\vk,\vp,-\vp) P_{11}(k) P_{11}(q) P_{11}(p)\ , \\
 \nn
P_{33}^{\rm (II)}(k) &= \int \!\! \frac{d^3 p}{(2\pi)^3} \int \!\! \frac{d^3 q}{(2\pi)^3} \,
 6 F_3^{(\rm s)}(\vkp,\vp,\vk-\vkp-\vp) F_3^{(\rm s)}(-\vkp,-\vp,-\vk+\vkp+\vp) \\
&\qquad\qquad\qquad\qquad\qquad\qquad \times P_{11}(q) P_{11}(p) P_{11}(|\vk-\vkp-\vp|)\ ,
\end{align}
and the kernels $F_{3,4,5}^{(\rm s)}$ are found from well-known recurrence relations, followed by symmetrization over all arguments. We repeat these relations here for convenience:
\bea
\nn
F_n(\vq_1,\dots,\vq_n) &=& \sum_{m=1}^{n-1} \frac{G_m(\vq_1,\dots,\vq_m)}{(2n+3)(n-1)}
\left[ (2n+1) \frac{\vk\cdot\vk_1}{k_1^2} F_{n-m}(\vq_{m+1},\dots,\vq_n) \right. \\
\nn
&&\qquad\qquad\qquad\qquad\qquad \left. +  \frac{k^2 (\vk_1\cdot\vk_2)}{k_1^2 k_2^2}
G_{n-m}(\vq_{m+1},\dots,\vq_n) \right], \\
\nn
G_n(\vq_1,\dots,\vq_n) &=& \sum_{m=1}^{n-1} \frac{G_m(\vq_1,\dots,\vq_m)}{(2n+3)(n-1)}
\left[ 3 \frac{\vk\cdot\vk_1}{k_1^2} F_{n-m}(\vq_{m+1},\dots,\vq_n) \right. \\
&&\qquad\qquad\qquad\qquad\qquad \left. + n \frac{k^2 (\vk_1\cdot\vk_2)}{k_1^2 k_2^2}
G_{n-m}(\vq_{m+1},\dots,\vq_n) \right]\ ,
\label{eq:sptrecurrence}
\eea
where $\vk_1=\vq_1+\cdots+\vq_m$, $\vk_2=\vq_{m+1}+\cdots+\vq_n$, $\vk=\vk_1+\vk_2$, and $F_n=G_n=1$.

The IR-safe formula is~\cite{Carrasco:2013sva}
\bea \nn
P_{\text{2-loop IR-safe}}(k) &=&  \int \!\! \frac{d^3 p}{(2\pi)^3} \int \!\! \frac{d^3 q}{(2\pi)^3}
\frac{1}{4}
\left[ \tilde{p}_\text{2-loop}(\vk,\vq,\vp) + \tilde{p}_\text{2-loop}(\vk,-\vq,\vp) \right. \\
&&\qquad\qquad\qquad\qquad
\left. + \tilde{p}_\text{2-loop}(\vk,\vq,-\vp) + \tilde{p}_\text{2-loop}(\vk,-\vq,-\vp) \right]\ ,
\eea
where
\bea
\nn
\tilde{p}_\text{2-loop}(\vk,\vq,\vp) &=& 
\left\{ 60 F_5^{(\rm s)}(\vk,\vkp,-\vkp,\vp,-\vp) P_{11}(k) P_{11}(q) P_{11}(p) \right. \\
\nn
&&\left. + 18 F_3^{(\rm s)}(\vk,\vkp,-\vkp) F_3^{(\rm s)}(-\vk,\vp,-\vp)
	P_{11}(k) P_{11}(q) P_{11}(p) \right. \\
\nn
&&\left. + 48F_2^{(\rm s)}(\vkp,\vk-\vkp) F_4^{(\rm s)}(-\vkp,\vkp-\vk, \vp,-\vp) 
P_{11}(q) P_{11}(p) P_{11}(|\vk-\vkp|) \, \Theta(|\vk-\vkp|-q) \right. \\
\nn
&&\left. + 48F_2^{(\rm s)}(\vp,\vk-\vp) F_4^{(\rm s)}(-\vp,\vp-\vk, \vq,-\vq) 
P_{11}(q) P_{11}(p) P_{11}(|\vk-\vp|) \, \Theta(|\vk-\vp|-p) \right. \\
\nn
&&\left. + 36 F_3^{(\rm s)}(\vkp,\vp,\vk-\vkp-\vp) F_3^{(\rm s)}(-\vkp,-\vp,-\vk+\vkp+\vp) \right. \\
&&\qquad \left. \times P_{11}(q) P_{11}(p) P_{11}(|\vk-\vkp-\vp|) \, \Theta(|\vk-\vq-\vp|-p)
\right\} \Theta(p-q)\ .
\label{eq:p2looplong}
\eea

\section{Diagrams Involving $K(a,a')$}\label{app:solutions}

\subsection{Solutions to Equations of Motion}

We wish to solve the following equations: 
\bea
&&a\H \delta'+\theta= - \! \int \frac{d^3q}{(2\pi)^3}\alpha(\vkp,\vk-\vkp)\delta(\vk-\vkp)\theta(\vkp)\ , \\
\nonumber
&& a\H \theta'+\H \theta+\frac{3}{2} \H_0^2 \, \Omm  \frac{a_0^3}{a} \delta
= - \! \int \frac{d^3q}{(2\pi)^3}\beta(\vkp,\vkkp)\theta(\vk-\vkp)\theta(\vkp) \\
\label{eq:eulerapp}
&&\qquad\qquad\qquad\qquad\qquad\qquad\quad
+ \epsilon\, k^2 \int \frac{da'}{a'\H(a')} K(a,a') [ \delta(a',\vx_{\rm fl})]_{\vk} \ .
\eea
with the ansatz
\bea
\label{eq:delansatz}
\delta(a,\vk) &=& \sum_{n=1}^\infty \, [D_1(a)]^n \delta^{(n)}(\vk)
+ \epsilon \sum_{n=1}^\infty \, [D_1(a)]^{n+2} \tilde{\delta}^{(n)}(\vk)\ , \\
\theta(a,\vk) &=& -\H(a) f \sum_{n=1}^\infty \, [D_1(a)]^n \theta^{(n)}(\vk)
-\epsilon \H(a) f \sum_{n=1}^\infty \, [D_1(a)]^{n+2} \tilde{\theta}^{(n)}(\vk)\ .
\eea
We have inserted a parameter $\epsilon$ to organize the powers of $K(a,a')$, or equivalently $c_n(a)$, that appear. Furthermore, we write the $\vk$-dependent solutions in the following manner:
\bea
\delta^{(n)}(\vk) &=& \int \! \frac{d^3 \vq_1}{(2\pi)^3} \cdots \int \! \frac{d^3 \vq_n}{(2\pi)^3}
(2\pi)^3 \delta_{\rm D}(\vk-\vq_{1\cdots n})
F_n(\vq_1,\dots,\vq_n)
\delta(\vq_1)\cdots\delta(\vq_n)\ , \\
\theta^{(n)}(\vk) &=& \int \! \frac{d^3 \vq_1}{(2\pi)^3} \cdots \int \! \frac{d^3 \vq_n}{(2\pi)^3}
(2\pi)^3 \delta_{\rm D}(\vk-\vq_{1\cdots n})
G_n(\vq_1,\dots,\vq_n)
\delta(\vq_1)\cdots\delta(\vq_n)\ ,
\label{eq:thetakansatz}
\eea
and analogously for $\tilde{\delta}^{(n)}(\vk)$ and $\tilde{\theta}^{(n)}(\vk)$.

Returning to position space for the moment, recall that the final integral in Eq.~(\ref{eq:eulerapp}) depends on $\delta$ evaluated at
\beq
\vx_{\rm fl}[\tau,\tau'] = \vx - \int_{\tau'}^{\tau} d\tau'' \vec{v}(\tau'',\vx_{\rm fl}[\tau,\tau'']).
\eeq
We can put $\delta(\tau',\vx_{\rm fl}[\tau,\tau'])$ into a tractable form by recursively Taylor-expanding $\vx_{\rm fl}$ around $\vx$:
\bea
\nn
&&\delta_\ell \! 
\lp \tau',\vx - \int_{\tau'}^{\tau} d\tau''  \, \vec{v} \! 
\lb \tau'',\vx - \int_{\tau''}^{\tau} d\tau''' \vec{v}(\vx+\cdots) \rb \rp \\
\nn
&&\quad
= 
\delta(\tau',\vx)
- \partial_i \delta(\tau',\vx) \! \int_{\tau'}^\tau d\tau'' v^i(\tau'',\vx)  \\
\nn
&&\quad\quad  
+ \partial_i \delta(\tau',\vx)
\int_{\tau'}^\tau d\tau'' \partial_j v^i(\tau'',\vx)
\int_{\tau''}^\tau d\tau'''  v^j(\tau''',\vx) \\
&&\quad\quad 
+ \frac{1}{2}  \partial_i \partial_j \delta(\tau',\vx)
\int_{\tau'}^\tau d\tau'' v^i(\tau'',\vx)
\int_{\tau'}^\tau d\tau'''  v^j(\tau''',\vx) + \cdots \ .
\label{eq:delfluidexpansion}
\eea
We will find below that these terms are sufficient to provide the diagrams that enter a two-loop calculation, and to make those diagrams IR-convergent. Using Eqs.~(\ref{eq:delansatz})-(\ref{eq:thetakansatz}) and (\ref{eq:delfluidexpansion}) in the equations of motion, it can be shown \cite{Bernardeau:2001qr} that if $\Omm(a)\approx f^2$, then the time-dependent factors drop out of the equations of motion in the SPT case (i.e.~$K(a,a')=0$) and recurrence relations for the kernels can be obtained. This also occurs in our case, under the additional assumption (mentioned in the main text) that {
\beq
c_n(a)  = \bar c_n (9 D_1(a)^{2} \H^2 f^2) \ ,
\eeq
where
\beq
c_n(a) \equiv \int \frac{da'}{a'\H(a')} K(a,a') \frac{D_1(a')^n}{D_1(a)^n} \ .
\eeq
These assumptions are not strictly necessary to make calculations in the EFTofLSS. We just use them because they simplify the algebra without affecting strongly the numerical results~\cite{Carrasco:2012cv}.

Under these assumptions, collecting the $\epsilon^0$ terms yields the standard SPT relations~(\ref{eq:sptrecurrence}) for $F_n$ and $G_n$, while the $\epsilon^1$ terms give equations for $\tilde{F}_n$ and $\tilde{G}_n$:
\begin{align}
\nn
\tilde{F}_n(\vq_1,\dots,\vq_n) &= \frac{1}{(n+1)(n+\frac{7}{2})}
\sum_{m=1}^{n-1} \lb (n+\frac{5}{2}) \alpha(\vk_1,\vk_2)
\mathcal{A}_m(\vq_1,\dots,\vq_n) + \beta(\vk_1,\vk_2) \mathcal{B}_m(\vq_1,\dots,\vq_n)\rb  \\
&\qquad\qquad\qquad\qquad - \frac{{ 9}}{(n+1)(n+\frac{7}{2})} 
\mathcal{S}_n(k), \\
\nn
\tilde{G}_n(\vq_1,\dots,\vq_n) &= \frac{1}{(n+1)(n+\frac{7}{2})}
\sum_{m=1}^{n-1} \lb \frac{3}{2} \alpha(\vk_1,\vk_2)
\mathcal{A}_m(\vq_1,\dots,\vq_n) + (n+2) \beta(\vk_1,\vk_2) \mathcal{B}_m(\vq_1,\dots,\vq_n) \rb \\
&\qquad\qquad\qquad\qquad - { 9} \frac{n+2}{(n+1)(n+\frac{7}{2})} 
 \mathcal{S}_n(k),
\end{align}
where
\bea
&&\mathcal{A}_m(\vq_1,\cdots\,\vq_n) \equiv \\ \nonumber
&&\qquad \tilde{G}_m(\vq_1,\dots,\vq_m)  F_{n-m}(\vq_{m+1},\dots,\vq_n)
+ G_m(\vq_1,\dots,\vq_m)  \tilde{F}_{n-m}(\vq_{m+1},\dots,\vq_n) \ , \\
&& \mathcal{B}_m(\vq_1,\cdots\,\vq_n) \equiv \\ \nonumber
&&\qquad \tilde{G}_m(\vq_1,\dots,\vq_m)  G_{n-m}(\vq_{m+1},\dots,\vq_n)
+ G_m(\vq_1,\dots,\vq_m)  \tilde{G}_{n-m}(\vq_{m+1},\dots,\vq_n) \ .
\eea

The ``source" functions $\mathcal{S}_n$ are given by
\begin{align}
\nn
\mathcal{S}_n(k) &= \bar c_n k^2 F_n(\vq_1,\dots,\vq_n)
+ k^2 \sum_{i=1}^{n-1} \frac{\vk_1 \cdot \vk_2}{k_2^2} \frac{1}{n-i} (\bar c_i-\bar c_n)
F_i(\vq_1,\dots,\vq_i) G_{n-i}(\vq_{i+1},\dots,\vq_n) \\
\nn
&\quad+ k^2
\sum_{i=1}^{n-2} \sum_{j=1}^{n-2} \frac{(\vk_1\cdot\vk_2) (\vk_2\cdot\vk_3)}{k_2^2 k_3^2}
 \frac{1}{n-i-j} \lb \frac{1}{j} (\bar c_i-\bar c_{i+j}) - \frac{1}{n-i} (\bar c_i-\bar c_n) \rb \\
\nn
&\qquad\qquad\qquad\qquad\qquad\qquad \times
F_i(\vq_1,\dots,\vq_i) G_j(\vq_{i+1},\dots,\vq_{i+j}) G_{n-i-j}(\vq_{i+j+1},\dots,\vq_n) \\
\nn
&\quad+ k^2
\sum_{i=1}^{n-1} \sum_{j=1}^{n-1}   \frac{1}{2} \frac{(\vk_1\cdot\vk_2) (\vk_1\cdot\vk_3)}{k_2^2 k_3^2}
\frac{1}{j(n-i-j)} \lb \bar c_i+\bar c_n-\bar c_{n-j} -\bar c_{i+j} \rb \\
&\qquad\qquad\qquad\qquad\qquad\qquad \times
F_i(\vq_1,\dots,\vq_i) G_j(\vq_{i+1},\dots,\vq_{i+j}) G_{n-i-j}(\vq_{i+j+1},\dots,\vq_n) \ ,
\label{eq:snequation}
\end{align}
where in the double sums we use $\vk_1=\vq_1+\cdots+\vq_i$, $\vk_2=\vq_{i+1}+\cdots+\vq_{i+j}$, and $\vk_3=\vq_{i+j+1}+\cdots+\vq_n$.

\subsection{New Diagrams}

Solving the above equations, the kernels we require are the following
\begin{align}
\tilde{F}_1(\vk) &= { - \bar c_1 k^2 } \ , \\
\nn
\tilde{F}_2^{\rm (s)}(\vq,\vk-\vq) &=
 \frac{3(\bar c_1-\bar c_2)k^2}{11} - \frac{6 \bar{c}_1 (\vkdq)}{11}
+ \frac{3 (-21\bar c_1+4\bar c_2) k^2 (\vkdq)}{154 q^2} + \frac{3 \bar c_1 (\vkdq)^2}{11q^2} \\
\nn
 &\quad +\frac{1}{11|\vk-\vq|^2}  \lb -\frac{7 \bar{c}_1 k^4}{2} - \frac{9 \bar{c}_2 k^4}{7}
+ \frac{13\bar{c}_1 k^2 (\vkdq)}{2} + \frac{36\bar{c}_2 k^2 (\vkdq)}{7} \right. \\
\nn
&\quad\left. - 9 \bar{c}_1 (\vkdq)^2 - (\bar{c}_1+3\bar{c}_2) k^2 q^2
+ 6\bar{c}_1 (\vkdq) q^2  \right. \\
&\quad\left. -\frac{\bar{c}_1 k^4 (\vkdq)}{q^2} -\frac{6\bar{c}_2 k^4 (\vkdq)}{7q^2}
+ \frac{2\bar{c}_1 k^2 (\vkdq)^2}{q^2}  \rb \ , 
\end{align}
and also $\tilde{F}_3^{\rm (s)}(\vk,\vq,-\vq)$, which is too lengthy to display here. However, note that $\tilde{F}_3^{\rm (s)}(\vk,\vq,-\vq)$ also contributes a UV divergence to $k^2 P_{11}$, which, as before, is degenerate with the corresponding counterterm, and so we can simply choose not to include that contribution in the first place. This can be accomplished by subtracting the UV-divergent terms from $\tilde{F}_3^{\rm (s)}(\vk,\vq,-\vq)$:
\beq
\tilde{F}_3^\text{(s, no UV)}(\vk,\vq,-\vq) = \tilde{F}_3^{\rm (s)}(\vk,\vq,-\vq)
 - \lim_{q\to\infty}{\tilde{F}_3^{\rm (s)}(\vk,\vq,-\vq)} \ .
\eeq

We can now begin to write the new diagrams containing single powers of $c_n$. At tree-level, there is one, coming from both $\langle \delta^{(1)} \tilde{\delta}^{(1)} \rangle$ and $\langle \tilde{\delta}^{(1)} \delta^{(1)} \rangle$:
\beq
P_{\rm tree}^{(c_{\rm s})} = - { 2 \bar c_1 k^2 P_{11}(k) } \ .
\eeq
The one-loop diagrams can be written in the following IR-safe form:
\begin{align}
\nn
P_{\text{1-loop IR-safe}}^{(c_{\rm s})} &= \int \frac{d^3\vq}{(2\pi)^3}
\left[ 6 P_{11}(k) P_{11}(q) \left\{ \tilde{F}_3^\text{(s, no UV)}(\vk,\vq,-\vq) 
+ \tilde{F}_1^{\rm (s)}(\vk) F_3^{(\rm s)}(\vk,\vq,-\vq) \right\} \right. \\
\nn
&\qquad\qquad\qquad \left.
+ 4 P_{11}(q) P_{11}(|\vk-\vq|) F_2^{(\rm s)}(\vq,\vk-\vq) \tilde{F}_2^{(\rm s)}(\vq,\vk-\vq) \Theta(|\vk-\vq|-q) \right. \\
&\qquad\qquad\qquad \left.
+ 4 P_{11}(q) P_{11}(|\vk+\vq|) F_2^{(\rm s)}(-\vq,\vk+\vq) \tilde{F}_2^{(\rm s)}(-\vq,\vk+\vq) \Theta(|\vk+\vq|-q) \right] \ .
\end{align}
For numerical computations, we define $\mu \equiv \text{cos}(\vkdq)$ and separate $P_{\text{1-loop IR-safe}}^{(c_{\rm s})}$ into three integrals (each of which is individually IR-safe), like so:
\beq
(2\pi) \co P_{\text{1-loop}}^{(c_{\rm s})}(k) \equiv
\bar c_1 \tilde P_1(k) +\bar c_2 \tilde P_2 (k) + \bar c_3 \tilde P_3(k) \ ,
\eeq
where 
\begin{align}
\nn
\tilde{P}_1(k) &= \frac{1}{(2\pi)^2} \int dq  \int d\mu
\frac{k^4 P_{11}(q)}{3003 \lb (k^2+q^2)^2 - 4k^2 q^2 \mu^2 \rb^2} \times \\
\nn
&\quad  \lb
2P_{11}(k) \left\{ 3003k^8 \mu^2 + k^6 q^2(-913+10991\mu^2-22090\mu^4)  \right.\right.\\
\nn
&\quad\quad\left.\left. + q^8(-913+2396\mu^2+6800\mu^4-5280\mu^6) \right.\right.\\
\nn
&\quad\quad\left.\left.+ k^4q^4(-2739+19021\mu^2-33296\mu^4+35032\mu^6)  \right.\right.\\
\nn
&\quad\quad\left.\left. + k^2q^6 (-2739+13429\mu^2-6062\mu^4-37760\mu^6+21120\mu^8) \right\} \right. \\
\nn
&\quad\left. -39 \, \Theta(\sqrt{k^2+q^2-2kq\mu}-q) P_{11}(\sqrt{k^2+q^2-2kq\mu}) \,
(k^2+q^2+2kq\mu)^2 \times \right. \\
\nn
&\quad\quad\left. 
\left\{ 77k^4\mu^2+k^3q(40\mu-306\mu^3) -8kq^3\mu(-4+31\mu^2+15\mu^4)+4q^4(-3+\mu^2+30\mu^4) \right.\right. \\
\nn
&\quad\quad\left.\left. + k^2q^2(3+46\mu^2+364\mu^4)\right\} \right. \\
\nn
&\quad\left. -39 \, \Theta(\sqrt{k^2+q^2+2kq\mu}-q) P_{11}(\sqrt{k^2+q^2+2kq\mu}) \,
(k^2+q^2-2kq\mu)^2 \times \right. \\
\nn
&\quad\quad\left.
\left\{ 77k^4\mu^2-k^3q(40\mu-306\mu^3) +8kq^3\mu(-4+31\mu^2+15\mu^4)+4q^4(-3+\mu^2+30\mu^4) \right.\right. \\
&\quad\quad\left.\left. + k^2q^2(3+46\mu^2+364\mu^4)\right\}
\rb \ ,
 \label{eq:ptilde1}
\end{align}
\begin{align}
\nn
\tilde{P}_2(k) &= -\frac{1}{(2\pi)^2} \int dq \, q \int d\mu
\frac{12k^4 P_{11}(q)}{539 \lb (k^2+q^2)^2 - 4k^2 q^2 \mu^2 \rb^2} \times \\
\nn
&\quad \lb \Theta(\sqrt{k^2+q^2-2kq\mu}-q) P_{11}(\sqrt{k^2+q^2-2kq\mu}) (k^2+q^2+2kq\mu)^2 \times \right. \\
\nn
&\quad\quad\left. \lp 7q^3[3-10\mu^2] + 7k^3\mu[5+2\mu^2] + 7kq^2\mu[1+20\mu^2] + k^2q[15-142\mu^2-20\mu^4] \rp \right. \\
\nn
&\quad\left.-\Theta(\sqrt{k^2+q^2+2kq\mu}-q) P_{11}(\sqrt{k^2+q^2+2kq\mu}) (k^2+q^2-2kq\mu)^2 \times \right. \\
&\quad\quad\left. \lp -7q^3[3-10\mu^2] + 7k^3\mu[5+2\mu^2] + 7kq^2\mu[1+20\mu^2] - k^2q[15-142\mu^2-20\mu^4] \rp
 \rb \ ,
 \label{eq:ptilde2}
\end{align}
and
\beq
\tilde{P}_3(k) = \frac{1}{(2\pi)^2} \int dq \, q^2 \int d\mu
\frac{12k^4(k^2+q^2)(1-\mu^2)^2 P_{11}(k) P_{11}(q)}
{91 \lb (k^2+q^2)^2 - 4k^2 q^2 \mu^2 \rb} \ .
 \label{eq:ptilde3}
\eeq
}
For the two-loop power spectrum, we also need the tree-level diagram that scales like $k^4 P_{11}$, which comes from $\langle \tilde{\delta}^{(1)} \tilde{\delta}^{(1)} \rangle$:
\beq
P_{\rm tree}^{(c_{\rm s},2)}  = { \bar c_1^2 \, k^4 P_{11}  }\ .
\eeq
The relationship between the $\bar{c}_n$ coefficients and the fit parameters $\co$ and $\ct$ is described in Sec.~\ref{sec:results} (see for example Eq.~(\ref{eq:mathcing})).

\section{Calculations with Renormalized Velocity Operator}\label{app:velocity}

In this appendix, we explicitly show that the two-loop calculation of $P_{\delta\delta}(k)$ does not depend on whether the equations of motion for $\delta$ and $v^i$ are written in terms of the bare or renormalized velocity. This validates the choice made in the body of the paper to use the bare velocity (for which the equations of motion more closely resemble the familiar ones of SPT) and in particular showing that those equations can be solved by Taylor expanding in $\vec v$. We do this by rewriting Eqs.~(\ref{eq:master}), in terms of the renormalized velocity, and showing that there are no new extra terms in~$P_{\delta\delta}$. 

The renormalized velocity $v^i_{\rm R}$ is written as the bare velocity $v^i$ plus an appropriate counterterm, whose form is fixed by the equivalence principle and diffeomorphism invariance (see Eq.~(\ref{eq:velocity_def_ren})):
\beq
\label{eq:vsub2}
v_{\rm R}^i = v^i - \int d\tau' \kappa_{\rm v}(\tau,\tau')
\d^i \d^2\phi(\tau',\vx_{\rm fl}[\tau,\tau'])\ .
\eeq
In terms of $v_{\rm R}^i$, the continuity and Euler equations are 
\begin{align}
\nn
&\dot{\delta} = -\frac{1}{a} \d_i \lp [1+\delta]v_{\rm R}^i \rp
- \frac{1}{a} \int \frac{da'}{a'\H(a')} K_{\rm v}(a,a') \d^2\delta(a',\vx_{\rm fl}[a,a']) \\
&\qquad\qquad
- \frac{1}{a} \d_i \lp \delta(a,\vx) \!\!
\int \frac{da'}{a'\H(a')} K_{\rm v}(a,a') \d^i \delta(a',\vx_{\rm fl}[a,a']) \rp\ , \\
\nn
&\dot{v}_{\rm R}^i+H v_{\rm R}^i+\frac{1}{a} v_{\rm R}^j\d_j v_{\rm R}^i+\frac{1}{a}\d^i\phi =
-\tfrac{1}{a} \tildegammai{}^i
-\H(a)\! \int \frac{da'}{a'\H(a')} K_{\rm v}(a,a') \d^i \frac{\d}{\d a} \delta(a',\vx_{\rm fl})  \\
\nn
&\qquad\qquad
- \frac{1}{a} \d_j v_{\rm R}^i(a,\vx) \!\! \int \frac{da'}{a'\H(a')}
K_{\rm v}(a,a') \d^j \delta(a',\vx_{\rm fl}) \\
&\qquad\qquad
- \frac{1}{a} v_{\rm R}^j(a,\vx) \!\! \int \frac{da'}{a'\H(a')}
K_{\rm v}(a,a') \d_j \d^i \delta(a',\vx_{\rm fl})\ .
\end{align}
where we have dropped a term of order $K_{\rm v}^2$ as it contributes at negligible order.  We have also defined the quantities
\beq
K_{\rm v}(a,a') \equiv \frac{3}{2} H_0^2 \, \Omm \frac{a_0^3}{a} \kappa_{\rm v}(a,a')\ ,
\eeq
and 
\bea\label{eq:ridef1}
&&\tildegammai{}^i = \gammai{}^i - (a \H  \partial_a + \H) \int \frac{da'}{a'\H(a')}
K_{\rm v}(a,a')  \d^i \delta(a',\vx_{\rm fl}) \\ \nonumber
&& \qquad\qquad+  a \H \int \frac{da'}{a'\H(a')} K_{\rm v}(a,a') \d^i \frac{\d}{\d a} \delta(a',\vx_{\rm fl})\nonumber \\ 
&&\qquad\equiv \epsilon \int \frac{da'}{a'\H(a')} \tilde K(a,a') \d^i [ \delta(a',\vx_{\rm fl})]_{\vk} \ .
\eea
By construction, the additional terms in $\tildegammai$ are equivalent to a redefinition of $K(a,a')$ because we have removed the $a$-derivative which acts on $\delta(a',\vx_{\rm fl})$ by adding in the definition of $\tildegammai$ the term in the second line of (\ref{eq:ridef1}).  Since the source in $\gammai$ is always $\d^i [ \delta(a',\vx_{\rm fl})]_{\vk}$, these additional terms can only modify the kernel.

In Fourier space, inserting factors of $\epsilon$ and $\epsilon_{\rm v}$ to distinguish the counterterms from $\gammai$ with those from the $v^i$ counterterm, we obtain
\begin{align}
\nn
&a\H \delta'+\theta_R= - \! \int \frac{d^3q}{(2\pi)^3}
\alpha(\vkp,\vk-\vkp)\delta(\vk-\vkp)\theta_R(\vkp)\ 
+ \epsilon_{\rm v} k^2  \int \frac{da'}{a'\H(a')} K_{\rm v}(a,a') [ \delta(a',\vx_{\rm fl})]_{\vk} \\
&\qquad\qquad\qquad
+\epsilon_{\rm v} \int \frac{da'}{a'\H(a')} K_{\rm v}(a,a')
\int \frac{d^3q}{(2\pi)^3} \vk\cdot(\vk-\vq) \delta(a,\vq) [ \delta(a',\vx_{\rm fl})]_{\vk-\vq}\ ,
\label{eq:contvelct}\\
\nn
& a\H \theta_R'+\H \theta_R+\frac{3}{2} \H_0^2 \, \Omm  \frac{a_0^3}{a} \delta
= - \! \int \frac{d^3q}{(2\pi)^3}\beta(\vkp,\vkkp)\theta_R(\vk-\vkp)\theta_R(\vkp) \\
\nn
&\qquad\qquad\qquad\qquad\qquad\qquad\quad
+ \epsilon k^2 \int \frac{da'}{a'\H(a')} \tilde K(a,a') [ \delta(a',\vx_{\rm fl})]_{\vk} \\
\nn
&\qquad\qquad\qquad\qquad\qquad\qquad\quad
+ \epsilon_{\rm v}\, a\H(a) k^2 \int \frac{da'}{a'\H(a')}
K_{\rm v}(a,a') \frac{\d}{\d a}[ \delta(a',\vx_{\rm fl})]_{\vk} \\
&\qquad\qquad\qquad\qquad\qquad\qquad\quad
+ \epsilon_{\rm v} k^2 \int \frac{da'}{a'\H(a')} K_{\rm v}(a,a') 
\int \frac{d^3q}{(2\pi)^3} \frac{\vq\cdot(\vk-\vq)}{q^2} \theta_R(a,\vq)
[ \delta(a',\vx_{\rm fl})]_{\vk-\vq}\ ,
\label{eq:eulervelct}
\end{align}
where $\theta_R$ is now the renormalized quantity $\theta_R=\d_i v^i_R$.

Using the expansion of $\delta(a,\vx_{\rm fl})$ up to third order in fields, as given by Eq.~(\ref{eq:delfluidexpansion}), we find that the last two lines of Eq.~(\ref{eq:eulervelct}) cancel, so that there are no new counterterms appearing in the Euler equation.

At this point, we use a modification of the ansatz from Sec.~\ref{app:solutions}:
\begin{align}
\label{eq:delansatznew}
\delta(a,\vk) &= \sum_{n=1}^\infty \, [D_1(a)]^n \delta^{(n)}(\vk)
+ \epsilon_{\rm v} \sum_{n=1}^\infty \, [D_1(a)]^{n+2} \hat{\delta}^{(n)}(\vk)+ \epsilon \sum_{n=1}^\infty \, [D_1(a)]^{n+2} \tilde{\delta}^{(n)}(\vk)\ , \\
\theta_R(a,\vk) &= -\H(a) f \sum_{n=1}^\infty \, [D_1(a)]^n \theta_R^{(n)}(\vk)
-\epsilon_{\rm v} \H(a) f \sum_{n=1}^\infty \, [D_1(a)]^{n+2} \hat{\theta}_R^{(n)}(\vk) \nonumber 
\\ & \qquad -\epsilon \H(a) f \sum_{n=1}^\infty \, [D_1(a)]^{n+2} \tilde{\theta_R}^{(n)}(\vk)\ ,
\end{align}
and
\bea
\hat{\delta}^{(n)}(\vk) &=& \int \! \frac{d^3 \vq_1}{(2\pi)^3} \cdots
\int \! \frac{d^3 \vq_n}{(2\pi)^3}
(2\pi)^3 \delta_{\rm D}(\vk-\vq_{1\cdots n})
\hat{F}_n(\vq_1,\dots,\vq_n)
\delta(\vq_1)\cdots\delta(\vq_n)\ , \\
\hat{\theta}_R^{(n)}(\vk) &=& \int \! \frac{d^3 \vq_1}{(2\pi)^3} \cdots
\int \! \frac{d^3 \vq_n}{(2\pi)^3}
(2\pi)^3 \delta_{\rm D}(\vk-\vq_{1\cdots n})
\hat{G}_n(\vq_1,\dots,\vq_n)
\delta(\vq_1)\cdots\delta(\vq_n)\ .
\label{eq:thetakansatz2}
\eea
Here the $\tilde \delta$ and $\tilde \theta_R$ are identical to the ones we determined in the previous section.  The quantities with hats correspond to the new contributions from the velocity counterterm. 

If we solve these equations using this ansatz, it is easy to see that there the new terms in the continuity equation and the Euler equation do not modify the $\delta$ correlation functions at linear order in $\epsilon_{\rm V}$.  Specifically, using this ansatz with the redefinition in (\ref{eq:vsub2}) implies that
\beq
\hat G_n =  -\frac{ k^2 F_n}{D_1(a)^2 \H f}  \int \frac{da'}{a'\H(a')} K_{\rm v}(a,a') \left[\frac{D_1(a')}{D_1(a)}\right]^{n} \ ,
\eeq
while leaving $\hat F = 0$.  However, when we made the definition of $\gammai \to \tildegammai$, we shifted the coefficients in $\tilde F$ and these must be compensated by new terms in $\hat F$.  In fact, we can predict the form of these new terms by recalling the definition of the $\bar{c}_n$ coefficients and defining $\bar{b}_n$ by analogy,
\begin{align}
&c_n(a) \equiv \int_{a_{\rm in}}^a
\frac{da'}{a'\H(a')} K(a,a') \frac{D_1(a')^n}{D_1(a)^n}, \qquad
c_n(a)  = \bar c_n (9 D_1(a)^{2} \H^2 f^2)\ ,
\label{eq:cndef2} \\ \nn
&b_n(a) \equiv \int_{a_{\rm in}}^a
\frac{da'}{a'\H(a')} K_{\rm v}(a,a') \frac{D_1(a')^n}{D_1(a)^n}, \qquad
b_n(a)  = \bar b_n (9 D_1(a)^{2} \H f)\ .
\end{align}
By construction, $\tildegammai$ corresponds to shifting $c_n (a) \to c_n(a) - (a \H \partial_a +\H + n ) b_n(a)$ or 
\beq
\bar c_n \to \bar c_n - (n+\tfrac{5}{2})\bar b_n \ ,
\eeq 
where we have used $f^2 \simeq \Omega_m$ (which is the same approximation that allows us to use this ansatz).  Therefore, we should find that $\hat F$ is equivalent to the same shift, but with the opposite sign.  We will now check this explicitly.  

\vskip 8pt
Collecting terms of order $\epsilon_{\rm v}^1$, we can find recurrence relations for $\hat{F}_n$ and $\hat{G}_n$:
\begin{align}
\nn
\hat{F}_n(\vq_1,\dots,\vq_n) &= \frac{1}{(n+1)(n+\frac{7}{2})}
\sum_{m=1}^{n-1} \lb (n+\frac{5}{2}) \alpha(\vk_1,\vk_2)
\hat{\mathcal{A}}_m(\vq_1,\dots,\vq_n) + \beta(\vk_1,\vk_2)
\hat{\mathcal{B}}_m(\vq_1,\dots,\vq_n)\rb  \\
&\qquad\qquad\qquad\qquad + \frac{9(n+5/2)}{(n+1)(n+\frac{7}{2})} 
\hat{\mathcal{S}}_n(\vq_1,\dots,\vq_n)\ , \\
\nn
\hat{G}_n(\vq_1,\dots,\vq_n) &= \frac{1}{(n+1)(n+\frac{7}{2})}
\sum_{m=1}^{n-1} \lb \frac{3}{2} \alpha(\vk_1,\vk_2)
\hat{\mathcal{A}}_m(\vq_1,\dots,\vq_n) + (n+2) \beta(\vk_1,\vk_2)
\hat{\mathcal{B}}_m(\vq_1,\dots,\vq_n) \rb \\
&\qquad\qquad\qquad\qquad + \frac{9}{(n+1)(n+\frac{7}{2})} \cdot
\frac{3}{2} \hat{\mathcal{S}}_n(\vq_1,\dots,\vq_n)\ ,
\end{align}
where
\bea
&&\hat{\mathcal{A}}_m(\vq_1,\cdots\,\vq_n) \equiv  \\ \nonumber
&& \qquad\hat{G}_m(\vq_1,\dots,\vq_m)  F_{n-m}(\vq_{m+1},\dots,\vq_n)
+ G_m(\vq_1,\dots,\vq_m)  \hat{F}_{n-m}(\vq_{m+1},\dots,\vq_n) \ , \\
&&\hat{\mathcal{B}}_m(\vq_1,\cdots\,\vq_n) \equiv  \\ \nonumber
&&\qquad \hat{G}_m(\vq_1,\dots,\vq_m)  G_{n-m}(\vq_{m+1},\dots,\vq_n)
+ G_m(\vq_1,\dots,\vq_m)  \hat{G}_{n-m}(\vq_{m+1},\dots,\vq_n) \ .
\eea
and
\begin{align}
\nn
\hat{\mathcal{S}}_n(\vq_1,\dots,\vq_n) &= \bar b_n { k^2} F_n(\vq_1,\dots,\vq_n)
+ \sum_{i=1}^{n-1} \bar{b}_{n-i} \, \vk\cdot\vk_2
F_i(\vq_1,\dots,\vq_i) F_{n-i}(\vq_{i+1},\dots,\vq_n) \\
\nn
&\quad + k^2 \sum_{i=1}^{n-1} \frac{\vk_1 \cdot \vk_2}{k_2^2} \frac{1}{n-i} (\bar b_i-\bar b_n)
F_i(\vq_1,\dots,\vq_i) G_{n-i}(\vq_{i+1},\dots,\vq_n) \\
\nn
&\quad+  k^2
\sum_{i=1}^{n-2} \sum_{j=1}^{n-2} \frac{(\vk_1\cdot\vk_2) (\vk_2\cdot\vk_3)}{k_2^2 k_3^2}
 \lb \frac{1}{n-i} \bar{b}_i - \frac{1}{j(n-i-j)} \bar{b}_{i+j} + \frac{1}{(n-i)(n-i-j)} \bar{b}_n \rb \\
\nn
&\qquad\qquad\qquad\qquad\qquad\qquad \times
F_i(\vq_1,\dots,\vq_i) G_j(\vq_{i+1},\dots,\vq_{i+j}) G_{n-i-j}(\vq_{i+j+1},\dots,\vq_n) \\
\nn
&\quad+  k^2
\sum_{i=1}^{n-2} \sum_{j=1}^{n-2} \frac{(\vk_1\cdot\vk_2) (\vk_1\cdot\vk_3)}{k_2^2 k_3^2}
 \frac{1}{2j(n-i-j)} \lb \bar{b}_i - \bar{b}_{i+j} - \bar{b}_{n-j} + \bar{b}_n \rb \\
\nn
&\qquad\qquad\qquad\qquad\qquad\qquad \times
F_i(\vq_1,\dots,\vq_i) G_j(\vq_{i+1},\dots,\vq_{i+j}) G_{n-i-j}(\vq_{i+j+1},\dots,\vq_n) \\
\nn
&\quad+ 
\sum_{i=1}^{n-2} \sum_{j=1}^{n-2}   \vk\cdot\vk_2 \frac{\vk_2\cdot\vk_3}{k_3^2}
\frac{1}{n-i-j} (\bar{b}_j-\bar{b}_{n-i}) \\
&\qquad\qquad\qquad\qquad\qquad\qquad \times
F_i(\vq_1,\dots,\vq_i) F_j(\vq_{i+1},\dots,\vq_{i+j}) G_{n-i-j}(\vq_{i+j+1},\dots,\vq_n)\ .
\label{eq:snequation2}
\end{align}

When both the $\gammai$ and $v^i$ counterterms are included, the corrections to $P_{\delta\delta}(k)$ will only involve $\tilde{F}_n$ and $\hat{F}_n$ in the combination $(\tilde{F}_n+\hat{F}_n)$. Moreover, by recalling the definition of the $\bar{c}_n$ and $\bar{b}_n$ coefficients in (\ref{eq:cndef2}),
we find after explicit calculations of the first three kernels that $(\tilde{F}_n+\hat{F}_n)$ becomes $\tilde{F}_n$ under the substitution
\beq
\bar{c}_n \to \bar{c}_n + \lp\frac{5}{2}+n\rp \bar{b}_n\ .
\label{eq:cnsub}
\eeq
This is precisely the form need to cancel the change in $\bar c_n$ from the redefinition of $ K \to \tilde K$.  We conclude that calculating with $\theta_R$ or $\theta$ will have no impact on the $\delta$ correlation functions (when using this ansatz).



\begin{thebibliography}{10}



\bibitem{Baumann:2010tm} 
  D.~Baumann, A.~Nicolis, L.~Senatore and M.~Zaldarriaga,
  ``Cosmological Non-Linearities as an Effective Fluid,''
  JCAP {\bf 1207}, 051 (2012)
  [\href{http://arxiv.org/abs/1004.2488}{{\tt arXiv:1004.2488}}] [astro-ph.CO].
  

\bibitem{Carrasco:2012cv} 
  J.~J.~M.~Carrasco, M.~P.~Hertzberg and L.~Senatore,
  ``The Effective Field Theory of Cosmological Large Scale Structures,''
  JHEP {\bf 1209}, 082 (2012)
 [\href{http://arxiv.org/abs/1206.2926}{{\tt arXiv:1206.2926}}] [astro-ph.CO].



\bibitem{Bernardeau:2001qr} 
  F.~Bernardeau, S.~Colombi, E.~Gaztanaga and R.~Scoccimarro,
  ``Large scale structure of the universe and cosmological perturbation theory,''
  Phys.\ Rept.\  {\bf 367}, 1 (2002)
  [\href{http://arxiv.org/abs/astro-ph/0112551}{{\tt astro-ph/0112551}}].
  

  
\bibitem{Crocce:2005xy}
  M.~Crocce and R.~Scoccimarro,
  ``Renormalized cosmological perturbation theory,''
  Phys.\ Rev.\ D {\bf 73} (2006) 063519
    [\href{http://arxiv.org/abs/astro-ph/0509418}{{\tt astro-ph/0509418}}].

\bibitem{Carlson:2009it} 
  J.~Carlson, M.~White and N.~Padmanabhan,
  ``A critical look at cosmological perturbation theory techniques,''
  Phys.\ Rev.\ D {\bf 80}, 043531 (2009)
  [\href{http://arxiv.org/abs/0905.0479}{{\tt arXiv:0905.0479}}] [astro-ph.CO].




\bibitem{senatoretalk}
L.~Senatore, talks at ICTP `Workshop on Large Scale Structures' Aug.~2012~
[\href{http://cdsagenda5.ictp.trieste.it/full_display.php?ida=a11314}{video}] \\
and at CERN Theory Institute `Theoretical Methods for non-linear cosmology' Sept.~2012~
[\href{http://indico.cern.ch/conferenceTimeTable.py?confId=198793#20120903}{slides}]   .



\bibitem{Pajer:2013jj}
  E.~Pajer and M.~Zaldarriaga,
  ``On the Renormalization of the Effective Field Theory of Large Scale Structures,''
  [\href{http://arxiv.org/abs/1301.7182}{{\tt arXiv:1301.7182}}] [astro-ph.CO].



\bibitem{Carrasco:2013sva}
  J.~J.~M.~Carrasco, S.~Foreman, D.~Green and L.~Senatore,
  ``The 2-loop matter power spectrum and the IR-safe integrand,''
   [\href{http://arxiv.org/abs/1304.4946}{{\tt arXiv:1304.4946}}] [astro-ph.CO].


\bibitem{Fitzpatrick:2009ci}
  A.~L.~Fitzpatrick, L.~Senatore and M.~Zaldarriaga,
  ``Contributions to the Dark Matter 3-Pt Function from the Radiation Era,''
  JCAP {\bf 1005} (2010) 004
     [\href{http://arxiv.org/abs/0902.2814}{{\tt arXiv:0902.2814}}] [astro-ph.CO].
  
  
\bibitem{Lewis:1999bs} 
  A.~Lewis, A.~Challinor and A.~Lasenby,
  ``Efficient computation of CMB anisotropies in closed FRW models,''
  Astrophys.\ J.\  {\bf 538}, 473 (2000)
 [\href{http://arxiv.org/abs/astro-ph/9911177}{{\tt astro-ph/9911177}}].
   
   
\bibitem{Jain:1995kx}
  B.~Jain and E.~Bertschinger,
  ``Self-similar evolution of cosmological density fluctuations,''
  Astrophys.\ J.\  {\bf 456} (1996) 43
  [\href{http://arxiv.org/abs/astro-ph/9503025}{{\tt astro-ph/9503025}}].

\bibitem{Scoccimarro:1995if}
  R.~Scoccimarro and J.~Frieman,
  ``Loop corrections in nonlinear cosmological perturbation theory,''
  Astrophys.\ J.\ Suppl.\  {\bf 105} (1996) 37
  [\href{http://arxiv.org/abs/astro-ph/9509047}{{\tt astro-ph/9509047}}].

\bibitem{Peloso:2013zw} 
  M.~Peloso and M.~Pietroni,
  ``Galilean invariance and the consistency relation for the nonlinear squeezed bispectrum of large scale structure,''
   [\href{http://arxiv.org/abs/1302.0223}{{\tt arXiv:1302.0223}}]  [astro-ph.CO].
   
   
\bibitem{Kehagias:2013yd} 
  A.~Kehagias and A.~Riotto,
  ``Symmetries and Consistency Relations in the Large Scale Structure of the Universe,''
  Nucl.\ Phys.\ B {\bf 873}, 514 (2013)
  [\href{http://arxiv.org/abs/1302.0130}{{\tt arXiv:1302.0130}}] [astro-ph.CO].
   
\bibitem{Bernardeau:2011vy}
  F.~Bernardeau, N.~Van de Rijt and F.~Vernizzi,
  ``Resummed propagators in multi-component cosmic fluids with the eikonal approximation,''
  Phys.\ Rev.\ D {\bf 85} (2012) 063509
   [\href{http://arxiv.org/abs/1109.3400}{{\tt arXiv:1109.3400}}]  [astro-ph.CO].

\bibitem{Anselmi:2012cn}
  S.~Anselmi and M.~Pietroni,
  ``Nonlinear Power Spectrum from Resummed Perturbation Theory: a Leap Beyond the BAO Scale,''
  JCAP {\bf 1212} (2012) 013
     [\href{http://arxiv.org/abs/1205.2235}{{\tt arXiv:1205.2235}}]  [astro-ph.CO].



\bibitem{Blas:2013bpa}
  D.~Blas, M.~Garny and T.~Konstandin,
  ``On the non-linear scale of cosmological perturbation theory,''
     [\href{http://arxiv.org/abs/1304.1546}{{\tt arXiv:1304.1546}}]  [astro-ph.CO].

   
   
\bibitem{Goroff:1986ep}
  M.~H.~Goroff, B.~Grinstein, S.~J.~Rey, and M.~B.~Wise,
  ``Coupling of Modes of Cosmological Mass Density Fluctuations,''
  Astrophys.\ J.\  {\bf 311} (1986) 6.

   
\bibitem{Heitmann:2008eq} 
  K.~Heitmann, M.~White, C.~Wagner, S.~Habib and D.~Higdon,
  ``The Coyote Universe I: Precision Determination of the Nonlinear Matter Power Spectrum,''
  Astrophys.\ J.\  {\bf 715}, 104 (2010)
  [\href{http://arxiv.org/abs/0812.1052}{{\tt arXiv:0812.1052}}]  [astro-ph].

\bibitem{Heitmann:2009cu} 
  K.~Heitmann, D.~Higdon, M.~White, S.~Habib, B.~J.~Williams and C.~Wagner,
  ``The Coyote Universe II: Cosmological Models and Precision Emulation of the Nonlinear Matter Power Spectrum,''
  Astrophys.\ J.\  {\bf 705}, 156 (2009)
  [\href{http://arxiv.org/abs/0902.0429}{{\tt arXiv:0902.0429}}]  [astro-ph.CO].

\bibitem{Lawrence:2009uk} 
  E.~Lawrence, K.~Heitmann, M.~White, D.~Higdon, C.~Wagner, S.~Habib and B.~Williams,
  ``The Coyote Universe III: Simulation Suite and Precision Emulator for the Nonlinear Matter Power Spectrum,''
  Astrophys.\ J.\  {\bf 713}, 1322 (2010)
    [\href{http://arxiv.org/abs/0912.4490}{{\tt arXiv:0912.4490}}]  [astro-ph.CO].

\bibitem{Heitmann:2013bra} 
  K.~Heitmann, E.~Lawrence, J.~Kwan, S.~Habib and D.~Higdon,
  ``The Coyote Universe Extended: Precision Emulation of the Matter Power Spectrum,''
   [\href{http://arxiv.org/abs/1304.7849}{{\tt arXiv:1304.7849}}]  [astro-ph.CO].
  
\bibitem{CUBA} 
  T.~Hahn,
  ``CUBA: A Library for multidimensional numerical integration,''
  Comput.\ Phys.\ Commun.\  {\bf 168}, 78 (2005)
  [\href{http://arxiv.org/abs/hep-ph/0404043v2}{{\tt hep-ph/0404043v2}}].
  
     
     
\bibitem{Mercolli:2013bsa}
  L.~Mercolli and E.~Pajer,
  ``On the Velocity in the Effective Field Theory of Large Scale Structures,''
  [\href{http://arxiv.org/abs/1307.3220v1}{{\tt arXiv:1307.3220v1}}]  [astro-ph.CO].
     
     
\bibitem{Okumura:2011pb} 
  T.~Okumura, U.~Seljak, P.~McDonald and V.~Desjacques,
  ``Distribution function approach to redshift space distortions. Part II: N-body simulations,''
  JCAP {\bf 1202}, 010 (2012)
 [\href{http://arxiv.org/abs/1109.1609}{{\tt arXiv:1109.1609}}]  [astro-ph.CO].
  
\bibitem{Pueblas:2008uv}
  S.~Pueblas and R.~Scoccimarro,
  ``Generation of Vorticity and Velocity Dispersion by Orbit Crossing,''
  Phys.\ Rev.\ D {\bf 80} (2009) 043504
  [\href{http://arxiv.org/abs/0809.4606}{{\tt arXiv:0809.4606}}] [astro-ph].



\bibitem{Creminelli:2005hu}
  P.~Creminelli, A.~Nicolis, L.~Senatore, M.~Tegmark and M.~Zaldarriaga,
  ``Limits on non-gaussianities from wmap data,''
  JCAP {\bf 0605} (2006) 004
    [\href{http://arxiv.org/abs/astro-ph/0509029}{{\tt astro-ph/0509029}}].

\bibitem{Senatore:2009gt}
  L.~Senatore, K.~M.~Smith and M.~Zaldarriaga,
  ``Non-Gaussianities in Single Field Inflation and their Optimal Limits from the WMAP 5-year Data,''
  JCAP {\bf 1001} (2010) 028
     [\href{http://arxiv.org/abs/0905.3746}{{\tt arXiv:0905.3746}}]  [astro-ph.CO].

\bibitem{Sefusatti:2007ih}
  E.~Sefusatti and E.~Komatsu,
  ``The bispectrum of galaxies from high-redshift galaxy surveys: Primordial non-Gaussianity and non-linear galaxy bias,''
  Phys.\ Rev.\ D {\bf 76} (2007) 083004
  [\href{http://arxiv.org/abs/0705.0343}{{\tt arXiv:0705.0343}}]  [astro-ph]. \\
  E.~Sefusatti, M.~Liguori, A.~P.~S.~Yadav, M.~G.~Jackson and E.~Pajer,
  ``Constraining Running Non-Gaussianity,''
  JCAP {\bf 0912} (2009) 022
    [\href{http://arxiv.org/abs/0906.0232}{{\tt arXiv:0906.0232}}]  [astro-ph.CO]. \\
  E. Sefusatti, private communications.


\bibitem{Ade:2013ydc}
  P.~A.~R.~Ade {\it et al.}  [Planck Collaboration],
  ``Planck 2013 Results. XXIV. Constraints on primordial non-Gaussianity,''
  [\href{http://arxiv.org/abs/1303.5084}{{\tt arXiv:1303.5084}}]  [astro-ph.CO].


\bibitem{Cheung:2007st}
  C.~Cheung, P.~Creminelli, A.~L.~Fitzpatrick, J.~Kaplan and L.~Senatore,
  ``The Effective Field Theory of Inflation,''
  JHEP {\bf 0803} (2008) 014
  [\href{http://arxiv.org/abs/0709.0293}{{\tt arXiv:0709.0293}}]  [hep-th].




   
\end{thebibliography}
\end{document}